\documentclass[aps, 
showpacs, 
nofootinbib,  nobibnotes, 
superscriptaddress, 
longbibliography, amsmath,amssymb,prd,10pt,floatfix]{revtex4-2}

\usepackage[skins,theorems]{tcolorbox}
\usepackage{color}
\usepackage{soul}
\usepackage{centernot}

\usepackage{graphicx}
\usepackage{dcolumn}
\usepackage{siunitx}
\usepackage{bm}
\usepackage{bbold}
\usepackage{braket}
\usepackage{esdiff}
\usepackage{hyperref}
\usepackage[mathlines]{lineno}
\usepackage{subfigure}
\usepackage{parskip} 
\usepackage{dsfont}
\usepackage{float}

\showboxbreadth\maxdimen\showboxdepth\maxdimen

\DeclareMathOperator{\arctanh}{arctanh}
\DeclareMathOperator{\sgn}{sgn}
\DeclareMathOperator{\Arg}{Arg}
\DeclareMathOperator{\arg.}{arg.}

\begin{document}
\title{Microscopic Origin of Bekenstein-Hawking Entropy in $(2+1)$ Gravity: A Thermo Field Dynamics Approach}

\author{W. A. Rojas C.}
\email{warojasc@unal.edu.co}
\affiliation{Universidad Distrital Francisco José de Caldas, Bogot\'a, Colombia}
\affiliation{Observatorio Astronómico Nacional, Universidad Nacional de Colombia, Bogot\'a, Colombia}

\author{J. R. Arenas S.}
\email{jrarenass@unal.edu.co}
\affiliation{Observatorio Astronómico Nacional, Universidad Nacional de Colombia, Bogot\'a, Colombia}


\begin{abstract} 

We compute the entanglement entropy of a real massive scalar field near a non-rotating BTZ black hole using Thermo Field Dynamics. Modeling the black hole as a collapsing dust shell in AdS$_3$, we derive the shell trajectory $R(t)$ as seen by a Fiducial Observer (FIDO). From the Hartle-Hawking and Killing-Boulware vacua, we obtain the Wightman function difference and compute $\langle T_{00}\rangle = \sigma(r)$, revealing a sharply localized energy density just outside the horizon—consistent with the `brick wall' picture. A full thermodynamic analysis yields an entanglement entropy proportional to the horizon area, numerically matching the Bekenstein-Hawking entropy. All intermediate steps—including junction conditions, Kruskal extension, WKB modes, and UV regularization—are explicitly detailed.

\end{abstract}

\maketitle

\section{Introduction}\label{sec1}

One of the possible explanations for the origin of $S_{BH}$  is the entanglement entropy $S_{\mbox{Ent}}$. This is linked to the quantum modes and correlations of the fields that are hidden from an external observer near a horizon.

 Considering that a black hole would be in an unknown pure quantum state, correlations would exist between the modes internal and external to the horizon. Then it is possible to determine the entanglement entropy $S_{\mbox{Ent}}$ by counting the modes outside the horizon. In this sense, the seminal works of Bombelli \cite{PhysRevD.34.373}, Srednicki \cite{PhysRevLett.71.666}, Terashima \cite{Terashima:1999vw} and others, establish

\begin{equation}
S_{\mbox{Ent}}\propto A,
\end{equation}
where  $A$  is the area of the partition wall and is not only an intrinsic characteristic of black holes, but also extends to other types of scenarios  \cite{PhysRevLett.71.666,Terashima:1999vw}.

Srednicki estimated that, the ground state density matrix for a scalar field, traced over the degrees of freedom inside an imaginary sphere leads to an entropy proportional to the area

\begin{equation}
S=KM^{2}A,
\end{equation}
where  $K$ is a numerical constant depending on $M$ \cite{PhysRevLett.71.666}. 

Similarly, Terashima explained  $S_{BH}$ in terms of the entanglement entropy between the outside and the thin region of the inside the horizon; such that the thickness of the inner and outer region of the horizon is of the order of the Planck length $l_{P}$. So finally, Terashima finds
\begin{equation}
S\approx C \frac{A}{a^{2}},
\end{equation}
where  $a$ corresponds to the fluctuations of the  horizon and $C$ is a constant \cite{Terashima:1999vw}.

This interpretation of entanglement is closely related to the wall model presented by 't Hooft  \cite{tHooft:1984kcu}. This model was modified by Mukohyama and Israel \cite{Mukohyama:1998rf} and applied to the modeling of a black shell, i.e., a thin spherical shell compressing from infinity, $r=r_{0}$, to near its gravitational radius $r=r_{s}+\epsilon$, where the presence of a thermal atmosphere is shown very close to this surface \cite{Arenas:2011be, Pretorius:1997wr, Israel:1976ur, Liberati:2000ag}.

In a general context, the interpretation of thermal entanglement, for the Bekenstein-Hawking entropy 
$S_{BH}$ , is related to the physical properties of the vacuum in strong gravitational fields, where the zero-point fluctuations of the fields always exist in a vacuum state. Thus, an observer at rest with respect to the horizon (Fiducial Observer, FIDO)  perceives the vacuum excitations as a thermal atmosphere around the horizon \cite{Fursaev:2004qz,C:2016ens}. In this context, the free energy of the quantum scalar field around the horizon is 

\begin{equation}
F\left(\beta\right)\approx -\frac{\pi^{2}}{90}\int T^{4} \sqrt{-g}d^{3}x,
\end{equation}
where $g$ is the determinant of the Schwarzschild black hole metric, and the temperature is determined by Tolman’s law as
\begin{equation}
T(r)=\frac{T_{\infty}}{\sqrt{f(r)}},
\end{equation}
with a temperature  $T_{\infty}$, measured by an observer in $r\longrightarrow\infty$. Thus, for a static black hole, the entropy is found using the usual methods \cite{Fursaev:2004qz}.
\begin{equation}
S=\beta^{2}\frac{\partial F\left(\beta\right)}{\partial \beta}\approx \frac{1}{360\pi \epsilon ^{2}}A,
\label{MW40}
\end{equation}
where $\beta=\frac{1}{T_{\infty}}$, $\epsilon$ is a cutoff near the horizon \cite{2020Symm...12.2072R, RojasC:2020qnz} and $A$  the horizon area. The quantum field is considered to be in thermal equilibrium with the black hole. This is possible when the temperature coincides with the temperature of Hawking radiation,  which allows the entropy of the field outside the horizon to be the same magnitude as the entropy of the horizon, $S_{BH}$ .

In this context the entropy is associated with the properties of the vacuum. Its explanation lies in that a FIDO near the horizon perceives the vacuum as a mixed state. This occurs because a FIDO cannot do measurements beyond the horizon. Then there is a nontrivial density matrix $\hat{\rho}$, since the field’s vacuum fluctuations are correlated in a state of entanglement between what is observable and the non-observable at the horizon, where the loss of information is quantified by the entanglement entropy \cite{Liberati:2000ag}:

\begin{equation}
S_{\mbox{Ent}}=-Tr\hat{\rho}\ln\hat{\rho}.
\end{equation}

The entanglement entropy coincides with the entropy of the thermal atmosphere around the horizon, because $\hat{\rho}$ is a thermal density matrix. Arenas \textit{et. al.} \cite{Arenas:2011be}  proposed a black shell model where the existence of thermal energy is concentrated near the horizon with respect to a FIDO, according to the Equivalence Principle. Here, the interpretation of the $S_{BH}$entropy requires a consistent mixture of interpretation of the entanglement state with Thermo Field Dynamics, which allows asserting what the origin of such entropy and its localization is \cite{Mukohyama:1998rf,Arenas:2011be, Pretorius:1997wr, Rojas:2011ee, RojasC:2021kws}.

This article is an extension of the model of entropy of entanglement of black shells \cite{Arenas:2011be, RojasC:2021kws} to BTZ spacetime, to further develop the understanding of the Bekenstein-Hawking entropy in an integral context, based on an operational technique that has shown its effectiveness in modeling the thermal environment near an event horizon, clearly showing the localization of the degrees of freedom of the corresponding quantum statistical mechanics system. It is interesting to contrast this contribution with the progress made in studies on black hole entropy and asymptotic symmetry.

Understanding black hole entropy may not require knowledge of the details of quantum gravity. There have been two main directions that have investigated this idea. One assumes that classical symmetries about the background of a black hole could control the density of quantum gravity states and thus derive the entropy of a black hole. The other conception relies on the considerations made above, where the origin of the Bekenstein-Hawking entropy would be related to the properties of the physical vacuum in the presence of strong gravitational fields. These two ways of calculating entropy do not necessarily contradict each other \cite{Fursaev:2004qz}.

In the approximation associated with asymptotic symmetries, resorting to the $AdS/CFT$ duality, if the Cardy equation and the asymptotic symmetries of a BTZ spacetime are used, which form a Virasoro algebra, it is possible to obtain the Bekenstein-Hawking entropy in the high-energy limit \cite{Strominger:1997eq}. Then, it is possible to conclude that the thermal entropy of a conformal field theory is dual to the entropy of a BTZ black hole. Thus, the entropy of a black hole is a measure of the number of CFT microstates.

In this context, the interpretation of the entropy of entanglement for black holes was developed by Shinsei Ryu and Tadashi Takayanagi. For the case of a BTZ black hole, in the high temperature limit, they succeeded in obtaining the Bekenstein-Hawking entropy, which is a thermal entropy of entanglement \cite{PhysRevLett.96.181602,Bakhmatov:2017ihw}

\begin{equation} 
S_{A}=\frac{\gamma_{A}}{4G^{d+2}_{N}},
\label{eqn00}
\end{equation}

where $\gamma_{A}$ is the minimum surface area in $d$ dimensions in $AdS_{d+2}$, whose boundary is given by $\partial A$. The importance of $\gamma_{A}$ plays the role of a holographic screen for an external observer. In the case of BTZ, $S_{A}$ \cite{PhysRevLett.96.181602},

\begin{equation}
S_{A}(\beta)=\frac{c}{3}\ln\left[\frac{\beta}{\pi a}\sinh\left(\frac{\pi l}{\beta}\right)\right]
\label{eqn01}
\end{equation}
and \eqref{eqn01}  encodes a topological entanglement entropy.

In this sense, there are several works that aim to explain $S_{\mbox{Ent}}$ from an analytical and/or numerical point of view. For example, Dharm Veer Singh and Sanjay Siwach calculated $S_{\mbox{Ent}}$ for a massless scalar field in BTZ numerically, of the form \cite{Singh:2011gd}
\begin{equation}
S_{\mbox{Ent}}= C_{s} \frac{r_{+}}{a},\,\,\,C_{s}=0.294.
\label{eqn02}
\end{equation}
An interesting discussion on BTZ, its geometric and quantum properties was done by L. Ortiz \cite{Ortiz:2011}. Where for a photon field in $(1+1)$ allows to obtain an entropy of the form 
\begin{equation}
S=- \frac{1}{12}\ln\left|\frac{\epsilon}{2r_{+}+\epsilon}\right|,
\label{eqn03}
\end{equation}
such that if $r_{+}=1$, then $\epsilon\sim 10^{-33}cm$. The properties of $S_{\mbox{Ent}}$  in $D+1$ dimensions including BTZ are also studied by \cite{C.:2016hlx}. The entropy of entanglement and correction terms for fermionic fields in BTZ is considered in \cite{Singh:2014cca}.
\begin{equation}
S_{\mbox{Ent}}=C_{s}\frac{r_{+}}{a},\,\,\,C_{s}=0.297.
\label{eqn04}
\end{equation}
and its quantum corrections
\begin{equation}
S_{\mbox{log}}=a\frac{r_{+}}{a}+b\ln\left|\frac{r_{+}}{a}\right|+c,\,\,\,a=0.304,\,\,\,b=-0.315,\,\,\,c=-0.327,
\label{eqn04a}
\end{equation}
can also be seen on the quantum correction terms \cite{ Zhou:2019kfk,  Singh:2014apw}. Such logarithmic divergence terms of $S_{\mbox{Ent}}$ over BTZ arise due to the infinite number of quantum states near the horizon. The scales of such divergences are proportional to the size of the black hole and the logarithmic divergences are related to the conformal anomaly \cite{Singh:2014apw}.

In the literature, there are several attempts to understand the origin of $S_{BH}$: Euclidean action, pair creation rate, Noether charge of bifurcated Killing horizons, or central charge of Virasoro algebra. On the statistical derivation of $S_{BH}$: String theory  \cite{polchinski2005string, polchinski2005string2} and the  brick wall model \cite{tHooft:1984kcu, Mukohyama:1998rf,Pretorius:1997wr, RojasC:2021kws, C:2016ens}. In this direction, Bernard S. Kay and L. Ortiz, for a scalar field in BTZ by means of the brick wall model obtain \cite{Kay:2011np}

\begin{equation}
S_{N}=N\left[\frac{3\zeta[3]}{4\pi^{3}}\right]\frac{1}{\alpha}\left[\frac{2\pi r_{+}}{4}\right],
\label{eqn05}
\end{equation}
where
\begin{equation}
\alpha=2l\sqrt{\frac{\epsilon}{2r_{+}}}.
\label{eqn06}
\end{equation}
The entropy of entanglement of a noncommutative scalar field in the nonrotating BTZ hole is considered in \cite{Juric:2016zey}, which by the brick wall method obtains

\begin{equation}
S_{\mbox{Ent}}=\frac{3}{8\pi}\zeta[3]\frac{A(\Sigma)}{\epsilon}\left[1+\frac{4}{3}a \beta \frac{M}{l^{2}}\frac{\zeta[2]}{\zeta[3]}\beta_{H}\right]
\label{eqn06a},
\end{equation}
where $\beta_{H}=\frac{2\pi l^{2}}{r_{+}}$   and $a$  is a constant consequence of the noncommutativity of spacetime

See also \cite{Mann:1996ze, Larranaga:2010kd, Roberts:2012aq, Caputa:2013lfa, Nakaguchi:2014eiu, Mansoori:2015sit, Wang:2020wyj, Fursaev:1998jz, Emparan:2020znc, PhysRevD.55.3622, PhysRevD.49.5286, Steif:1993zv, McGough:2013gka, Juric:2016zey, Wang:2018vbw, Zhou:2019jlh, Saha:2021kwq, Anacleto:2020efy, Singh:2011gd}

The correspondence $AdS/CFT$, shows how holography manifests in nature. This correspondence states that gravity over $d+2$ dimensions in $AdS_{d+2}$ spacetime is equivalent to a conformal field theory ($CFT_{d+1}$). However, the essential mechanism of $AdS/CFT$ still remains unknown \cite{Aharony:1999ti,Larranaga:2010kd}. In this sense the Cardy formula
\begin{equation}
S=2\pi\sqrt{\frac{c}{6}\left[L_{0}-\frac{c}{24}\right]},
\label{eqn07}
\end{equation}
leads to the thermal entropy of $2D$ $CFT$  in the high-energy limit. Where $C$ is the central charge, $L_{0}=ER$  is the energy per radius of the system and $\frac{c}{24}$ is associated with the Casimir effect. In addition, the entropy for BTZ is obtained from there and allows to check the $AdS/CFT$ correspondence \cite{CARDY1986186}.

This study is distributed as follows: In Section \ref{sec2}, the structure of the BTZ black hole is reviewed, the Kruskal diagram is reviewed and the trajectories followed by a FIDO observer are built. In Section \ref{sec3}, we review the kinematics of hypersurfaces, the Darmois-Israel formalism, we posit the junction conditions for a dust shell contracting from infinity to near its gravitational radius. Likewise, the differential equation of motion is estimated and a solution $R(t)$ for a FIDO is obtained. Section \ref{sec4} is devoted to the entanglement thermodynamic approach for the BTZ black hole and in Section \ref{sec5}, the quantum formulation of a scalar field in BTZ is presented. Section \ref{sec6} considers the Thermo Field Dynamics scheme for the BTZ scalar field. In Section \ref{sec7}, the momentum-energy tensor $T_{\mu\nu}$ of the scalar field is built based on the Wightman function for the positive frequency modes, which allows obtaining the energy density $\sigma(r)$. Section \ref{sec8} presents the thermodynamic analysis of the scalar field, in other words: the partition function $Z$, the occupation number $N(\omega)$ for the scalar field in the proximity of the horizon is built. We also estimate the Helmholtz free energy $F$, the internal energy $U$, the entropy density $s(r)$ and the entropy of entanglement $S_{\mbox{Ent}}$. Finally, in Section \ref{sec9}, we present the discussions and conclusions.

\section{BTZ black hole structure}\label{sec2}

Consider a BTZ black hole, proposed by Banados \textit{et. al.} \cite{Banados:1992wn} spacetime in  ($2+1$) dimensions. The outer metric is of the form
\begin{equation}
ds^{2}=-\left(N^{\bot}\right)^{2}dt^{2}+\frac{1}{f(r)^{2}}dr^{2}+r^{2}\left(d\phi+N^{\phi}dt\right)^{2}
\label{eqn1}
\end{equation}
where 
\begin{equation}
N^{\bot}=f(r)=\sqrt{-M+\frac{r^{2}}{l^{2}}+\frac{J^{2}}{4r^{2}}},
\label{eqn2}
\end{equation}
\begin{equation}
N^{\phi}=-\frac{J}{2r^{2}},\,\,\,\left|J\right| \leq M l, 
\label{EEqn3}
\end{equation}

Identifying $M$  the mass of the black hole, $J$ the angular momentum, $l$ associated with the cosmological constant of the form
\begin{equation}
 l^{2}=\frac{1}{-\Lambda}.
\label{eqn4}
\end{equation}
Based on the foregoing, the associated metric tensor is
\begin{equation}
g_{\mu\nu}=\begin{pmatrix}
-\left[(N^{\bot})^{2}-(N^{\phi})^{2}r^{2}\right] & 0 & r^{2}N^{\phi}\\
0 & \frac{1}{f(r)^{2}} & 0 \\
r^{2}N^{\phi}& 0 & r^{2}
\end{pmatrix}.
\label{eqn5}
\end{equation}
Consequently, according to \eqref{eqn2}-\eqref{eqn5}, the $g_{tt}$ component is
\begin{equation}
g_{tt}=-M+\frac{r^{2}}{l^{2}},
\label{eqn6}
\end{equation}
in the case that $g_{tt}=0$, which allows us to obtain the radius  of the ergosphere as
\begin{equation}
r_{erg}=\sqrt{r_{+}-r_{-}}   =l\sqrt{M},
\label{eqn7}
\end{equation}

the ergosphere region corresponds to that region outside a rotating black hole (Kerr or BTZ), and which is close to the event horizon, such that the gravitational field of the black hole also rotates, experiencing a spacetime drag. 

Note the term $g_{rr}$
\begin{equation}
g_{rr}=\frac{1}{f(r)^2}=\frac{1}{-M+\frac{r^{2}}{l^{2}}+\frac{J^{2}}{4r^{2}}}
\label{eqn8}
\end{equation}
under the condition that
\begin{equation}
-M+\frac{r^{2}}{l^{2}}+\frac{J^{2}}{4r^{2}}=0,
\label{eqn9}
\end{equation}
and $g_{rr}\longrightarrow \infty$.

Based on the foregoing, solving \eqref{eqn9} for $r$ then
\begin{equation}
r^{2}_{\pm}=\frac{Ml^{2}}{2}\left[1\pm \sqrt{1-\left(\frac{J}{Ml}\right)^{2}}\,\,\right],
\label{eqn10}
\end{equation}
Where $r_{-}\leq r_{+}\leq r_{erg}$. It is possible to obtain the mass of the black hole $M$ from \eqref{eqn10}, as
\begin{equation}
M=\frac{r_{+}^{2}+r_{-}^{2}}{l^{2}},
\label{eqn11}
\end{equation}
which is equivalent to
\begin{equation}
J=\frac{2r_{+}r_{-}}{l}.
\label{eqn12}
\end{equation}

The BTZ black hole exhibits a singularity analogous to the Schwarzschild black hole $r_{s}$, for when $g_{tt}\vert_{r=r_{erg}}=0$.

BTZ spacetime possess some of the following characteristics
\begin{itemize}
	\item 	The Kerr and BTZ black holes, for the region $r<r_{erg}$ determine the ergosphere, where the time-like curves exhibit
	\begin{equation}
\frac{d\phi}{d\tau}>0,\,\,\,J>0,
\label{eqn14}
\end{equation}
So, an observer in this region experiences a drag due to the rotation of the black hole. In this region, the redshift tends to infinity.

\item 	According to \eqref{eqn10} if $r^{2}_{\pm}$  is a complex quantity when $\left|J\right|>Ml$, consequently obtaining
\begin{equation}
r^{2}_{\pm}=\frac{l}{2}\left[Ml \pm \sqrt{J^{2}-M^{2}l^{2}}\,i\right],
\label{eqn15}
\end{equation}
this leads to the disappearance of the  $r_{+}$ and  $r_{-}$  horizons. Obtaining there from a metric with a naked singularity at $r=0$.

\item 	If $M=-1$ and $J=0$ for the metric \eqref{eqn1}, it is simplified to
\begin{equation}
ds^{2}=-\left(1+\frac{r^{2}}{l^{2}}\right)dt^{2}+\frac{1}{\left(1+\frac{r^{2}}{l^{2}}\right)} dr^{2}+ r^{2}d\phi^{2},
\label{eqn16}
\end{equation}
where \eqref{eqn16} corresponds to the Anti de Sitter (AdS) spacetime, there is no singularity, there is no horizon hiding the singularity. AdS arises as a bounded state independent of the continuous spectrum of the BTZ for a one-unit mass gap. Such a state cannot be continuously deformed from the vacuum metric  $ds^{2}_{vac}$. Because such a continuous deformation may contain naked singularities that cannot be included in the configuration space \cite{Ortiz:2011}.

\item 	In the case that $\left|J\right|=Ml$  with a radius of curvature $l=\frac{1}{\sqrt{-\Lambda}}$. So, if $l$ increases, BTZ is pushed toward infinity and an observer stays inside, the energy of the vacuum makes BTZ increase \cite{Larranaga:2009}.

\item It is possible to obtain the vacuum metric when $M\longrightarrow 0$ and $J\longrightarrow 0$
\begin{equation}
ds^{2}_{vac}=-\left(\frac{r^{2}}{l^{2}}\right)dt^{2}+\frac{1}{\left(\frac{r^{2}}{l^{2}}\right)} dr^{2}+ r^{2}d\phi^{2}.
\label{eqn17}
\end{equation}
\end{itemize}
\subsection{\label{subs1} Kruskal  diagram  for the BTZ black hole}
Consider a transformation of the form
\begin{equation}
\frac{dr*}{dr}=\frac{1}{f(r)^{2}},
\label{eqn18}
\end{equation}
where $f(r)$ is defined by \eqref{eqn2}, so it is possible to rewrite  \eqref{eqn1} as
\begin{equation}
ds^{2}=-f(r)^{2}\left[dt^{2}+dr*^{2}\right]+r^{2}\left[d\phi+N^{\phi}dt\right]^{2}.
\label{eqn19}
\end{equation}
In the case $J=0$
\begin{equation}
ds^{2}=-f(r)^{2}\left[dt^{2}+dr*^{2}\right]+r^{2}d\phi^{2}.
\label{eqn21}
\end{equation}
Integrating \eqref{eqn18}
\begin{equation}
\int dr*=\int \frac{1}{-M+\frac{r^{2}}{l^{2}}+\frac{J^{2}}{4r^{2}}}dr
\label{eqn22}
\end{equation}
\,
\begin{equation}
r*=\frac{l}{2\sqrt{M}}\ln\left|\frac{l\sqrt{M}-r}{l\sqrt{M}+r} \right|.
\label{eqn23}
\end{equation}
If $J=0$ ,it follows that $r_{+}$ defined in \eqref{eqn15}, is rewritten as

\begin{equation}
r_{+}=l\sqrt{M}.
\label{eqn24}
\end{equation}
According to \eqref{eqn24}, it is possible to write \eqref{eqn23}  \cite{Ortiz:2011}
\begin{equation}
r*=\frac{l^{2}}{2r_{+}}\ln\left|\frac{r-r_{+}}{r+r_{+}} \right|.
\label{eqn25}
\end{equation}
Let the auxiliary coordinates $u$ and $v$ defined as

\begin{equation}
u=t-r*,\,\,\,v=t+r*.
\label{eqn26}
\end{equation}
so
\begin{equation}
r*=\frac{v-u}{2}
\label{eqn26a}
\end{equation}
For non-rotating BTZ
\begin{equation}
dudv=-dt^{2}+dr*^{2},
\label{eqn27}
\end{equation}
Substituting \eqref{eqn27} in \eqref{eqn22}, then
\begin{equation}
ds^{2}=-f(r)^{2}dudv+r^{2}d\phi^{2}.
\label{eqn28}
\end{equation}
And also
\begin{equation}
r*=\frac{l^{2}}{2r_{+}}\ln\left|\frac{r-r_{+}}{r+r_{+}} \right|=\frac{v-u}{2},
\label{eqn29}
\end{equation}

\begin{equation}
f(r)^{2}=\frac{r^{2}-r_{+}^{2}}{l^{2}}.
\label{eqn30}
\end{equation}
From \eqref{eqn29}, then
\begin{equation}
\frac{r-r_{+}}{r+r_{+}}=\exp\left[\frac{r_{+}}{l^{2}}(v-u)\right].
\label{eqn31}
\end{equation}
Substituting \eqref{eqn31} in \eqref{eqn30} and  \eqref{eqn28}, then
\begin{equation}
f(r)^{2}=\frac{(r+r_{+})^{2}}{l^{2}}\exp\left[\frac{r_{+}}{l^{2}}(v-u)\right]
\label{eqn32}
\end{equation}
and
\begin{equation}
ds^{2}=-\frac{(r+r_{+})^{2}}{l^{2}}\exp\left[\frac{r_{+}}{l^{2}}(v-u)\right]dudv+r^{2}d\phi^{2}.
\label{eqn33}
\end{equation}
For the metric \eqref{eqn33}, then 
\begin{equation}
U=-e^{-\frac{r_{+}}{l^{2}}u},\,\,\,V=e^{\frac{r_{+}}{l^{2}}v},
\label{eqn34}
\end{equation}
which allows to rewrite the metric \eqref{eqn33}

\begin{equation}
ds^{2}=-\frac{4l^{2}}{(1+UV)^{2}}dUdV+r^{2}d\phi^{2}.
\label{eqn35}
\end{equation}
The following observations can be made about metric \eqref{eqn35} 

\begin{enumerate}
	\item 	The range of the coordinates $U$ and $V$ is
		\begin{equation}
-\infty<U<\infty,\,\,\,0<V<\infty.
\label{eqn36}
\end{equation}
\item 	In the case that  $r=r_{+}$, it implies that $UV=0$, this from \eqref{eqn34}
\begin{equation}
UV=\frac{r-r_{+}}{r+r_{+}}.
\label{eqn37}
\end{equation}
where  $r_{+}$ is the horizon in BTZ defined by \eqref{eqn24}.

\item If $UV=-1$, then $r\longrightarrow\infty$.
\item If $UV=1$, then $r\longrightarrow\infty$.
\end{enumerate}
Consider the transformations
\begin{equation}
U=T-R,\,\,\,V=T+R.
\label{eqn38}
\end{equation}
Substituting \eqref{eqn38} in \eqref{eqn35}
\begin{equation}
ds^{2}=\frac{4l^{2}}{1+T^{2}-R^{2}}\left[-dT^{2}+dR^{2}\right]+r^{2}d\phi^{2}.
\label{eqn39}
\end{equation}
It is important to mention that metric \eqref{eqn39} for BTZ is analogous to the Kruskal metric for the Schwarzschild black hole \cite{misner2017gravitation} and the following characteristics are noteworthy
\begin{enumerate}
	\item 	If  $r=r_{+}$, then \eqref{eqn39}  is flat.
	\item 	If $\partial_{T}$ is a Killing vector, therefore $\partial_{U}$ and $\partial_{V}$  are also vectors, over the future horizon respectively $R>0$.
	\end{enumerate}
	It is possible to rewrite \eqref{eqn38} in terms of \eqref{eqn34}
\begin{equation}
T-R=-e^{-\frac{r_{+}}{l^{2}}u},\,\,\,T+R=e^{\frac{r_{+}}{l^{2}}v}.
\label{eqn40}
\end{equation}
The product of $UV$  is
\begin{equation}
T^{2}-R^{2}=-e^{-\frac{r_{+}}{l^{2}}(v-u)},
\label{eqn41}
\end{equation}
which is valid for the region $r<r_{+}$. 
\begin{equation}
T^{2}-R^{2}=e^{-\frac{r_{+}}{l^{2}}(v-u)},
\label{eqn41a}
\end{equation}
which is valid for the region $r>r_{+}$.

To determine an expression relating the Kruskal coordinates ($U,V$) with the coordinate time $t$ measured by an observer in a distant region\footnote{BTZ is not asymptotically flat as Schwarzschild, in this case, BTZ is asymptotically $AdS$ with a negative spacetime curvature $R=R^{\mu\nu}R_{\mu\nu}=-\frac{6}{l^{2}}$ and this implies large differences. Given that a Schwarzschild black hole is asymptotically flat, which corresponds to Minkowski spacetime, several observers in this region can define a single Killing vector $t$, which functions as a time parameter and define a single vacuum state for a quantum field in this region\cite{wald1994quantum}.} , then

\begin{equation}
\ln\left|\frac{V}{-U}\right|=\ln\left|\frac{T+R}{R-T}\right|=\ln\left|\frac{e^{\frac{r_{+}}{l^{2}}v}}{e^{-\frac{r_{+}}{l^{2}}u}}\right|
\label{eqn42}
\end{equation}

\begin{equation}
\ln\left| \frac{T+R}{R-T}\right|=\frac{r_{+}}{l^{2}} (v+u).
\label{eqn43}
\end{equation}
The relationship between coordinate time $t$  and the auxiliary coordinates  $u,v$  is established \eqref{eqn26}

\begin{equation}
u+v=2t.
\label{eqn44}
\end{equation}

Substituting \eqref{eqn44} in \eqref{eqn43}, then
 \begin{equation}
t=\frac{l^{2}}{2r_{+}} \ln\left| \frac{T+R}{R-T}\right|.
\label{eqn45}
\end{equation}

\begin{equation}
\frac{2r_{+}t}{l^{2}}= \ln\left| \frac{1+T/R}{1-T/R}\right|.
\label{eqn45a}
\end{equation}
Let  $x=T/R$, then
\begin{equation}
\tanh^{-1}x=\frac{1}{2}\ln\left| \frac{1+x}{1-x}\right|, 
\label{eqn46}
\end{equation} 
Therefore, \eqref{eqn45} is simplified to
\begin{equation}
t=\frac{l^{2}}{r_{+}}\tanh^{-1}\left[\frac{T}{R}\right].
\label{eqn47}
\end{equation}
Recalling
\begin{equation}
T=\pm\sqrt{R^{2}-e^{-\frac{r_{+}}{l^{2}}(v-u)}},
\label{eqn48}
\end{equation}
in addition, from \eqref{eqn26}
\begin{equation}
2r*=v-u.
\label{eqn49}
\end{equation}
Also \eqref{eqn29}, allows concluding that \eqref{eqn48} can be transformed into
\begin{equation}
T=\pm\sqrt{R^{2}+\frac{r-r_{+}}{r+r_{+}}}.
\label{eqn50}
\end{equation}

It is useful to mention the following characteristics of \eqref{eqn50}

\begin{enumerate}
	\item 	When $r=r_{+}$, it follows that \eqref{eqn50} is simplified to
	\begin{equation}
T=\pm R
\label{eqn51}
\end{equation}
which allows defining the horizon.
\item With $r=3r_{+}$, it follows that \eqref{eqn50}  is simplified to
	\begin{equation}
\frac{T^{2}}{1/2}-\frac{R^{2}}{1/2}=1.
\label{eqn52} 
\end{equation}
This is essential given that a FIDO moves on a hyperbolic trajectory in BTZ spacetime  \cite{Banados:1992wn,1986bhmp.book.....T,2003qugr.book.....C}.

\item Likewise, taking \eqref{eqn47} and \eqref{eqn24}
\begin{equation}
T=R\tanh \left[\frac{t\sqrt{M}}{l}\right]
\label{eqn53}
\end{equation}
\end{enumerate}
\begin{figure}[H]
\centering
		\includegraphics[width=0.4\textwidth]{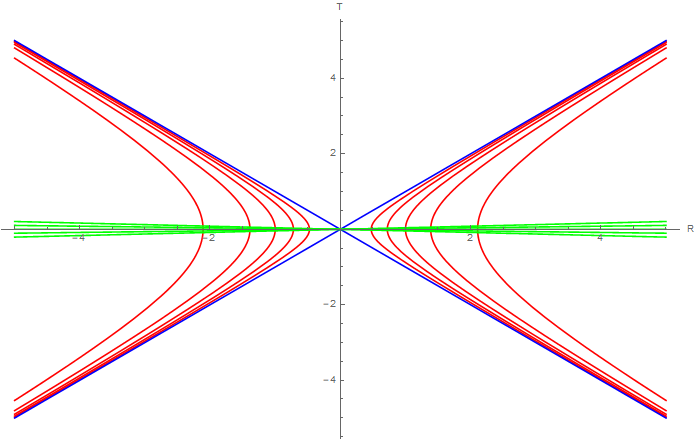}
\caption{Kruskal diagram for the BTZ black hole}
\label{Fig1}
\end{figure}
\section{KINEMATICS OF HYPERSURFACES }\label{sec3}

Consider a  manifold $\mathcal{M}$ of dimension $(2+1)$ and within it a hypersurface $\Sigma$, with the condition that $\Sigma\subset \mathcal{M}$ and can be time-like, space-like or null. A specific hypersurface  $\Sigma$ can be chosen when the coordinates $x^{\alpha}$ of variety  $\mathcal{M}$  are constrained of the form
\begin{equation}
\Phi\left(x^{\alpha}\right)=0.
\end{equation}
Thus, $\Sigma$  can be specified with a constraint on the coordinates whose parametric equations are of the form 

\begin{equation}
x^{\alpha}=x^{\alpha}(y^{a}),\,\,\,x^{\alpha}\in \mathcal{M},\,\,\,y^{a}\in \Sigma\,\,\,\mbox{y}\,\,\,\Sigma\subset \mathcal{M},
\label{A1}
\end{equation}
where $y^{a}$  corresponds to intrinsic coordinates in $\Sigma$. Also, the hypersurface $\Sigma$ is determined by its normal vector  $n_{\alpha}$, such that the unit normal vector is defined of the form
\begin{equation}
n^{\alpha}n_{\alpha}=\epsilon =\pm 1,
\label{A3}
\end{equation}
is  $-1$,  if $\Sigma$  is space-like and $+1$,  if $\Sigma$  is time-like.
\begin{equation}
 n_{\alpha}=\frac{\epsilon \partial_{\alpha }\Phi}{\sqrt{g^{\mu\nu}\partial_{\mu }\Phi\partial_{\nu }\Phi}}
\label{A3w}
\end{equation}

\subsection{First Fundamental Form: Induced Metric}
The induced metric on  $\Sigma$  is obtained when the displacements are limited to such a hypersurface of the form
\begin{equation}
ds^{2}_{\Sigma}=h_{ab}dy^{a}dy^{b}, 
\label{A4}
\end{equation}

\begin{equation}
h_{ab}=g_{\alpha \beta}e^{\alpha}_{a}e^{\beta}_{b}.
\label{A44}
\end{equation}
Where $h_{ab}$ is known as the induced metric or first fundamental form and the tangent vectors to the integral curves contained in $\Sigma$ are of the form
\begin{equation}
e^{\alpha}_{a}=\frac{\partial x^{\alpha}}{\partial y^{a}}.
\label{A5}
\end{equation}

\subsection{Second Fundamental Form: Extrinsic Curvature}
While the intrinsic curvature of the manifold $\mathcal{M}$ is completely determined by the Riemann tensor

\begin{equation}
R^{\alpha}_{\beta\gamma\delta}=\frac{\partial \Gamma^{\alpha}_{\delta\beta}}{\partial x^{\gamma}}+\frac{\partial \Gamma^{\alpha}_{\gamma\beta}}{\partial x^{\delta}}+\Gamma^{\alpha}_{\gamma\sigma}\Gamma^{\sigma}_{\delta\beta}-\Gamma^{\alpha}_{\sigma\delta}\Gamma^{\sigma}_{\delta\beta}
\end{equation}
The extrinsic curvature or Second Fundamental Form $K_{ab}$, defines how the hypersurface $\Sigma$  is curved with respect to $\mathcal{M}$ in which it is embedded.

Let $K_{ab }$ be closely related to the normal derivative of the metric tensor $g_{\alpha\beta}$

\begin{equation}
K_{ab}=\frac{1}{2}\left[\mathcal{L}_{n}g_{\mu\nu}\right]e^{\alpha}_{a}e^{\beta}_{b}=n_{\alpha;\beta} e^{\alpha}_{a}e^{\beta}_{b}, 
\label{AW6}
\end{equation}

\subsection{Darmois-Israel Formalism}
Let hypersurface $\Sigma$ divide spacetime into two regions: $M^{+}$ and $M^{-}$, such that $g^{+}_{\alpha\beta}\in M^{+}$ and $g^{-}_{\alpha\beta}\in M^{-}$

\begin{figure}[H]
\centering
	\includegraphics[width=0.4\textwidth]{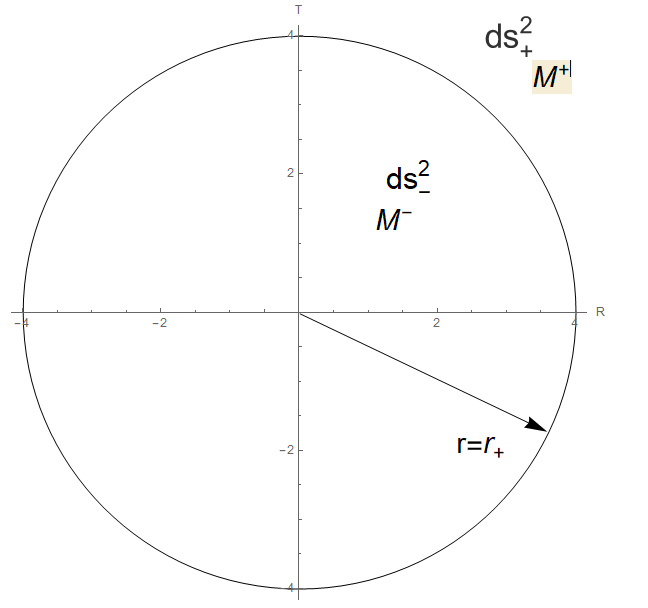}
\caption{Two spacetime regions meeting at a common boundary. Taken from \cite{poisson2004relativist}.}
\label{Fig2}
\end{figure}
\subsection{First Junction Condition}
The first junction condition states that: The induced metric must be the same on both sides of $\Sigma$, which allows establishing the continuity of the first fundamental form
\begin{equation}
\left[h_{ab}\right]=h_{ab}^{+}|_{\Sigma}-h_{ab}^{-}|_{\Sigma}=0.
\label{A7}
\end{equation}

\subsection{Second Junction Condition}
The second junction condition states that the extrinsic curvature must be the same on both sides of the hypersurface, which ensures continuity of the second fundamental form

\begin{equation}
\left[K_{ab}\right]=K_{ab}^{+}|_{\Sigma}-K_{ab}^{-}|_{\Sigma}=0.
\label{A8}
\end{equation}
Both conditions are independent of $x^{\alpha}$, if the second junction condition is violated, the spacetime is singular in $\Sigma$,, which must be associated with the presence of matter on the hypersurface.
\begin{equation}
S_{ab}=-\frac{\epsilon}{8\pi}\left(\left[K_{ab}\right]-\left[K\right]h_{ab}\right),\,\,\,K\equiv h^{ab}K_{ab}=n^{\alpha}_{;\alpha}.
\label{A9}
\end{equation}
Which relates the momentum-energy tensor to the jump in the extrinsic curvature from one side of $\Sigma$ to another side  \cite{poisson2004relativist}. The momentum-energy tensor of the surface layer is determined as
\begin{equation}
T^{\alpha\beta}_{\Sigma}=\delta(l)S^{ab}e^{\alpha}_{\,a}e^{\beta}_{\,b}.
\end{equation}
\subsection{Junction Conditions for BTZ}
This section describes a shell (dust ring), in ($2+1$)  dimensions. Such that it contracts from infinity to near its gravitational radius  $r=r_{+}+\epsilon$. To an external observer, the shell looks like a BTZ black hole. In a first approximation with $J=0$, let \eqref{eqn1}
\begin{equation}
ds^{2}_{+}=-f(r)dt^{2}+\frac{1}{f(r)}dr^{2}+r^{2}d\phi^{2},
\label{eqn60}
\end{equation}
where  
\begin{equation}
f(r)=\left(-M+\frac{r^{2}}{l^{2}}\right)
\label{eqn61}
\end{equation}
for the outer shell. And for the inner shell, there is a Minkowski spacetime
\begin{equation}
ds^{2}_{-}=-dt^{2}+dr^{2}+r^{2}d\phi^{2},
\label{eqn62}
\end{equation}
\subsection{Inner Solution }
The coordinates for the inner solution of the shell \eqref{eqn62} are
\begin{equation}
t=\bar{T}(\tau),\,\,\,r=R(\tau).
\label{eqn63}
\end{equation}
From the foregoing, \eqref{eqn62} simplifies to
\begin{equation}
ds^{2}_{-}=-\left[\dot{\bar{T}}^{2}+\dot{R}^{2}\right]d\tau^{2}+R^{2}d\phi^{2}.
\label{eqn64}
\end{equation}

The extrinsic coordinates defined on $\Sigma\subset M^{-}$  are
\begin{equation}
y^{a}_{-}=(\tau,\phi)\in \Sigma
\label{eqn65}
\end{equation}
and the intrinsic coordinates  $x^{\alpha}_{-}\in M$
\begin{equation}
x^{\alpha}_{-}=(\bar{T}(\tau),R(\tau),\phi).
\label{eqn66}
\end{equation}
From the foregoing, the relationship between the coordinates between $\Sigma$ and $M^{-}$ is
\begin{equation}
e^{\alpha}_{a}=\frac{\partial x^{\alpha}}{\partial y^{a}}.
\label{eqn67}
\end{equation}
According to \eqref{eqn67}, then
\begin{equation}
e^{\alpha}_{\tau}=u^{\alpha}_{-}=\left[\dot{\bar{T}},\dot{R},0\right],
\label{eqn68}
\end{equation}
where  $u^{\alpha}$, defines the 3-velocity for an observer falling radially with the shell (FFO). Also,
\begin{equation}
e^{\alpha}_{\phi}=u^{\beta}_{-}=\left[0,0,0\right].
\label{eqn69}
\end{equation}
The normal vector $n^{-}_{\alpha} \perp \Sigma$ is defined as
\begin{equation}
n^{-}_{\alpha} =\left[n_{\tau},n_{r},n_{\phi} \right].
\label{eqn70}
\end{equation}
The orthonormality condition
\begin{equation}
n^{\pm}_{\alpha} u^{\alpha}_{\pm}=0,
\label{eqn71}
\end{equation}
and the normalization condition
\begin{equation}
n^{\pm}_{\alpha} n^{\alpha}_{\pm}=1.
\label{eqn72}
\end{equation}
The foregoing makes it possible to conclude that $n^{-}_{\alpha}$  is a time-like vector and that $\Sigma$ corresponds to a space-like hypersurface. From \eqref{eqn71}
\begin{equation}
n^{-}_{\alpha} u^{\alpha}_{-}=n^{-}_{\tau}\dot{\bar{T}}+n^{-}_{r}\dot{R}=0,
\label{eqn73}
\end{equation}
therefore, obtaining that $n^{-}_{\phi}=0$. From the condition of normalization condition  \eqref{eqn72}
\begin{equation}
 g^{\alpha\alpha}n^{-}_{\alpha}n^{-}_{\alpha}=1.
\label{eqn74}
\end{equation}
\begin{equation}
 g^{tt}\left[n^{-}_{t}\right]^{2}+ g^{rr}\left[n^{-}_{r}\right]^{2}=1.
\label{eqn75}
\end{equation}
Where the coefficients $ g^{tt}=-1$ and  $g^{rr}=1$ for the metric \eqref{eqn62}, then
\begin{equation}
-\left[n^{-}_{t}\right]^{2}+ \left[n^{-}_{r}\right]^{2}=1.
\label{eqn76}
\end{equation}
Which allows determining the system of equations
\begin{equation}
n^{-}_{\tau}\dot{\bar{T}}+n^{-}_{r}\dot{R}=0, \,\,\,-\left[n^{-}_{t}\right]^{2}+ \left[n^{-}_{r}\right]^{2}=1.
\label{eqn77}
\end{equation}
Once \eqref{eqn77} has been solved, the normal vector  $n^{-}_{\alpha}$ is
\begin{equation}
n^{-}_{\alpha}=\left[-\dot{R},\dot{\bar{T}},0\right],
\label{eqn78}
\end{equation}
with the condition that \cite{poisson2004relativist}
\begin{equation}
\dot{\bar{T}}^{2}-\dot{R}^{2}=1
\label{eqn78a}
\end{equation}
\subsection{Outer Solution }
The outer solution of the Shell is defined by \eqref{eqn60} and \eqref{eqn61}. Let
\begin{equation}
F(R)=f(r),
\label{eqn79}
\end{equation}
then \eqref{eqn60} transforms into
\begin{equation}
ds^{2}_{+}=-F(R)dt^{2}+\frac{1}{F(R)}dr^{2}+r^{2}d\phi^{2}.
\label{eqn79a}
\end{equation}
Such that  $\Sigma\in M^{+}$ is obtained
\begin{equation}
x^{\alpha}_{+}=(t,r,\phi)\in M^{+}.
\label{eqn80}
\end{equation}
\begin{equation}
y^{a}_{+}=(\tau,\phi)\in \Sigma.
\label{eqn81}
\end{equation}
It is useful to express

\begin{equation}
t=T(\tau),\,\,\,r=R(\tau)\,\,\,\mbox{and}\,\,\,F=F(R)=M+\frac{R(\tau)^{2}}{l^{2}}.
\label{eqn82}
\end{equation}
From \eqref{eqn82}, the metric \eqref{eqn79a} is simplified to
\begin{equation}
ds^{2}_{+}=-\left[F\dot{T}-\frac{\dot{T}}{F}\right]d\tau^{2}+\dot{R}^{2}d\phi^{2}.
\label{eqn83}
\end{equation}
The intrinsic coordinates $x^{\alpha}_{+}\in M^{+}$
\begin{equation}
x^{\alpha}_{+}=(T(\tau),R(\tau),\phi).
\label{eqn84}
\end{equation}
Again,  from \eqref{eqn67}, it is possible to obtain
\begin{equation}
u^{\alpha}_{+}= \left[\dot{T},\dot{R},0\right],
\label{eqn85}
\end{equation}
where $u^{\alpha}_{+}$ is the 3-velocity measured from $M^{+}$.

On the other hand, the normal vector $n^{+}_{\alpha}$  is determined as
\begin{equation}
n^{+}_{\alpha}=\left[n^{+}_{\alpha},n^{+}_{r},n^{+}_{\phi}\right].
\label{eqn86}
\end{equation}

As in the inner solution, the components of the normal vector $n^{+}_{\alpha}$  must be obtained from conditions \eqref{eqn71} and \eqref{eqn72}. Therefore,
\begin{equation}
n^{+}_{\alpha}u^{\alpha}_{+}=0,\,\,\,n^{+}_{\alpha}n^{\alpha}_{+}=g^{\alpha\alpha}\left[n^{+}_{\alpha}\right]^{2}=1,
\label{eqn87}
\end{equation}
explicitly developing the orthonormality condition
\begin{equation}
n^{+}_{\alpha}u^{\alpha}_{+}=n^{+}_{t}u^{t}_{+}+n^{+}_{r}u^{r}_{+}+n^{+}_{\phi}u^{\phi}_{+}=0,
\label{eqn88}
\end{equation}
then
\begin{equation}
n^{+}_{t}=-n^{+}_{r}\frac{\dot{R}}{\dot{T}}.
\label{eqn89}
\end{equation}
And the normalization condition
\begin{equation}
g^{tt}\left[n^{+}_{t}\right]^{2}+g^{rr}\left[n^{+}_{r}\right]^{2}+g^{\phi\phi}\left[n^{+}_{\phi}\right]^{2}=1
\label{eqn90}
\end{equation}
For the $g^{\mu\nu}$ components of the metric \eqref{eqn79}, explicitly they are
\begin{equation}
g^{\mu\nu}_{+}=
\begin{pmatrix}
-\frac{1}{F(R)} & 0 & 0\\
0 & F(R) & 0 \\
0& 0 &\frac{1}{R^{2}}. 
\end{pmatrix}
\label{eqn91}
\end{equation}
This allows \eqref{eqn90}
\begin{equation}
-\frac{1}{F}\left[n^{+}_{t}\right]^{2}+F\left[n^{+}_{r}\right]^{2}=1.
\label{eqn92}
\end{equation}
From   \eqref{eqn89}, the radial component $n^{+}_{r}$  is obtained, which is
\begin{equation}
n^{+}_{r}=\frac{\dot{T}}{\sqrt{F\dot{T}^{2}-\frac{\dot{R}^{2}}{F}}}.
\label{eqn93}
\end{equation}
and the time component, $n^{+}_{t}$   is
\begin{equation}
n^{+}_{t}=\frac{-\dot{R}}{\sqrt{F\dot{T}^{2}-\frac{\dot{R}^{2}}{F}}}.
\label{eqn94}
\end{equation}
This leads to
{\begin{equation}
n^{+}_{\alpha}=\frac{1}{\sqrt{F\dot{T}^{2}-\frac{\dot{R}^{2}}{F}}}\left[-\dot{R},-\dot{T},0\right]=\left[-\dot{R},-\dot{T},0\right]
\label{eqn95}
\end{equation}
where \cite{poisson2004relativist}.
\begin{equation}
\sqrt{F\dot{T}^{2}-\frac{\dot{R}^{2}}{F}}=1
\label{eqn96a}
\end{equation}
\subsection{External estimation of Extrinsic Curvature}
It is possible to determine the extrinsic curvature $K^{+}_{\alpha\beta}$  from \eqref{AW6} for the metric \eqref{eqn79} and the normal vector $n^{+}_{\alpha}$  defined in \eqref{eqn95}
\begin{equation}
K^{\pm}_{\alpha\beta}=n^{\pm}_{\alpha;\beta}.
\label{eqn96}
\end{equation}

To determine the components of the extrinsic curvature $K^{+}_{\alpha\beta}$, it is necessary to determine the Christoffel symbols. A direct calculation allows to find
\begin{equation}
\Gamma^{r}_{rr}=\frac{F'}{2F},\,\,\,\Gamma=^{\phi}_{r\phi}=\frac{1}{r},\,\,\,\Gamma^{t}_{rt}=\frac{F'}{2F},
\label{eqn97}
\end{equation}
\begin{equation}
\Gamma^{r}_{\phi\phi}=-rF\,\,\,\mbox{and} \,\,\,\Gamma^{r}_{tt}=\frac{FF'}{2}.
\label{eqn98}
\end{equation}
The components of the extrinsic curvature are
\begin{equation}
K^{+}_{00}=-\frac{1}{2}FF'\dot{T},\,\,\,K^{+}_{01}=-\frac{1}{2}\frac{F'\dot{R}}{F},
\label{eqn99}
\end{equation}
\begin{equation}
K^{+}_{10}=\frac{1}{2}\frac{F'\dot{R}}{F},\,\,\,K^{+}_{11}=\frac{1}{2}\frac{F'\dot{T}}{F},
\label{eqn100}
\end{equation}
\begin{equation}
K^{+}_{22}=r\dot{T}F.
\label{eqn101}
\end{equation}
The contraction of $K^{+\alpha}_{\alpha}$ is
\begin{equation}
K^{\pm\alpha}_{\alpha}=g^{\alpha\beta}_{\pm}K^{\pm}_{\alpha\beta}.
\label{eqn102}
\end{equation}
This is 
\begin{align}
K^{+\alpha}_{\alpha}&=K^{+0}_{0}+ K^{+1}_{1} K^{+2}_{2}   
\notag\\
&=\frac{F'\dot{T}}{2}+\frac{F\dot{T}}{2}+\frac{F\dot{T}}{R}.
&\hspace{0.3cm}
\label{eqn103}
\end{align}
Let
\begin{equation}
\beta_{+}=\dot{T}F,\,\,\,\dot{\beta_{+}}=F'\dot{R}\dot{T}.
\label{eqn104}
\end{equation}
Therefore, from  \eqref{eqn103} and \eqref{eqn104}
\begin{equation}
K^{+0}_{0}=\frac{\dot{\beta}_{+}}{2\dot{R}},\,\,\,K^{+1}_{1}=\frac{\beta_{+}}{2}\,\,\,\mbox{and}\,\,\,K^{+2}_{2}=\frac{\beta_{+}}{R}
\label{eqn105}
\end{equation}
\subsection{Internal	estimation of the extrinsic curvature seen}

Similarly, the extrinsic curvature $K^{-}_{\alpha\beta}$ is estimated from \eqref{AW6}  for the metric \eqref{eqn62} and the normal vector $n^{-}_{\alpha}$  defined in \eqref{eqn78}. The Christoffel symbols are
\begin{equation}
\Gamma^{\phi}_{r\phi}=\frac{1}{r}\,\,\,\mbox{and}\,\,\,\Gamma^{r}_{\phi\phi}=-r.
\label{eqn106}
\end{equation}
From \eqref{eqn96} and \eqref{eqn106}, the components of the extrinsic curvature $K^{-}_{\alpha\beta}$ are obtained
\begin{equation}
K^{-}_{22}=r\dot{\bar{T}}. 
\label{eqn107}
\end{equation}
The contraction of the extrinsic curvature \eqref{eqn102} leads to
\begin{equation}
K^{-\phi}_{\phi}=\frac{\sqrt{1+\dot{R}^{2}}}{R}=\frac{\beta_{-}}{R},
\label{eqn108}
\end{equation}
where from \eqref{eqn78a}, $\beta_{-}$ is defined
\begin{equation}
\beta_{-}=\dot{\bar{T}}=\sqrt{1+\dot{R}^{2}}.
\label{eqn109}
\end{equation}
\subsection{Shell motion equation}
Recalling \eqref{A9}
\begin{equation}
S_{ab}=-\frac{\epsilon}{8\pi}\left(\left[K_{ab}\right]-\left[K\right]h_{ab}\right),
\label{eqn110}
\end{equation}
where $h_{ab}\vert_{\Sigma}$, is the metric induced on the hypersurface $\Sigma$. Also,
\begin{equation}
\left[K_{ab}\right]=K^{+}_{ab}\vert_{\Sigma}-K^{-}_{ab}\vert_{\Sigma},
\label{eqn111}
\end{equation}
\begin{equation}
\left[K\right]=K^{+}\vert_{\Sigma}-K^{-}\vert_{\Sigma}.
\label{eqn112}
\end{equation}
Explicitly \eqref{eqn110}
\begin{equation}
S_{ab}=-\frac{\epsilon}{8\pi}\left[\left(K^{+}_{ab}-K^{-}_{ab}\right)-\left(K^{+}-K^{-}\right)h_{ab}\right],
\label{eqn113}
\end{equation}
the matter component is written as  $S_{ab}$
\begin{equation}
S^{b}_{a}=S_{ac}h^{cb},
\label{eqn114}
\end{equation}
where the indices  $a,b$ run on the coordinates defined on the coordinates $(\tau,\phi)$. That is
\begin{align}
S^{a}_{a}&=S_{ac}h^{cb}   
\notag\\
&=S_{\tau}^{\tau}+S_{\phi}^{\phi}.
&\hspace{0.3cm}
\label{eqn115}
\end{align}
Inserting \eqref{eqn115} in \eqref{eqn113}
\begin{equation}
S^{b}_{a}=-\frac{\epsilon}{8\pi}\left[\left(K^{+b}_{a}-K^{-b}_{a}\right)-\left(K^{+}-K^{-}\right)\delta^{b}_{a}\right],
\label{eqn116}
\end{equation}
Where $\delta^{b}_{a}=h^{cb}h_{ac}$. For the component $\tau$ in \eqref{eqn116}, it is
\begin{widetext} 
\begin{equation}
S^{\tau}_{\tau}=-\frac{\epsilon}{8\pi}\left[\left(K^{+\tau}_{\tau}-K^{-\tau}_{\tau}\right)-\left(K^{+\tau}_{\tau}-K^{-\tau}_{\tau}+K^{+\phi}_{\phi}-K^{-\phi}_{\phi}\right)\delta^{\tau}_{\tau}\right].
\label{eqn117}
\end{equation}
\end{widetext}
It must be required that $[h_{ab}]=h^{+}_{ab}\vert_{\Sigma}-h^{+}_{ab}\vert_{\Sigma}=0$, where in addition $h_{ab}$  and $e^{\mu}_{a}$ are defined by \eqref{A44} and \eqref{eqn67}. Since $x^{\alpha_{-}}=\left[\bar{T},R,\phi\right]\in M^{+}$ and $y^{a}_{-}=\left[\tau,\phi\right]\in\Sigma$.  This allows obtaining
\begin{equation}
e^{\alpha}_{a}=[e^{\alpha}_{\tau},e^{\alpha}_{\phi}].
\label{eqn118}
\end{equation}
this is
\begin{equation}
e^{\alpha}_{\tau}=[\dot{\bar{T}},\dot{R},0],
\label{eqn119}
\end{equation}
\begin{equation}
e^{\alpha}_{\phi}=[0,0,1].
\label{eqn120a}
\end{equation}
The foregoing allows calculating the induced metric from \eqref{A44}
\begin{equation}
h_{ab}=\begin{pmatrix}
\dot{\bar{T}}+\dot{R}& 0 
 \\
 0 & R^{2}
\end{pmatrix}.
\label{eqn120}
\end{equation}
From \eqref{eqn120}, it follows that \eqref{eqn117}  is simplified to
\begin{equation}
S^{\tau}_{\tau}=\frac{\epsilon}{8\pi}\left[\frac{\beta_{+}}{R}-\frac{\beta_{-}}{R}\right].
\label{eqn121}
\end{equation}
Consider now the surface tensor $S^{ab}$ defined over $\Sigma$
\begin{equation}
S^{ab}=\lambda u^{a}u^{b}
\label{eqn122}
\end{equation}
where $u^{\alpha}_{-}$ is defined by \eqref{eqn68}, then
\begin{equation}
S^{a}_{b}=\lambda u^{a}u_{b}
\label{eqn123}
\end{equation}
\begin{equation}
S^{\tau}_{\tau}=-\lambda, 
\label{eqn124}
\end{equation}
where $\lambda$ is the linear mass density of the Shell in (2+1) dimensions and $u^{a}u_{a}=-1$. Introducing \eqref{eqn124} into \eqref{eqn121}
\begin{equation}
-\lambda=\frac{\epsilon}{8\pi R}\left[\beta_{+}-\beta_{-}\right].
\label{eqn125}
\end{equation}
With the condition that the hypersurface $\Sigma$  is a time-like surface $\epsilon=1$. Similarly,
\begin{equation}
S^{\phi}_{\phi}=-\frac{\epsilon}{8\pi}\left[-K^{+\tau}_{\tau}+K^{-\tau}_{\tau}\right],
\label{eqn126}
\end{equation}
\begin{equation}
-K^{+\tau}_{\tau}+K^{-\tau}_{\tau}=0.
\label{eqn127}
\end{equation}
Where the following has been considered for \eqref{eqn126}
\begin{equation}
S^{\phi}_{\phi}=S_{\phi\phi}h^{\phi\phi},
\label{eqn128}
\end{equation}
\begin{equation}
S^{\phi}_{\phi}=\lambda u^{\phi}u^{\phi}=0.
\label{eqn129}
\end{equation}
So, for \eqref{eqn127}
\begin{equation}
K^{+\tau}_{\tau}=\frac{\dot{\beta}_{+}}{2\dot{R}},\,\,\,K^{-\tau}_{\tau}=0\rightarrow \dot{\beta}_{-}=0
\label{eqn130}
\end{equation}
\begin{equation}
\frac{d}{d\tau}\left[\beta_{+}-\beta_{-}\right]=0.
\label{eqn131}
\end{equation}
Therefore,
\begin{equation}
\dot{\beta}_{+}=\frac{d}{d\tau}\left[\beta_{+}-\beta_{-}\right].
\label{eqn131a}
\end{equation}
Taking \eqref{eqn125} and time-like $\Sigma$ , it is possible to obtain
\begin{equation}
-\lambda(8\pi R)=\left[\beta_{+}-\beta_{-}\right],
\label{eqn132}
\end{equation}
Inserting \eqref{eqn132} in \eqref{eqn131}

\begin{equation}
\dot{\beta}_{+}=\frac{d}{d\tau}\left[-8\pi R\lambda\right],\,\,\,R=R(\tau).
\label{eqn133}
\end{equation}
\begin{equation}
\dot{\beta}_{+}=-8\pi \lambda\frac{dR}{d\tau}.
\label{eqn134}
\end{equation}
Considering \eqref{eqn104}, it allows to simplify \eqref{eqn134}
\begin{equation}
\frac{dF}{dR}\frac{dT}{d\tau}=-8\pi\lambda.
\label{eqn135}
\end{equation}
Considering \eqref{eqn82}
\begin{equation}
F'(R)=\frac{2R}{l^{2}}.
\label{eqn136}
\end{equation}
Substituting \eqref{eqn136} in \eqref{eqn135}, results in

\begin{equation}
\frac{R}{l^{2}}\dot{T}=-4\pi\lambda.
\label{eqn137}
\end{equation}
The condition of normalization of $n^{+}_{\alpha}$ given in \eqref{eqn96a}, allows obtaining

\begin{equation}
\dot{T}=\sqrt{\frac{1}{F}\left(1+\frac{\dot{R}^{2}}{F}\right)}.
\label{eqn138}
\end{equation}
Inserting \eqref{eqn138} in \eqref{eqn137}

\begin{equation}
\frac{\dot{R}^{2}}{F}=4(2\pi \lambda)^{2}l^{2}\left(\frac{l^{2}}{R^{2}}\right)F-1.
\label{eqn139}
\end{equation}
It is possible to define the mass of the Shell as

\begin{equation}
\frac{M}{R}=2\pi \lambda,
\label{eqn140}
\end{equation}

therefore \eqref{eqn139}

\begin{equation}
\frac{\dot{R}^{2}}{F}=4\left(\frac{M}{R}\right)^{2}l^{2}\left(\frac{l^{2}}{R^{2}}\right)F-1.
\label{eqn139a}
\end{equation}
From \eqref{eqn82}

\begin{equation}
R=\sqrt{l^{2}(F-M)},
\label{eqn140a}
\end{equation}

its differential is
\begin{equation}
dR=\frac{l}{2\sqrt{F-M}}dF.
\label{eqn141}
\end{equation}

Inserting \eqref{eqn140a} in \eqref{eqn139a}

\begin{equation}
\frac{dR}{d\tau}=\sqrt{\left[\frac{4M^{2}F}{(F-M)^{2}}-1\right]F}.
\label{eqn142}
\end{equation}

A very important observation is that \eqref{eqn142} corresponds to the shell motion equation contracting in spacetime BTZ \cite{1966NCimB..44....1I,Arenas:2011be,2020Symm...12.2072R,RojasC:2021kws}, whose motion is measured by a shell-comoving observer with proper time $\tau$ , according to Figure \ref{Fig1}.

\subsection{Solution to the shell motion equation}
This section discusses a possible solution to equation  \eqref{eqn142}.

Let 
\begin{equation}
a=\frac{M}{\mu}=\frac{2\pi \lambda R}{2\pi \lambda R_{0}}=1,
\label{eqn143}
\end{equation}
where $\mu$ is the mass of the shell in $R_{0}$ and $M$ is the mass of the shell in $R$, with the condition that $R_{0}\gg R$.

\begin{equation}
\frac{\mu}{R_{0}}=2\pi \lambda.
\label{eqn144}
\end{equation}
Considering \eqref{eqn82}, \eqref{eqn141} and \eqref{eqn143}, then  \eqref{eqn139} is simplified to

\begin{equation}
\frac{dF}{d\tau}=\frac{2}{l}\sqrt{(F-M)F\left[\frac{4M^{2}l^{2}F}{a^{2}R_{0}^{2}(F-M)}-1\right]}.
\label{eqn145}
\end{equation}
It is useful to express the proper time $\tau$ in terms of the coordinate time $t$, as

\begin{equation}
d\tau= \sqrt{F}dt.
\label{eqn146}
\end{equation}
Therefore, 
\begin{equation}
\frac{dF}{dt}=\alpha \sqrt{F^{2}\left[F-\vartheta(F-M)\right]}.
\label{eqn147}
\end{equation}
where
\begin{equation}
\alpha=\frac{4M}{aR_{0}},\,\,\,\vartheta=\left[\frac{aR_{0}}{2Ml}\right]^{2}.
\label{eqn148}
\end{equation}
Integrating \eqref{eqn147} yields

\begin{equation}
\int_{F_{0}}^{F}\frac{1}{\sqrt{F^{2}\left[F-\vartheta(F-M)\right]}}=\alpha \int^{t}_{t_{0}}dt
\label{eqn149}
\end{equation}

\begin{equation}
\arctanh \left[\sqrt{\frac{F+\vartheta M-FM}{\vartheta M}}\right]^{F}_{F_{0}}=-\frac{\alpha\sqrt{\vartheta M}}{2}(t-t_{0}),
\label{eqn151}
\end{equation}
where 
\begin{equation}
\arctanh \left[x\right]=\frac{1}{2}\ln\left|\frac{1+x}{1-x}\right|=x+\frac{x^{3}}{3}+\frac{x^{5}}{7}+\ldots,
\label{eqn152}
\end{equation}

\begin{equation}
F=M+\left(\frac{R}{l}\right)^{2},
\label{eqn153}
\end{equation}
\begin{equation}
F_{0}=M+\left(\frac{R_{0}}{l}\right)^{2},
\label{eqn154}
\end{equation}
and
\begin{equation}
\mathbf{A}=\frac{\sqrt{\vartheta M}+\sqrt{F_{0}+\vartheta M-F_{0}\vartheta}}{\sqrt{\vartheta M}-\sqrt{F_{0}+\vartheta M-F_{0}\vartheta}}
\label{eqn155}
\end{equation}
allows us to obtain
\begin{equation}
R(t)=\sqrt{\frac{l^{2}\vartheta M}{1-\vartheta}\left\{\left[\frac{\mathbf{A} e^{-\alpha \sqrt{\vartheta M}(t-t_{0})}-1}{\mathbf{A} e^{-\alpha \sqrt{\vartheta M}(t-t_{0})}+1}\right]^{2}+1\right\}}.
\label{eqn156}
\end{equation}

Let \eqref{eqn156} correspond to the shell motion equation in BTZ spacetime, as seen by a FIDO observer measuring a coordinate time $t$. Let $r_{+}$  correspond to the position of the horizon, defined by \eqref{eqn24}, which allows establishing that the horizon not only depends on the mass of the BTZ black hole, but also on the value of the cosmological constant \eqref{eqn4} \cite{poisson2004relativist,RojasC:2021kws,RojasC:2020qnz,2020Symm...12.2072R,Castro:2020lyn,1966NCimB..44....1I,Musgrave:1995ka,gron2007einstein,PhysRev.153.1388,Frauendiener1995,Kijowski2006,Oppenheimer:1939ue}.

\begin{figure}[H]
\centering
		\includegraphics[width=0.4\textwidth]{ 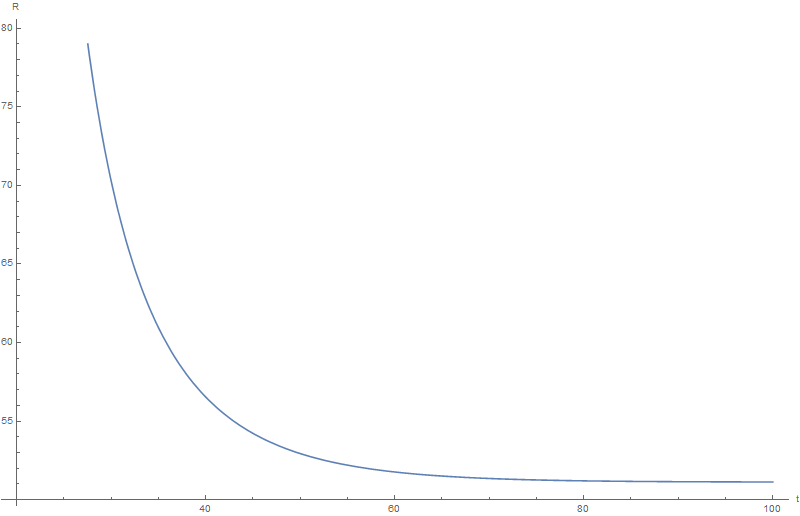}
\caption{ Representation of $R(t)$ given by \eqref{eqn156}.}
\label{Fig3}
\end{figure}
\section{ENTANGLEMENT THERMODYNAMICS FOR BTZ}\label{sec4}
Considering a metric of the form \eqref{eqn21}, it can be rewritten as
\begin{equation}
ds^{2}=\frac{l^{4}}{r_{+}^{2}}\frac{(r+r_{+})^{2}}{l^{2}}dUdV+r^{2}d\phi.
\label{eqn157}
\end{equation}
where $r_{+}$  is determined by \eqref{eqn24}, the coordinates $(U,V)$ given by \eqref{eqn34}. And furthermore, the geometry of the BTZ black hole is described by Figure \ref{Fig1}. Under such conditions, let a scalar field $\Phi$, whose Lagrangian density is \cite{Birrell:1982ix}
\begin{equation}
\mathcal{L}_{M}=\frac{1}{2}\sqrt{-g}\left[ g^{\mu\nu}\partial_{\mu}\Phi\partial_{\nu}\Phi-m^{2}\Phi^{2}\right].
\label{eqn158}
\end{equation}
The action that includes the space-time component and the matter fields is \cite{Banados:1992wn}
\begin{equation}
S=\int \left[\frac{1}{2\pi}\sqrt{-g}\left(R+ \frac{2}{l^{2}}\right)+\mathcal{L}_{M}\right]d^{3}x.
\label{eqn159}
\end{equation}
The variation $\delta S$ leads to the Klein-Gordon equation
\begin{equation}
\left[\square-m^{2}\right]\Phi=0,\,\,\,\square=\frac{1}{\sqrt{-g}}\partial_{\mu}\left[\sqrt{-g}g^{\mu\nu}\partial_{\nu}\right],
\label{eqn160}
\end{equation}
where $\mu,\nu=0,1,2$. Consider a possible solution of \eqref{eqn160}

\begin{equation}
\Phi(t,r,\phi)=\frac{\varphi_{\Omega}(r)}{\sqrt{2\omega}}e^{-i\omega}e^{i\mathfrak{m}\phi},
\label{eqn161}
\end{equation}
where $\mathfrak{m}$ corresponds to the magnetic quantum number, associated to the angular part of the scalar field $\Phi$. These orthogonal modes under Klein-Gordon inner product
\begin{equation}
\left(\Phi_{1},\Phi_{2}\right)=-i\int\left[\Phi_{1}\left(\frac{\partial}{\partial t}\Phi^{*}_{2}\right)-\left(\frac{\partial}{\partial t}\Phi_{1}\right)\Phi^{*}_{2}\right]d^{n-1}x.
\label{eqn162}
\end{equation}
According to \eqref{eqn161} and \eqref{eqn162}, it is possible to obtain \cite{Birrell:1982ix,greiner2013field}
\begin{equation}
\delta(\Omega_{1}-\Omega_{2})=\int \varphi_{\Omega _{1}}(r)\varphi^{*}_{\Omega _{2}}(r)dr
\label{eqn163}
\end{equation}
\begin{equation}
\delta(\mathfrak{m}_{1}-\mathfrak{m}_{2})=\int e^{i(\mathfrak{m}_{1}-\mathfrak{m}_{2})\phi}d\phi.
\label{eqn164}
\end{equation}
Expanding \eqref{eqn161} into \eqref{eqn160} leads to
 
\begin{equation}
\frac{1}{r}\frac{\partial}{\partial r}\left[r f(r)\frac{\partial \varphi_{\Omega}(r)}{\partial r}\right]+\varphi_{\Omega}(r)\left[\frac{\omega^{2}}{f(r)}-\frac{\mathfrak{m}^{2}}{r}-m^{2}\right]=0,
\label{eqn165}
\end{equation}
with
\begin{equation}
dr*=\frac{1}{f(r)}dr,
\label{eqn166}
\end{equation}
allows simplifying \eqref{eqn165}

\begin{equation}
\frac{d}{dr*}\left[\frac{d\varphi_{\Omega}(r)}{dr*}\right]+\mathbf{T}(\omega,\mathfrak{m},m,r)\varphi_{\Omega}(r)=0, 
\label{eqn167}
\end{equation}
where 
\begin{equation}
\mathbf{T}(\omega,\mathfrak{m},m,r)=\omega^{2}-\left(m^{2}+\frac{\mathfrak{m}}{r}\right)f(r).
\label{eqn168}
\end{equation}
Under the WKB approximation \cite{mathews1970mathematical}, consider a harmonic solution of the form
\begin{equation}
\varphi_{\Omega}(r)=e^{-i\phi(r)},
\label{eqn169}
\end{equation}
under the condition that  $\mathbf{T}(\omega,\mathfrak{m},m,r)$ varies very slightly, which is why $\phi''(r)$is very small. Substituting \eqref{eqn166} in \eqref{eqn164} leads to
\begin{equation}
\phi(r)=\int \sqrt{\mathbf{T}(\omega,\mathfrak{m},m,r)}dr.
\label{eqn170}
\end{equation}
\eqref{eqn167} is satisfied when
\begin{equation}
\phi''(r)\cong \frac{1}{2}\left|\frac{\mathbf{T}'}{\sqrt{\mathbf{T}}}\right|\ll \mathbf{T},\,\,\,\mathbf{T}=\mathbf{T}(\omega,\mathfrak{m},m,r).
\label{eqn171}
\end{equation}
\subsection{Modes $HH^{*}$ and $KB^{*}$}
From the foregoing, the most general solution for \eqref{eqn167} is
\begin{equation}
\varphi_{\Omega}(r)=\frac{1}{\sqrt[4]{\mathbf{T}}}\left[C_{1}e^{+i\int\sqrt{\mathbf{T}}dr}+C_{2}e^{-i\int\sqrt{\mathbf{T}}dr}  \right].
\label{eqn172}
\end{equation}
Considering one of the solutions of \eqref{eqn169}, it is possible to write
\begin{equation}
\Phi(t,r,\phi)=\frac{e^{-i\left(\int\sqrt{\mathbf{T}}dr-\mathfrak{m}\phi\right)} e^{-i\omega t}}{\sqrt[4]{4\omega^{2}\mathbf{T}}}.
\label{eqn173}
\end{equation}
If, in addition, $\underline{x}=r,\phi$ and $\Omega=\omega,\mathfrak{m}$

\begin{equation}
\Phi_{\Omega}(\underline{x})=\frac{e^{-i\left(\int\sqrt{\mathbf{T}}dr-\mathfrak{m}\phi\right)}}{\sqrt[4]{4\omega^{2}\mathbf{T}}}.
\label{eqn174}
\end{equation}
\begin{equation}
\Phi_{\Omega}(t,\underline{x})=\Phi_{\Omega}(\underline{x})e^{-i\omega t}.
\label{eqn175}
\end{equation}
Using tortoise Coordinates
\begin{equation}
t=\pm r*,\,\,\,r*=r +2M\ln\left|\frac{r}{2M}-1\right|,
\label{eqn176}
\end{equation}
\begin{equation}
\Phi(t,\underline{x})=\Phi_{\Omega}(r*,\underline{x})=\Phi_{\Omega}(r*,\underline{x})e^{-i\omega r*}.
\label{eqn177}
\end{equation}
For \eqref{eqn177}, it is possible to define the incoming and outgoing modes of the scalar field over BTZ space-time as
\begin{equation}
\Phi^{(+)}_{\Omega}(r*,\underline{x})=\Phi_{\Omega}(\underline{x})e^{i\omega r*},\,\,\,\mbox{incoming modes. }.
\label{eqn178}
\end{equation}
\begin{equation}
\Phi^{(-)}_{\Omega}(r*,\underline{x})=\Phi_{\Omega}(\underline{x})e^{-i\omega r*},\,\,\,\mbox{outgoing modes}.
\label{eqn179}
\end{equation}
The relationship between $r*$  and the auxiliary coordinates $u,v$ is defined by \eqref{eqn26}. Thus. the modes \eqref{eqn177} 
\begin{equation}
\Phi(u,v,\underline{x})=\Phi_{\Omega}(\underline{x})e^{-\frac{i\omega v}{2}}e^{\frac{i\omega u}{2}}.
\label{eqn180}
\end{equation}
For $J=0$, the Carter-Penrose diagram is \cite{Ortiz:2011}

\begin{figure}[H]
\centering
		\includegraphics[width=0.4\textwidth]{ 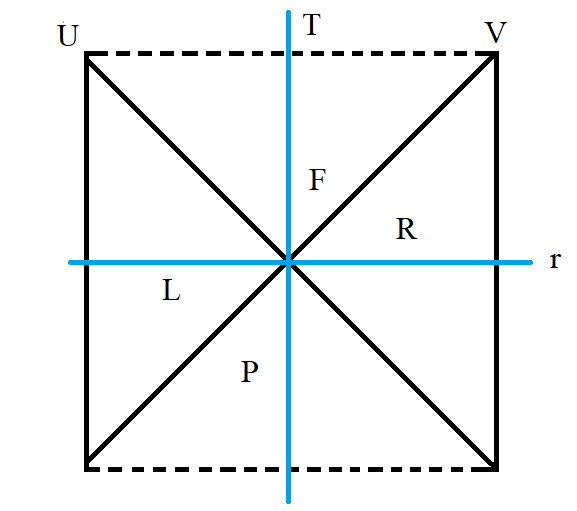}
\caption{Carter-Penrose diagram for a BTZ black hole.}
\label{Fig6}
\end{figure}
Figure \ref{Fig6} shows a Carter-Penrose diagram for a BTZ black hole. From \eqref{eqn180}, the outgoing modes with $u=0$ and the incoming modes $v=0$ are defined
\begin{equation}
\Phi^{\epsilon}(t,\underline{x})=\Phi^{in}_{\Omega}(\underline{x})e^{-\frac{i\omega v}{2}}=\Phi^{in}_{\Omega}(v,\underline{x}).
\label{eqn180a}
\end{equation}\begin{equation}
\Phi^{\epsilon}(t,\underline{x})=\Phi^{out}_{\Omega}(\underline{x})e^{\frac{i\omega u}{2}}=\Phi^{out}_{\Omega}(u,\underline{x}).
\label{eqn180b}
\end{equation}
\begin{figure}[H]
\centering
		\includegraphics[width=0.45\textwidth]{ 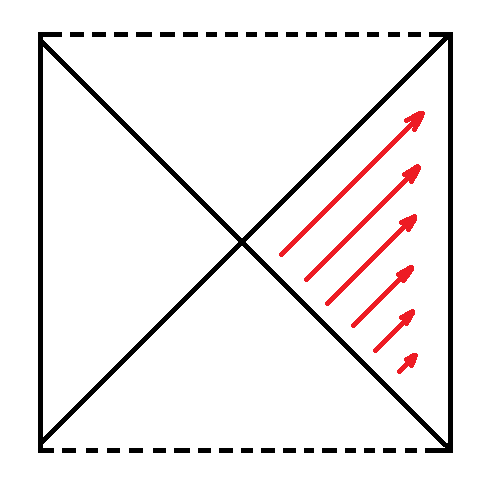}
\caption{Outgoing modes of the scalar field in the Carter-Penrose diagram for a BTZ black hole.}
\label{Fig7}
\end{figure}
\begin{figure}[H]
\centering
		\includegraphics[width=0.45\textwidth]{ 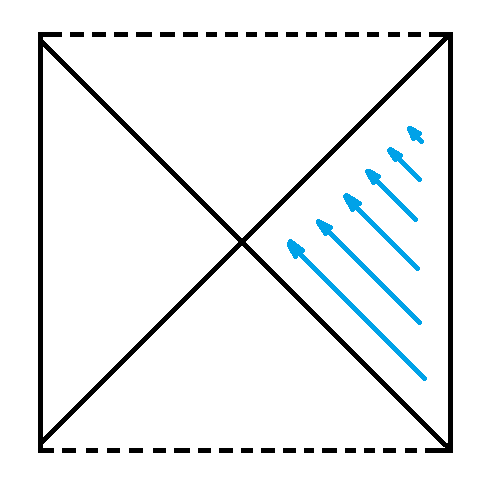}
\caption{Incoming modes of the scalar field in the Carter-Penrose diagram for a BTZ black hole.}
\label{Fig8}
\end{figure}
The modes $\Phi^{in}_{\Omega}(v,\underline{x})$ and $\Phi^{out}_{\Omega}(v,\underline{x})$ are written in terms of the Kruskal null coordinates $U,V$ as

\begin{equation}
\Phi^{\epsilon}_{\Omega}\left(U,\underline{x}\right)=\Theta (\epsilon,V)\Phi^{in}_{\Omega}(v,\underline{x})=\Theta (\epsilon,V)\Phi^{in}_{\Omega}(\underline{x})e^{-\frac{i\omega v}{2}}
\label{eqn181}
\end{equation}
\begin{equation}
\Phi^{\epsilon}_{\Omega}\left(V,\underline{x}\right)=\Theta (-\epsilon,V)\Phi^{out}_{\Omega}(u,\underline{x})=\Theta (-\epsilon,U)\Phi^{out}_{\Omega}(\underline{x})e^{-\frac{i\omega u}{2}},
\label{eqn182}
\end{equation}
where the function $\Theta_{\epsilon}(x)$ is
\begin{equation}
\Theta_{\epsilon}(x)=\frac{1}{2}\left[\Theta(-\epsilon U)+\Theta(\epsilon V)\right].
\label{eqn183}
\end{equation}
This allows defining  $(+)\in \mathbb{R}$ and  $(-)\in \mathbb{L}$ of the Carter-Penrose diagram in Figure (\ref{Fig6}). Consequently, the complete modes for the region $R$ are
\begin{align}
\Phi^{(+)}_{\Omega}\left(u,v,\underline{x}\right)&=\Phi^{(+)}_{\Omega}\left(v,\underline{x}\right)+\Phi^{(+)}_{\Omega}\left(u,\underline{x}\right)
\notag\\
&=\Phi^{(+)}_{\Omega}\left(\underline{x}\right)e^{\frac{-i\omega v}{2}}+\Phi^{(+)}_{\Omega}\left(\underline{x}\right)e^{\frac{-i\omega u}{2}}
\notag\\
&=\Phi^{(+)in}_{\Omega}\left(v,\underline{x}\right)+\Phi^{(+)out}_{\Omega}\left(u,\underline{x}\right).
&\hspace{0.3cm}
\label{eqn184}
\end{align}
And the complete modes for the region $L$  are
\begin{align}
\Phi^{(-)}_{\Omega}\left(u,v,\underline{x}\right)&=\Phi^{(-)}_{\Omega}\left(v,\underline{x}\right)+\Phi^{(-)}_{\Omega}\left(u,\underline{x}\right)
\notag\\
&=\Phi^{(-)}_{\Omega}\left(\underline{x}\right)e^{\frac{-i\omega v}{2}}+\Phi^{(-)}_{\Omega}\left(\underline{x}\right)e^{\frac{-i\omega u}{2}}
\notag\\
&=\Phi^{(-)in}_{\Omega}\left(v,\underline{x}\right)+\Phi^{(-)out}_{\Omega}\left(u,\underline{x}\right).
&\hspace{0.3cm}
\label{eqn185}
\end{align}
Consequently, it is possible to obtain the modes of the scalar field in BTZ, which are contained in the $R$ and $L$ regions as
\begin{equation}
\Phi_{\Omega}\left(u,v,\underline{x}\right)=\Phi^{(-)}_{\Omega}\left(u,v,\underline{x}\right)+\Phi^{(+)}_{\Omega}\left(u,v,\underline{x}\right).
\label{eqn186}
\end{equation}
Therefore, there are two representations in terms of the Killing-Boulware modes ($KB^{*}$)
\begin{equation}
\Phi^{(\epsilon)}_{\Omega}(U,\underline{x}),\,\,\,\Phi^{(\epsilon)}_{\Omega}(V,\underline{x}),
\label{eqn187}
\end{equation}
that are seen by a FIDO observer. And another representation of Hartle-Hawking modes
\begin{equation}
\Psi^{(\epsilon)}_{\Omega}(U,\underline{x}),\,\,\,\Psi^{(\epsilon)}_{\Omega}(V,\underline{x}),
\label{eqn187a}
\end{equation}

where such modes are seen by a FFO observer\footnote{ The $KB^{*}$ modes have been defined as those modes of the scalar field that are measured by a FIDO observer. This type of observer is accelerated to a fixed distance above the BTZ spacetime horizon. And which are not the $KB$ modes of the quantum field for Schwarzschild space-time that were originally defined. The $HH^{*}$ modes are the modes of the quantum field, which are seen by a FFO observer, falling radially in the direction of the black hole in BTZ spacetime. And not to be confused with the $HH$ modes defined for Schwarzschild spacetime .} \cite{1986bhmp.book.....T,Mukohyama:1998rf,Arenas:2011be,RojasC:2021kws,Fursaev:2004qz} . The modes are orthogonal for a FIDO observer when

\begin{align}
    \left(\Phi^{(\epsilon)+}_{\Omega}(\underline{x}),\Phi^{(\epsilon')+}_{\Omega'}(\underline{x}')\right)	&=  \left(\Phi^{(\epsilon)-}_{\Omega}(\underline{x}),\Phi^{(\epsilon')-}_{\Omega'}(\underline{x}')\right) 
\notag\\
	&=\epsilon(\omega)\delta_{\Omega\Omega'}\delta_{\epsilon\epsilon'}
\label{eqn188}
\end{align}
And they are null when \cite{Arenas_2005}
\begin{equation}
\left(\Phi^{(\epsilon)\pm}_{\Omega}(\underline{x}),\Phi^{(\epsilon')\mp}_{\Omega'}(\underline{x}')\right)=0.
\label{eqn189}
\end{equation}
\begin{figure}[H]
\centering
		\includegraphics[width=0.45\textwidth]{ 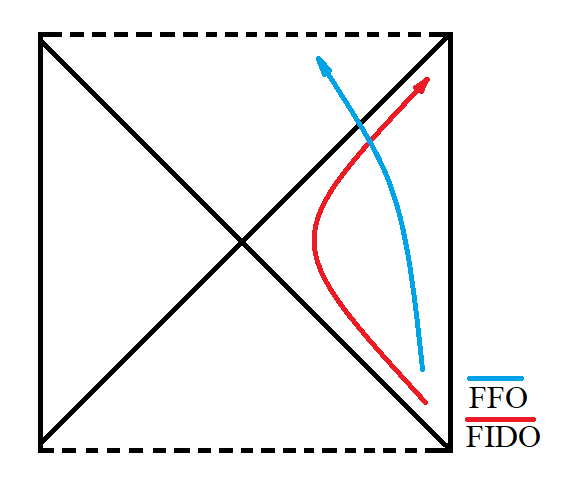}
\caption{Types of FIDO and FFO observers in the Carter-Penrose diagram for a BTZ black hole..}
\label{Fig9}
\end{figure}
And, for a FFO observer it is

\begin{equation}
\left(\Psi^{(\epsilon)+}_{\Omega}(\underline{x}),\Psi^{(\epsilon')+}_{\Omega'}(\underline{x}')\right)=  \left(\Psi^{(\epsilon)-}_{\Omega}(\underline{x}),\Psi^{(\epsilon')-}_{\Omega'}(\underline{x}')\right)=\epsilon(\omega)\delta_{\Omega\Omega'}\delta_{\epsilon\epsilon'}
\label{eqn190}
\end{equation}
And they are null when \cite{Arenas_2005}
\begin{equation}
\left(\Psi^{(\epsilon)\pm}_{\Omega}(\underline{x}),\Psi^{(\epsilon')\mp}_{\Omega'}(\underline{x}')\right)=0.
\label{eqn191}
\end{equation}
The relationship between the $KB^{*}$ and $HH^{*}$ modes is mediated by a Bogoliubov transformation of the form \cite{Birrell:1982ix}
\begin{equation}
\Psi^{(\epsilon)}_{\Omega}(\underline{x})=\Phi^{(\epsilon)}_{\Omega}(\underline{x})\cosh(x)+\Phi^{(-\epsilon)}_{\Omega}(\underline{x})\sinh(x).
\label{eqn192}
\end{equation}
To this end, consider the maximally extended BTZ space-time as

\begin{figure}[H]
\centering
		\includegraphics[width=0.45\textwidth]{ 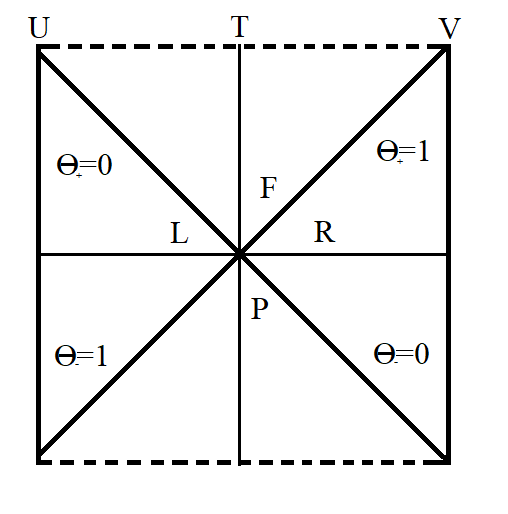}
\caption{  The Heaviside step function $\Theta_{\epsilon(x)}$ for the Carter-Penrose diagram   of a BTZ black hole.}
\label{Fig10}
\end{figure}

Figure \ref{Fig10} defines the Heaviside step function $\Theta_{\epsilon(x)}$ for the Carter-Penrose diagram   of a BTZ black hole. And this function is defined as 
\begin{equation}
\Theta_{\epsilon}(x)=\Theta_{-\epsilon}(x)=\frac{1}{2}\left[\Theta(-\epsilon U)+\Theta(\epsilon V)\right],
\label{eqn194}
\end{equation}
which must satisfy
\begin{equation}
\Theta_{\epsilon}+\Theta_{-\epsilon}=1.
\label{eqn195}
\end{equation}
For the Heaviside step function
\begin{equation}
H(-x)=1-H(x),
\label{eqn196}
\end{equation}
allows writing
\begin{equation}
\Theta_{\epsilon}-\Theta_{-\epsilon}=\Theta(\epsilon V)-\Theta(\epsilon U).
\label{eqn197}
\end{equation}
Furthermore, the Heaviside step function $H(x)$  and the function $\sgn(x)$
\begin{equation}
H(x)=\frac{1}{2}\left[1+\sgn(x)\right].
\label{eqn198}
\end{equation}
That allows writing
\begin{equation}
\Theta_{\epsilon}-\Theta_{-\epsilon}=\frac{1}{2}\left[\frac{\left|\epsilon V\right|}{\epsilon V}-\frac{\left|\epsilon U\right|}{\epsilon U}\right].
\label{eqn199}
\end{equation}
On the other hand, the logarithm of a complex number $Z=re^{i\theta}$, which leads to $W=\ln\left|Z\right|+n(2\pi i)$, with $n=1,2\ldots$. That is
\begin{equation}
\ln\left|Z\right|=\ln\left|r\right|+i\Arg Z,
\label{eqn200}
\end{equation}
where $Z=x+iy$,, the Euler identity $e^{i\phi}=\cos\phi+i\sin\phi$ and $\left|Z\right|=\sqrt{x^{2}+y^{2}}$. Obtaining

\begin{equation}
\phi=\Arg Z=\arctan \left(\frac{y}{x}\right), 
\label{eqn201}
\end{equation}
where
\begin{equation}
\arg. z=\Arg Z+2\pi n,\,\,\,n\in Z.
\label{eqn202}
\end{equation}
From the foregoing, consider the modes of positive frequency $\omega>0$  and the spectrum of the Fourier transform containing only positive frequencies. If such a function is regular and bounded in the middle of the complex plane, it is possible to write \cite{Israel2003}, $\ln_{+}\left|z\right|$ is real on the lower imaginary axis and branch cut on the lower half-plane.

From the foregoing, they are frequency functions \cite{Pulido_2008}

\begin{equation}
e^{\pm i\alpha t_{\epsilon}}=
 \begin{array}{rcl}
\ \omega>0,\,\,\, \epsilon=+1.
 \\
\omega<0,\,\,\, \epsilon=-1.
\end{array}
\label{eqn204}
\end{equation}

Defined for the null times $U,V$ and for $\alpha \in R$. The positive frequency modes ($\omega>0$) are those where the Fourier transform is null, in other words

\begin{equation}
\phi(\omega, \bar{x})=\frac{1}{2\pi}\int^{\infty}_{-\infty}e^{i\omega t}\phi(t, \vec{x})dt=0,
\label{eqn205}
\end{equation}
where $\phi(t, \vec{x})$ is an analytic and bounded function in the lower half-plane, then

\begin{equation}
e^{\pm i\alpha \ln_{+}\left|x\right|}=\int^{\infty}_{0}A_{\pm}(\omega)e^{-i\omega x}d\omega.
\label{eqn206}
\end{equation}

This allows extending \eqref{eqn205} as

\begin{equation}
\ln_{+}\left|Z\right|=\ln\left|Z\right|+i\left(\arg. Z +\frac{\pi}{2}\right),
\label{eqn207}
\end{equation}
\begin{equation}
\frac{-3\pi}{2}<\arg. Z<\frac{3\pi}{2},
\label{eqn207a}
\end{equation}
where $\ln_{+}\left|Z\right|$ is a regular function in the lower half-plane of the complex plane and $e^{\pm i\alpha \ln_{+}\left|Z\right|}$ is regular and bounded in the lower plane. Therefore

\begin{equation}
A_{\pm}(\omega)=\frac{1}{2\pi}\int^{\infty}_{-\infty}e^{\pm i\alpha \ln_{+}\left|Z\right|}e^{i\omega z} dz=0,\,\,\,\omega<0.
\label{eqn208}
\end{equation}
From the foregoing, it is possible to obtain \cite{Pulido_2008,Arenas_2005,Arenas:2011be,Israel2003,RojasC:2021kws}

\begin{equation}
\ln_{\epsilon}\left|x\right|=\ln\left|x\right|+\frac{i\pi}{2}\epsilon(x)\epsilon,\,\,\,-\infty<x<\infty,\,\,\,\epsilon=\pm 1.
\label{eqn209}
\end{equation}

For a BTZ black hole with $J=0$, the metric is defined as \eqref{eqn60} and \eqref{eqn61}. The surface gravity is \cite{Jiang:2007pn,Vagenas:2001sm}

\begin{equation}
\kappa_{0}=\frac{1}{2}\left(\frac{df(r)}{dr}\right)_{r_{+}}=\left(\frac{r}{l^{2}}\right)_{r_{+}}=\frac{\sqrt{M}}{l}.
\label{eqn210}
\end{equation}
The Hawking temperature for BTZ is

\begin{equation}
T_{H}=\frac{\kappa_{0}}{2\pi}=\frac{1}{2\pi}\frac{\sqrt{M}}{l}=\frac{\sqrt{\Lambda M}}{2\pi}.
\label{eqn211}
\end{equation}

An interesting aspect of BTZ black hole field solutions is that they admit a cosmological constant $\Lambda<0$, which is asymptotically Anti de Sitter $AdS$. Like black holes that are asymptotically flat, these solutions also admit thermal properties such as temperature and entropy. But they differ from the former because a black hole in $AdS$ has a minimum temperature, which occurs when its size is the AdS characteristic radius. In other words, for very large black holes, the redshift of the measured temperature at infinity is very large $T_{H}|_{r\longrightarrow\infty}\longrightarrow\infty$ \cite{Hawking:1982dh}. In other words, the Hawking radiation emitted by a BTZ Hole, gains more energy as it propagates. In contrast, the Hawking radiation emitted by a Schwarzschild black hole decreases with $1/r^{2}$. This is due to a possible radiation-gravitational field interaction.

\begin{figure}[H]
\centering
		\includegraphics[width=0.4\textwidth]{ 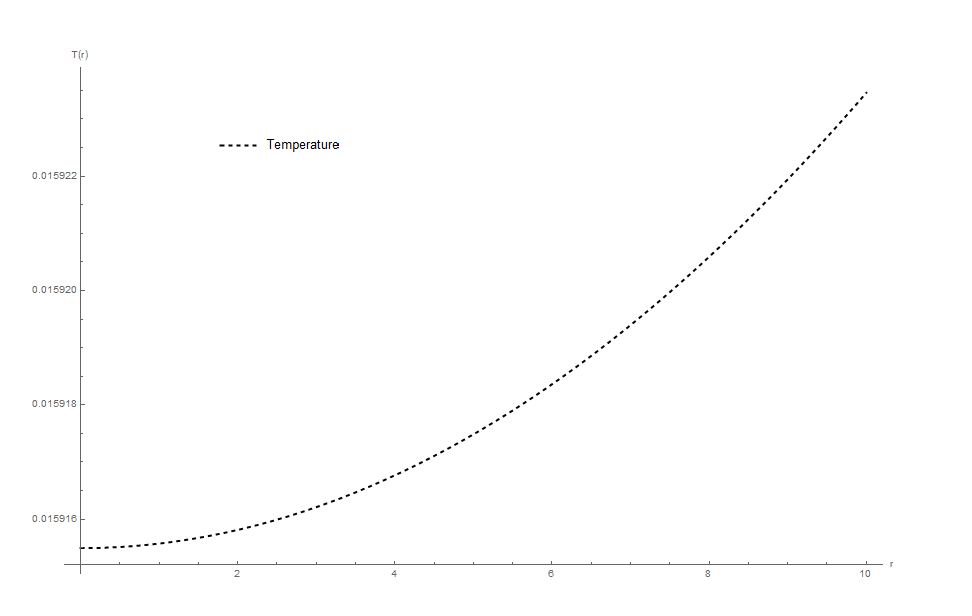}
\caption{Temperature behavior of BTZ when $r\longrightarrow\infty$ \cite{Hawking:1982dh}.}
\label{Fig1w}
\end{figure}

Taking the null coordinates $U,V$ given by \eqref{eqn34}, which are rewritten in terms of the surface gravity  \eqref{eqn210}

\begin{equation}
U=-e^{-\kappa_{0}u},\,\,\,V=e^{\kappa_{0} v}
\label{eqn211a}
\end{equation}

which are equivalent to
\begin{equation}
u=-\frac{1}{\kappa_{0}}\ln\left|U\right|,\,\,\,v=\frac{1}{\kappa_{0}}\ln\left|V\right|,
\label{eqn212}
\end{equation}
if, in addition, we consider the auxiliary coordinates $(u,v)$, see equation \eqref{eqn44}. Obtaining for \eqref{eqn212}

\begin{equation}
2\kappa_{0}t=\ln\left|V\right|-\ln\left|U\right|=\ln\left|\frac{V}{U}\right|.
\label{eqn213}
\end{equation}
Considering \eqref{eqn209}, it is possible to write

\begin{equation}
\ln_{\epsilon}\left|V\right|=\ln\left|V\right|+\frac{i\pi}{2}\epsilon(V)\epsilon,
\label{eqn214}
\end{equation}
\begin{equation}
\ln_{\epsilon}\left|U\right|=\ln\left|U\right|+\frac{i\pi}{2}\epsilon(U)\epsilon.
\label{eqn215}
\end{equation}
In other words,
\begin{equation}
2\kappa_{0}t=\ln\left|\frac{V}{U}\right|+i\pi \epsilon \frac{1}{2}\left[\epsilon (V)-\epsilon (U)\right].
\label{eqn216}
\end{equation}
Defining
\begin{equation}
\frac{1}{2}\epsilon \left[\epsilon (V)-\epsilon (U)\right]=\Theta_{\epsilon}-\Theta_{-\epsilon}.
\label{eqn217}
\end{equation}
Then \eqref{eqn216} is simplified to

\begin{equation}
2t_{\epsilon}\kappa_{0}=\ln\left|\frac{V}{U}\right|+i\pi \left[\Theta_{\epsilon}-\Theta_{-\epsilon}\right].
\label{eqn218}
\end{equation}
From Figure \ref{Fig10}, it follows that for region $R$, the values of the Heaviside step function, $\Theta_{\epsilon}=1$ and $\Theta_{-\epsilon}=0$. Therefore, \eqref{eqn218} is simplified to

\begin{equation}
2t_{+}\kappa_{0}=\ln\left|\frac{V}{U}\right|+i\pi,
\label{eqn219}
\end{equation}
moreover, if \eqref{eqn213} is considered, it is possible to rewrite \eqref{eqn219} as

\begin{equation}
t_{+}=t+\frac{i\pi}{2\kappa_{0}},
\label{eqn220}
\end{equation}
\begin{equation}
t_{\epsilon}=\left[t+\frac{i\pi}{2\kappa_{0}}\right]\Theta_{\epsilon}\in R.
\label{eqn221}
\end{equation}
Likewise, for region $L$

\begin{equation}
t_{-}=t-\frac{i\pi}{2\kappa_{0}},
\label{eqn222}
\end{equation}
\begin{equation}
t_{\epsilon}=\left[t-\frac{i\pi}{2\kappa_{0}}\right]\Theta_{-\epsilon}\in L.
\label{eqn223}
\end{equation}
In general, the global time $t_{\epsilon}$, can be written as a composition of the times for regions $R$ and $L$. Then

\begin{equation}
t_{\epsilon}=t_{+}\vert_{R}+t_{-}\vert_{L}.
\label{eqn224}
\end{equation}
Taking \eqref{eqn221}, multiplying by $i\omega$  and applying an exponential, allows to obtain

\begin{equation}
e^{-i\omega t_{\epsilon}}=e^{-i\omega t}e^{\frac{\pi\omega}{2\kappa_{0}}}\Theta_{\epsilon},
\label{eqn225}
\end{equation}
where \eqref{eqn225} corresponds to the positive frequency modes $\omega>0$ for region $R$. Similarly for region $L$

\begin{equation}
e^{-i\omega t_{\epsilon}}=e^{-i\omega t}e^{\frac{-\pi\omega}{2\kappa_{0}}}\Theta_{-\epsilon},
\label{eqn226}
\end{equation}
therefore, \eqref{eqn226} corresponds to the positive frequency modes $\omega>0$ for region $L$. From the foregoing, it follows that the positive frequency modes of the maximally extended BTZ scalar field correspond to the composition of \eqref{eqn225} and \eqref{eqn226}

\begin{equation}
e^{-i\omega t_{\epsilon}}=e^{-i\omega t}\left[e^{\frac{\pi\omega}{2\kappa_{0}}}\Theta_{\epsilon}+e^{\frac{-\pi\omega}{2\kappa_{0}}}\Theta_{-\epsilon}\right].
\label{eqn227}
\end{equation}
Define a discrete-valued function for the positive frequency modes

\begin{equation}
\epsilon'=\epsilon(\omega),
\label{eqn228}
\end{equation}
so that at time $t_{\epsilon}$, it depends on region $R$ or $L$ and on the possible discrete values for the modes of the scalar field in BTZ. This allows rewriting \eqref{eqn227}  as
\begin{equation}
e^{-i\omega t_{\epsilon\epsilon'}}=e^{-i\omega t}\left[e^{\frac{\pi\omega\epsilon'}{2\kappa_{0}}}\Theta_{\epsilon}+e^{\frac{-\pi\omega\epsilon'}{2\kappa_{0}}}\Theta_{-\epsilon}\right].
\label{eqn229}
\end{equation}
Let
\begin{equation}
e^{-\frac{\pi\left|\omega\right|}{\kappa_{0}}}=\tanh \chi,
\label{eqn230}
\end{equation}
then
\begin{equation}
e^{-\frac{\pi\left|\omega\right|}{2\kappa_{0}}}=\sqrt{\frac{\sinh \chi}{\cosh\chi}}.
\label{eqn231}
\end{equation}
So \eqref{eqn229}  is simplified to
\begin{equation}
e^{-i\omega t_{\epsilon\epsilon'}}=\frac{e^{-i\omega t}}{\sqrt{\sinh \chi \cosh \chi}}\left[ \cosh \chi\,\,\Theta_{\epsilon}+ \sinh \chi\,\,\Theta_{-\epsilon}\right]
\label{eqn232}
\end{equation}
\begin{equation}
e^{-i\omega t_{\epsilon\epsilon(\omega)}}\sqrt{\sinh \chi \cosh \chi}=e^{-i\omega t}\left[ \cosh \chi\,\,\Theta_{\epsilon}+ \sinh \chi\,\,\Theta_{-\epsilon}\right].
\label{eqn233}
\end{equation}
When considering the modes of the field by regions  $R$  and $L$, according to \eqref{eqn177}, it follows that
\begin{equation}
\Phi_{\Omega}^{(\epsilon)}=\Phi_{\Omega}(\underline{x})e^{-i\omega t}\Theta_{\epsilon}(x)\,\,\in R.
\label{eqn234}
\end{equation}
\begin{equation}
\Phi_{\Omega}^{(-\epsilon)}=\Phi_{\Omega}(\underline{x})e^{-i\omega t}\Theta_{-\epsilon}(x)\,\,\in L.
\label{eqn235}
\end{equation}
From the foregoing, it follows that \eqref{eqn233} is simplified to
\begin{equation}
e^{-i\omega t_{\epsilon\epsilon(\omega)}}\sqrt{\sinh \chi \cosh \chi}\,\,\Phi_{\Omega}=\Phi_{\Omega}^{(\epsilon)} \cosh \chi+ \Phi_{\Omega}^{(-\epsilon)}\sinh \chi.
\label{eqn232a}
\end{equation}
For \eqref{eqn232a}, the $HH^{*}$ modes are recognized for the scalar field in BTZ as
\begin{equation}
\Psi^{\epsilon}_{\Omega}(\underline{x})=e^{-i\omega t_{\epsilon\epsilon(\omega)}}\sqrt{\sinh \chi \cosh \chi}\,\,\Phi_{\Omega}(\underline{x}).
\label{eqn233b}
\end{equation}
From the foregoing, it is possible to obtain the Bogoliubov transformation between $HH^{*}$ modes and $KB^{*}$ modes defined as \cite{Arenas:2011be,Pulido_2008,Israel2003,RojasC:2021kws,Birrell:1982ix}
\begin{equation}
\Psi^{\epsilon}_{\Omega}(\underline{x})=\Phi^{\epsilon}_{\Omega}(\underline{x})\cosh\chi+\Phi^{-\epsilon}_{\Omega}(\underline{x})\sinh\chi.
\label{eqn233c}
\end{equation}

\section{QUANTUM FORMULATION}\label{sec5}
In this section, a quantum approximation to the scalar field in BTZ spacetime is considered. Let the scalar field modes be of the form \cite{klauber2013student}
\begin{equation}
\Phi_{\Omega}(t,\underline{x})=\frac{e^{-i\left[\omega t-\int\sqrt{\mathbf{T}}dr+\mathfrak{m}\phi\right]}}{\sqrt[4]{4\omega^{2}\mathbf{T}V^{2}}},
\label{eqn234a}
\end{equation}
where $V$ is the 2-Volume for BTZ spacetime. From \eqref{eqn234a}, it is possible to define
\begin{equation}
F_{\Omega}=\frac{e^{-i\left[\omega t-\int\sqrt{\mathbf{T}}dr+\mathfrak{m}\phi\right]}}{\sqrt[4]{4\omega^{2}\mathbf{T}V^{2}}},
\label{eqn235a}
\end{equation}
and its complex conjugate as
\begin{equation}
F^{*}_{\Omega}=\frac{e^{i\left[\omega t-\int\sqrt{\mathbf{T}}dr+\mathfrak{m}\phi\right]}}{\sqrt[4]{4\omega^{2}\mathbf{T}V^{2}}}.
\label{eqn236}
\end{equation}
Therefore, the field operator $\Phi_{\Omega}(t,\underline{x})$, is rewritten as
\begin{align}
\Phi_{\Omega}(t,\underline{x})&=\sum_{\Omega}{\left[a_{\Omega}F_{\Omega}+b^{\dagger}_{\Omega}F^{*}_{\Omega}\right]}   
\notag\\
&=\sum_{\Omega}{a_{\Omega}\frac{e^{-i\left[\omega t-\int\sqrt{\mathbf{T}}dr+\mathfrak{m}\phi\right]}}{\sqrt[4]{4\omega^{2}\mathbf{T}V^{2}}}+b^{\dagger}_{\Omega}\frac{e^{i\left[\omega t-\int\sqrt{\mathbf{T}}dr+\mathfrak{m}\phi\right]}}{\sqrt[4]{4\omega^{2}\mathbf{T}V^{2}}}}
&\hspace{0.3cm}
\label{eqn237}
\end{align}
and its conjugate hermitian
\begin{align}
\Phi^{*}_{\Omega}(t,\underline{x})&=\sum_{\Omega}{\left[a^{\dagger}_{\Omega}F^{*}_{\Omega}+b_{\Omega}F_{\Omega}\right]}   
\notag\\
&=\sum_{\Omega}{a^{\dagger}_{\Omega}\frac{e^{i\left[\omega t-\int\sqrt{\mathbf{T}}dr+\mathfrak{m}\phi\right]}}{\sqrt[4]{4\omega^{2}\mathbf{T}V^{2}}}+b_{\Omega}\frac{e^{-i\left[\omega t-\int\sqrt{\mathbf{T}}dr+\mathfrak{m}\phi\right]}}{\sqrt[4]{4\omega^{2}\mathbf{T}V^{2}}}}
&\hspace{0.3cm}
\label{eqn238}.
\end{align}
The field operators can be rewritten as
\begin{equation}
\Phi_{\Omega}(t,\underline{x})=\sum_{\Omega}\frac{1}{\sqrt[4]{4\omega^{2}\mathbf{T}V^{2}}}\left[a_{\Omega}e^{-ikx}+b^{\dagger}_{\Omega}e^{ikx}\right],
\label{eqn239}
\end{equation} 
and its conjugate hermitian
\begin{equation}
\Phi^{*}_{\Omega}(t,\underline{x})=\sum_{\Omega}\frac{1}{\sqrt[4]{4\omega^{2}\mathbf{T}V^{2}}}\left[a^{\dagger}_{\Omega}e^{ikx}+b_{\Omega}e^{-ikx}\right].
\label{eqn240}
\end{equation}
Where $\mu=0,1,2$, in other words
\begin{align}
kx&=k_{\mu}x^{\mu}= k_{0}x^{0}+k_{1}x^{1}+k_{2}x^{2}
\notag\\
&=g^{\mu\nu}k_{\mu}x_{\nu}=g^{00}k_{0}x_{0}+g^{ij}k_{i}x_{j}
&\hspace{0.3cm}
\notag\\
&=\omega t-\int\sqrt{\mathbf{T}}dr+\mathfrak{m}\phi
\label{eqn241}.
\end{align}
Such that for \eqref{eqn241} is valid if
\begin{equation}
g^{00}k_{0}x_{0}=\omega t,\,\,\,g^{ij}k_{i}x_{j}=-\int\sqrt{\mathbf{T}}dr+\mathfrak{m}\phi.
\label{eqn242}
\end{equation}
So, it is possible to write the 3-wave vector as
\begin{equation}
kx=\omega t-g^{ij}k_{i}x_{j}.
\label{eqn243}
\end{equation}
Then,  \eqref{eqn239} and \eqref{eqn240} are simplified to
\begin{widetext} 
\begin{equation}
\Phi_{\Omega}(t,\underline{x})=\sum_{\Omega}\frac{1}{\sqrt[4]{4\omega^{2}\mathbf{T}V^{2}}}\left[a_{\Omega}e^{-i\left(\omega t-g^{ij}k_{i}x_{j}\right)}+b^{\dagger}_{\Omega}e^{i\left(\omega t-g^{ij}k_{i}x_{j}\right)}\right],
\label{eqn244}
\end{equation}
\begin{equation}
\Phi^{*}_{\Omega}(t,\underline{x})=\sum_{\Omega}\frac{1}{\sqrt[4]{4\omega^{2}\mathbf{T}V^{2}}}\left[a^{\dagger}_{\Omega}e^{i\left(\omega t-g^{ij}k_{i}x_{j}\right)}+b_{\Omega}e^{-i\left(\omega t-g^{ij}k_{i}x_{j}\right)}\right].
\label{eqn245}
\end{equation}
\end{widetext}
The Lagrangian density must be considered for a complex scalar field as
\begin{widetext} 
\begin{align}
\mathcal{L}&=\frac{1}{2} \sqrt{-g}\left[g^{00}(\partial_{0}\Phi)^{2}\right]+\frac{1}{2}\sqrt{-g}g^{ij}(\partial_{i}\Phi)^{2}-m^{2}\Phi^{2} 
\notag\\
&=\frac{1}{2} \sqrt{-g}\left[g^{00}\partial_{0}\Phi^{\dagger}\partial_{0}\Phi\right]+\frac{1}{2}\sqrt{-g}g^{ij}\partial_{i}\Phi^{\dagger}\partial_{j}\Phi-m^{2}\Phi^{2}
&\hspace{0.3cm}
\notag\\
&=\frac{1}{2} \sqrt{-g}\left[g^{00}\dot{\Phi}^{\dagger}\dot{\Phi}\right]+\frac{1}{2}\sqrt{-g}g^{ij}\partial_{i}\Phi^{\dagger}\partial_{j}\Phi-m^{2}\Phi^{2}.
&\hspace{0.3cm}
\label{eqn246}
\end{align}
\end{widetext}
The Hamiltonian density $\mathcal{H}$, is obtained from a Legendre transformation
\begin{align}
\mathcal{H}&=\frac{\partial\mathcal{L}}{\partial\dot{\Phi}^{r}}  \dot{\Phi}^{r}-\mathcal{L}
\notag\\
&=\Pi  \dot{\Phi}+\Pi^{\dagger}  \dot{\Phi}^{\dagger}-\mathcal{L},
&\hspace{0.3cm}
\label{eqn247}
\end{align}
where canonically conjugate moments are
\begin{align}
 \Pi&=\frac{\partial\mathcal{L}}{\partial\dot{\Phi}}  
\notag\\
&=\frac{1}{2}\sqrt{-g}g^{00}\dot{\Phi}^{\dagger},
&\hspace{0.3cm}
\label{eqn248}
\end{align}
\begin{align}
 \Pi^{\dagger}&=\frac{\partial\mathcal{L}}{\partial\dot{\Phi}}  
\notag\\
&=\frac{1}{2}\sqrt{-g}g^{00}\dot{\Phi}.
&\hspace{0.3cm}
\label{eqn249}
\end{align}
Inserting \eqref{eqn248} and \eqref{eqn249} in \eqref{eqn247}
\begin{equation}
\mathcal{H}=\frac{1}{2}\sqrt{-g}g^{00}\dot{\Phi}^{\dagger}\dot{\Phi}+ \frac{1}{2}\sqrt{-g}g^{ij}\partial_{i}\Phi^{\dagger}\partial_{j}\Phi         +m^{2}\Phi^{\dagger}\Phi.
\label{eqn250}
\end{equation}
To obtain the complete Hamiltonian, the integral over 2-space must be considered
\begin{widetext} 
\begin{align}
H&=\int d^{2}x \mathcal{H} 
\notag\\
&=\int d^{2}x \left[\frac{1}{2}\sqrt{-g}g^{00}\dot{\Phi}^{\dagger}\dot{\Phi}+ \frac{1}{2}\sqrt{-g}g^{ij}\partial_{i}\Phi^{\dagger}\partial_{j}\Phi +m^{2}\Phi^{\dagger}\Phi\right]
&\hspace{0.3cm}
\notag\\
&=\frac{1}{2}\int d^{2}x  \left[\sqrt{-g}g^{00}\dot{\Phi}^{\dagger}\dot{\Phi}\right]+\frac{1}{2}\int d^{2}x\left[\sqrt{-g}g^{ij}\partial_{i}\Phi^{\dagger}\partial_{j}\Phi\right]+\int d^{2}x \,\,\left[m^{2}\Phi^{\dagger}\Phi\right].
\label{eqn251}
\end{align}
\end{widetext}
Calculating the first contribution in \eqref{eqn251}, which corresponds to the time component according to the field operator given in \eqref{eqn237} and \eqref{eqn238}, 
\begin{widetext} 
\[\frac{1}{2}\int d^{2}x  \left[\sqrt{-g}g^{00}\dot{\Phi}^{\dagger}\dot{\Phi}\right] = \frac{1}{2} \sum_{\Omega,\Omega'}  a^{\dagger}_{\Omega}a_{\Omega'}\int d^{2}x\sqrt{-g}g^{00}\frac{\partial F^{*}_{\Omega}}{\partial t}\frac{\partial F_{\Omega'}}{\partial t}+ a^{\dagger}_{\Omega}b^{\dagger}_{\Omega'}\int d^{2}x\sqrt{-g}g^{00}\frac{\partial F^{*}_{\Omega}}{\partial t}\frac{\partial F^{*}_{\Omega'}}{\partial t}
\]
\begin{equation}	+ b_{\Omega}a_{\Omega'}\int d^{2}x\sqrt{-g}g^{00}\frac{\partial F_{\Omega}}{\partial t}\frac{\partial F_{\Omega'}}{\partial t}+b_{\Omega}b^{\dagger}_{\Omega'}\int d^{2}x\sqrt{-g}g^{00}\frac{\partial F_{\Omega}}{\partial t}\frac{\partial F^{*}_{\Omega'}}{\partial t}.
 \label{eqn252}
\end{equation}
\end{widetext}
If we also take advantage of \eqref{eqn241}, it is possible to rewrite \eqref{eqn235a} and \eqref{eqn236}
\begin{equation}
F_{\Omega}=\frac{e^{-i\left[\omega t-g^{ij}k_{i}x_{j}\right]}}{\sqrt[4]{4\omega^{2}\mathbf{T}V^{2}}},
\label{eqn253}
\end{equation}
and its complex conjugate as
\begin{equation}
F^{*}_{\Omega}=\frac{e^{i\left[\omega t-g^{ij}k_{i}x_{j}\right]}}{\sqrt[4]{4\omega^{2}\mathbf{T}V^{2}}}.
\label{eqn254}
\end{equation}
Therefore, developing  \eqref{eqn252} 
\begin{equation}
\int d^{2}x\sqrt{-g}g^{00}\frac{\partial F^{*}_{\Omega}}{\partial t}\frac{\partial F_{\Omega'}}{\partial t}=\frac{\omega^{2}}{2\omega\sqrt{\mathbf{T}}}\frac{1}{\sqrt{f(r)}}
\label{eqn255}
\end{equation}
\begin{equation}
\int d^{2}x\sqrt{-g}g^{00}\frac{\partial F^{*}_{\Omega}}{\partial t}\frac{\partial F^{*}_{\Omega'}}{\partial t}=\frac{-\omega^{2}}{2\omega\sqrt{\mathbf{T}}}\frac{e^{2i\omega t}}{\sqrt{f(r)}}
\label{eqn256}
\end{equation}
\begin{equation}
\int d^{2}x\sqrt{-g}g^{00}\frac{\partial F_{\Omega}}{\partial t}\frac{\partial F_{\Omega'}}{\partial t}=\frac{-\omega^{2}}{2\omega\sqrt{\mathbf{T}}}\frac{e^{-2i\omega t}}{\sqrt{f(r)}}
\label{eqn257}
\end{equation}
\begin{equation}
\int d^{2}x\sqrt{-g}g^{00}\frac{\partial F_{\Omega}}{\partial t}\frac{\partial F^{*}_{\Omega'}}{\partial t}=\frac{\omega^{2}}{2\omega\sqrt{\mathbf{T}}}\frac{1}{\sqrt{f(r)}}.
\label{eqn258}
\end{equation}
Inserting \eqref{eqn255}-\eqref{eqn258} in \eqref{eqn252}, corresponds to the time portion of the Hamiltonian $H$ in \eqref{eqn251}
\begin{widetext} 
\begin{equation}
\frac{1}{2}\int d^{2}x  \left[\sqrt{-g}g^{00}\dot{\Phi}^{\dagger}\dot{\Phi}\right]=\frac{1}{2}\sum_{\Omega}\frac{\omega^{2}}{2\omega\sqrt{\mathbf{T}}}\frac{1}{\sqrt{f(r)}}\left[-a^{\dagger}_{\Omega}a_{\Omega}+a^{\dagger}_{\Omega}b^{\dagger}_{\Omega}e^{2i\omega t}+b_{\Omega}a_{\Omega}e^{-2i\omega t}-b_{\Omega}b^{\dagger}_{\Omega}\right].
\label{eqn259}
\end{equation}
\end{widetext}
Where 
\begin{equation}
\int d^{2}x\sqrt{-g}g^{00}=\int \frac{r}{f(r)}drd\phi
\label{eqn260}
\end{equation}
Is considered as the 2-volume differential element.

Consider the following term in \eqref{eqn251}, which corresponds to the spatial components
\begin{widetext} 

	\[
	\frac{1}{2}\int d^{2}x\left[\sqrt{-g}g^{ij}\partial_{i}\Phi^{\dagger}\partial_{j}\Phi\right]=\frac{1}{2} \sum_{\Omega,\Omega'}  a^{\dagger}_{\Omega}a_{\Omega'}\int d^{2}x\sqrt{-g}g^{ij}\frac{\partial F^{*}_{\Omega}}{\partial x^{i}}\frac{\partial F_{\Omega'}}{\partial x^{j}}+ a^{\dagger}_{\Omega}b^{\dagger}_{\Omega'}\int d^{2}x\sqrt{-g}g^{ij}\frac{\partial F^{*}_{\Omega}}{\partial x^{i}}\frac{\partial F^{*}_{\Omega'}}{\partial x^{j}}
\]

\begin{equation}
+ b_{\Omega}a_{\Omega'}\int d^{2}x\sqrt{-g}g^{ij}\frac{\partial F_{\Omega}}{\partial x^{i}}\frac{\partial F_{\Omega'}}{\partial x^{j}}+b_{\Omega}b^{\dagger}_{\Omega'}\int d^{2}x\sqrt{-g}g^{ij}\frac{\partial F_{\Omega}}{\partial x^{i}}\frac{\partial F^{*}_{\Omega'}}{\partial x^{j}}.
\label{eqn261}
\end{equation}

\end{widetext}
Developing \eqref{eqn261}

\begin{equation}
\int d^{2}x\sqrt{-g}g^{ij}\frac{\partial F^{*}_{\Omega}}{\partial x^{i}}\frac{\partial F_{\Omega'}}{\partial x^{j}}=\frac{k^{2}}{2\omega\sqrt{\mathbf{T}}}f^{3/2}(r)
\label{eqn262}
\end{equation}
\begin{equation}
\int d^{2}x\sqrt{-g}g^{ij}\frac{\partial F^{*}_{\Omega}}{\partial x^{i}}\frac{\partial F^{*}_{\Omega'}}{\partial x^{j}}=\frac{k^{2}e^{2i\omega t}}{2\omega\sqrt{\mathbf{T}}}f^{3/2}(r)
\label{eqn263}
\end{equation}
\begin{equation}
\int d^{2}x\sqrt{-g}g^{ij}\frac{\partial F_{\Omega}}{\partial x^{i}}\frac{\partial F_{\Omega'}}{\partial x^{j}}=\frac{k^{2}e^{-2i\omega t}}{2\omega\sqrt{\mathbf{T}}}f^{3/2}(r)
\label{eqn264}
\end{equation}
\begin{equation}
\int d^{2}x\sqrt{-g}g^{ij}\frac{\partial F_{\Omega}}{\partial x^{i}}\frac{\partial F^{*}_{\Omega'}}{\partial x^{j}}=\frac{k^{2}}{2\omega\sqrt{\mathbf{T}}}f^{3/2}(r).
\label{eqn265}
\end{equation}
Inserting \eqref{eqn262}-\eqref{eqn265} in \eqref{eqn261}, it corresponds to the geometric portion of the Hamiltonian $H$ in \eqref{eqn251}
\begin{widetext} 
\begin{equation}
\frac{1}{2}\int d^{2}x\left[\sqrt{-g}g^{ij}\partial_{i}\Phi^{\dagger}\partial_{j}\Phi\right]=\frac{1}{2} \sum_{\Omega} \frac{k^{2}}{2\omega\sqrt{\mathbf{T}}}f^{3/2}(r) \left[a^{\dagger}_{\Omega}a_{\Omega}+a^{\dagger}_{\Omega}b^{\dagger}_{\Omega}e^{2i\omega t}+b_{\Omega}a_{\Omega}e^{-2i\omega t}+b_{\Omega}b^{\dagger}_{\Omega}\right].
\label{eqn266}
\end{equation}
\end{widetext}
Consider the third term in \eqref{eqn251}, which corresponds to the mass term
\begin{widetext} 
\begin{equation}
\int d^{2}x \,\,m^{2}\Phi^{\dagger}\Phi=m^{2}\sum_{\Omega,\Omega'}\left[a_{\Omega}a^{\dagger}_{\Omega'}\int d^{2}x F_{\Omega}F^{*}_{\Omega'}+a_{\Omega}b_{\Omega'}\int d^{2}x F_{\Omega}F_{\Omega'}+b^{\dagger}_{\Omega}a^{\dagger}_{\Omega'}\int d^{2}x F^{*}_{\Omega}F^{*}_{\Omega'}+b^{\dagger}_{\Omega}b_{\Omega'}\int d^{2}x F^{*}_{\Omega}F_{\Omega'}\right].
\label{eqn267}
\end{equation}
\end{widetext}
Developing \eqref{eqn267}
\begin{equation}
\int d^{2}x F_{\Omega}F^{*}_{\Omega'}=\frac{1}{2\omega \sqrt{\mathbf{T}}}
\label{eqn268}
\end{equation}
\begin{equation}
\int d^{2}x F_{\Omega}F_{\Omega'}=\frac{e^{-2i\omega t}}{2\omega \sqrt{\mathbf{T}}}
\label{eqn269}
\end{equation}
\begin{equation}
\int d^{2}x F^{*}_{\Omega}F^{*}_{\Omega'}=\frac{e^{2i\omega t}}{2\omega \sqrt{\mathbf{T}}}
\label{eqn270}
\end{equation}
\begin{equation}
\int d^{2}x F^{*}_{\Omega}F_{\Omega'}=\frac{1}{2\omega \sqrt{\mathbf{T}}}
\label{eqn271}
\end{equation}
Inserting \eqref{eqn268}-\eqref{eqn271} in \eqref{eqn267}, it corresponds to the mass portion of the Hamiltonian $H$ in \eqref{eqn251}
\begin{widetext} 
\begin{equation}
\int d^{2}x \,\,m^{2}\Phi^{\dagger}\Phi=m^{2}\sum_{\Omega}\frac{1}{2\omega \sqrt{\mathbf{T}}}\left[a_{\Omega}a^{\dagger}_{\Omega}+a_{\Omega}b_{\Omega}e^{-2i\omega t}+b^{\dagger}_{\Omega}a^{\dagger}_{\Omega}e^{2i\omega t}+b^{\dagger}_{\Omega}b_{\Omega}\right].
\label{eqn272}
\end{equation}
\end{widetext}
Substituting  \eqref{eqn259}, \eqref{eqn266} and \eqref{eqn272}  into (290)
\begin{widetext} 
\begin{align}
H&=\frac{1}{4\omega\sqrt{\mathbf{T}}}\sum_{\Omega}\frac{\omega^{2}}{\sqrt{f(r)}}\left[a^{\dagger}_{\Omega}a_{\Omega}-a^{\dagger}_{\Omega}b^{\dagger}_{\Omega}e^{2i\omega t}-b_{\Omega}a_{\Omega}e^{-2i\omega t}+b_{\Omega}b^{\dagger}_{\Omega}\right]
\notag\\
&+ k^{2}f^{3/2}(r) \left[a^{\dagger}_{\Omega}a_{\Omega}+a^{\dagger}_{\Omega}b^{\dagger}_{\Omega}e^{2i\omega t}+b_{\Omega}a_{\Omega}e^{-2i\omega t}+b_{\Omega}b^{\dagger}_{\Omega}\right]
&\hspace{0.3cm}
\notag\\
&+m^{2}  \left[ a_{\Omega}a^{\dagger}_{\Omega}+a_{\Omega}b_{\Omega}e^{-2i\omega t}+b^{\dagger}_{\Omega}a^{\dagger}_{\Omega}e^{2i\omega t}+b^{\dagger}_{\Omega}b_{\Omega}\right],
\label{eqn273}
\end{align}
\end{widetext}
where $f(r)$ is determined by \eqref{eqn61},
\begin{equation}
f(r)=\left(-M+\frac{r^{2}}{l^{2}}\right).
\label{eqn274}
\end{equation}
If in addition, surface gravity \eqref{eqn210} is considered
\begin{equation}
\kappa_{0}=\frac{1}{2}\left(\frac{df(r)}{dr}\right)_{r_{+}},
\label{eqn275}
\end{equation}
it is possible to rewrite \eqref{eqn274}  in terms of the surface gravity $\kappa_{0}$ \cite{Mukohyama:1998rf}
\begin{equation}
f(r)\approx 2\kappa_{0}\epsilon=const.
\label{eqn276}
\end{equation}
Where, for \eqref{eqn276} 

\begin{itemize}
	\item $\kappa_{0}$ is constant over the horizon.
	\item $\epsilon$  is the cutoff measured by a FIDO  \cite{RojasC:2020qnz,2020Symm...12.2072R}. 
\end{itemize}
The foregoing makes it possible to simplify \eqref{eqn273}
\begin{widetext} 
\begin{align}
H&=\frac{1}{4\omega\sqrt{\mathbf{T}}}\sum_{\Omega}\frac{\omega^{2}}{\sqrt{2\kappa_{0}\epsilon}}\left[a^{\dagger}_{\Omega}a_{\Omega}-a^{\dagger}_{\Omega}b^{\dagger}_{\Omega}e^{2i\omega t}-b_{\Omega}a_{\Omega}e^{-2i\omega t}+b_{\Omega}b^{\dagger}_{\Omega}\right]
\notag\\
&+ k^{2}(2\kappa_{0}\epsilon)^{3/2} \left[a^{\dagger}_{\Omega}a_{\Omega}+a^{\dagger}_{\Omega}b^{\dagger}_{\Omega}e^{2i\omega t}+b_{\Omega}a_{\Omega}e^{-2i\omega t}+b_{\Omega}b^{\dagger}_{\Omega}\right]
&\hspace{0.3cm}
\notag\\
&+m^{2}  \left[ a_{\Omega}a^{\dagger}_{\Omega}+a_{\Omega}b_{\Omega}e^{-2i\omega t}+b^{\dagger}_{\Omega}a^{\dagger}_{\Omega}e^{2i\omega t}+b^{\dagger}_{\Omega}b_{\Omega}\right].
\label{eqn277}
\end{align}
\end{widetext}
With the condition that
\begin{equation}
\omega^{2}=k^{2}+m^{2},
\label{eqn288}
\end{equation}
which makes it possible to write
\begin{equation}
H=\sum_{\Omega}\frac{1}{2\sqrt{\mathbf{T}}}2\omega\left[ a_{\Omega}a^{\dagger}_{\Omega}+b_{\Omega}b^{\dagger}_{\Omega}\right].
\label{eqn289}
\end{equation}
Considering the commutation relations for the operators $\left[a_{\Omega},a^{\dagger}_{\Omega}\right]=\left[b_{\Omega},b^{\dagger}_{\Omega}\right]=1$, it is possible to obtain
\begin{equation}
H=\sum_{\Omega}\frac{1}{\sqrt{\mathbf{T}}}\omega\left[ a^{\dagger}_{\Omega}a_{\Omega}+b^{\dagger}_{\Omega}b_{\Omega}\right]+Z.P.E.,
\label{eqn290}
\end{equation}
where $Z.P.E$ is recognized as the zero-point energy, $N= a^{\dagger}_{\Omega}a_{\Omega}$ and $\bar{N}= b^{\dagger}_{\Omega}b_{\Omega}$. Thus, the Hamiltonian operator has been estimated for the scalar field in the proximity of the BTZ hole \cite{RojasC:2021kws,wald1994quantum,Ortiz:2011,Birrell:1982ix,greiner2013field,Arenas:2011be}.
\section{Thermo  Field  Dynamics on BTZ black hole}\label{sec6}
This section discusses the implications of Thermo Field Dynamics (TFD) in the proximity of a BTZ black hole \cite{Terashima:1999vw,Terashima:1999vw,Biswas:1987cp,Israel2003,Takahashi:1996zn,Israel:1976ur,Pulido_2008,Arenas_2005,Arenas:2011be,RojasC:2021kws}.

 Considering that the BTZ spacetime was written as \eqref{eqn157}, where $r_{+}$ is given by \eqref{eqn24} and the null coordinates $U,V$  by  \eqref{eqn34}. Which allows building the Carter-Penrose diagram as \cite{Banados:1992wn,Ortiz:2011,Larranaga:2009,2003qugr.book.....C}, see Figure \ref{Fig6}. Where, $(+)\in R$ and $(-)\in L$. In this context, the scalar field $\Phi(t,\underline{x})$ has a Hamiltonian of eigenvalues constituting the eigenvalues as

\begin{equation}
H^{(+)},\Ket{n}^{(+)}\in R,\,\,\,H^{(-)},\Ket{n}^{(-)}\in L,
\label{eqn291}
\end{equation}
and the eigenvalue equation

\begin{equation}
H^{(+)}\Ket{n}^{(+)}=E^{(+)}\Ket{n}^{(+)}\,\,\,\in R, 
\label{eqn292}
\end{equation}
\begin{equation}
H^{(-)}\Ket{n}^{(-)}=E^{(-)}\Ket{n}^{(-)}\,\,\,\in L.
\label{eqn292a}
\end{equation}
The TFD technique establishes the field Hamiltonian as a state of entanglement between the field 
\begin{equation}
\Phi^{(+)}(t,\underline{x})\in R,\,\,\, \Phi^{(-)}(t,\underline{x})\in L.
\label{eqn293}
\end{equation}
Consequently, the Hamiltonian $H$ of the complete field is determined as
\begin{widetext} 
\begin{align}
H&=H^{(+)}-H^{(+)}   
\notag\\
&=\sum_{\Omega}\omega\left[ a^{\dagger(+)}_{\Omega}a^{(+)}_{\Omega}+b^{\dagger(+)}_{\Omega}b^{(+)}_{\Omega}-a^{\dagger(-)}_{\Omega}a^{(-)}_{\Omega}-b^{\dagger(-)}_{\Omega}b^{(-)}_{\Omega}\right]\omega
&\hspace{0.3cm}
\notag\\
&=\sum_{\Omega}\left[N^{(+)}_{\Omega}-N^{(-)}_{\Omega}+\bar{N}^{(+)}_{\Omega}-\bar{N}^{(-)}_{\Omega}\right]\omega.
\label{eqn295}
\end{align}
\end{widetext}
Where the modes of the field $\Phi_{\Omega}(t,\underline{x})$ are expressed by regions $R,L$ in terms of the creation and annihilation operators of the particles and corresponding antiparticles \eqref{eqn237}

\begin{equation}
\Phi^{(+)}_{\Omega}(t,\underline{x})=\sum_{\Omega}{\left[a^{(+)}_{\Omega}F_{\Omega}+b^{\dagger(+)}_{\Omega}F^{*}_{\Omega}\right]} 
\label{eqn296}
\end{equation}
\begin{equation}
\Phi^{(-)}_{\Omega}(t,\underline{x})=\sum_{\Omega}{\left[a^{(-)}_{\Omega}F_{\Omega}+b^{\dagger(-)}_{\Omega}F^{*}_{\Omega}\right]}.
\label{eqn297}
\end{equation}
From the foregoing, eight modes are necessary to describe the scalar field $\Phi_{\Omega}(t,\underline{x})$ in BTZ. Consequently, having the commutation relations for the operators of creation and annihilation per region $R,L$ defined as

\begin{equation}
\left[ a^{(+)}_{\Omega},a^{\dagger(+)}_{\Omega}\right]=\left[b^{(+)}_{\Omega},b^{\dagger(+)}_{\Omega}\right]=1
\end{equation}

\begin{equation}
\left[a^{(-)}_{\Omega},a^{\dagger(-)}_{\Omega}\right]=\left[b^{(-)}_{\Omega},b^{\dagger(-)}_{\Omega}\right]=1,
\label{eqn297b}
\end{equation}

and the other possible combinations are null.

On the other hand, the normalization condition on the quantum states of the field is

\begin{equation}
\Ket{m^{(+)},n^{(+)},m^{(-)},n^{(-)}}=\Ket{m^{(+)}}\Ket{n^{(+)}}\Ket{m^{(-)}}\Ket{n^{(-)}},
\label{eqn298}
\end{equation}
for modes $\omega<0$ and  $\omega>0$.

 The completeness relation
\begin{widetext} 
\begin{equation}
I=\sum_{m^{(\pm)},n^{(\pm)}}=\Ket{m^{(+)},n^{(+)},m^{(-)},n^{(-)}}\Bra{m^{(+)},n^{(+)},m^{(-)},n^{(-)}}.
\label{eqn299}
\end{equation}
\end{widetext}
In addition, the empty states per region are defined as

\begin{equation}
a^{(+)}_{\Omega}\Ket{0}^{(+)+}_{B}=0,
\label{eqn300}
\end{equation}
vacuum state for particles $\omega>0$in region R.
\begin{equation}
b^{(+)}_{\Omega}\Ket{0}^{(+)-}_{B}=0,
\label{eqn301}
\end{equation}
vacuum state for particles $\omega<0$ in region R.
\begin{equation}
a^{(-)}_{\Omega}\Ket{0}^{(-)-}_{B}=0,
\label{eqn302}
\end{equation}
vacuum state for particles $\omega>0$  in region L.
\begin{equation}
b^{(-)}_{\Omega}\Ket{0}^{(-)+}_{B}=0,
\label{eqn301a}
\end{equation}
vacuum state for particles $\omega<0$  in region L.

From the foregoing, the vacuum state that includes the regions $R$ and $L$ is

\begin{equation}
\Ket{0}_{B}=\Ket{0}^{(+)+}_{B}\Ket{0}^{(+)-}_{B}\Ket{0}^{(-)-}_{B}\Ket{0}^{(-)+}_{B},
\label{eqn302a}
\end{equation}
such that the vacuum state $_{B}\left\langle 0\vert0\right\rangle_{B}=1$. The vacuum state is of the form
\begin{equation}
a^{(+)}_{\Omega}a^{(-)}_{\Omega}\Ket{0^{(+)},0^{(-)}}^{+}_{B}=0,
\label{eqn303}
\end{equation}
\begin{equation}
b^{(+)}_{\Omega}b^{(-)}_{\Omega}\Ket{0^{(+)},0^{(-)}}^{-}_{B}=0
\label{eqn304}
\end{equation}
This makes it possible to write the temperature-dependent vacuum state as
\begin{equation}
\Ket{0(\beta)}^{+}_{B}=\Ket{0^{(+)}(\beta),0^{(-)}(\beta)}^{+}_{B},
\label{eqn305}
\end{equation}
\begin{equation}
\Ket{0(\beta)}^{-}_{B}=\Ket{0^{(+)}(\beta),0^{(-)}(\beta)}^{-}_{B}.
\label{eqn305a}
\end{equation}
Consequently, the thermal vacuum state is

\begin{equation}
\Ket{0(\beta)}^{+}_{B}=\sum_{n}\frac{e^{\frac{-\beta E_{n}}{2}}}{n!\sqrt{Z(\beta)}}\left[a^{\dagger(+)}_{\Omega}\right]^{n}\left[a^{\dagger(-)}_{\Omega}\right]^{n}\Ket{0^{(+)},0^{(-)}}^{+}_{B},
\label{eqn306}
\end{equation}
\begin{equation}
\Ket{0(\beta)}^{-}_{B}=\sum_{m}\frac{e^{\frac{-\beta E_{m}}{2}}}{m!\sqrt{Z(\beta)}}\left[b^{\dagger(+)}_{\Omega}\right]^{m}\left[b^{\dagger(-)}_{\Omega}\right]^{m}\Ket{0^{(+)},0^{(-)}}^{-}_{B}.
\label{eqn306a}
\end{equation}
and the thermal vacuum state complete for  $R$ and $L$ regions on BTZ spacetime
\begin{widetext} 
\begin{align}
\Ket{0(\beta)}_{B}&=\Ket{0(\beta)}^{+}_{B}  \Ket{0(\beta)}^{-}_{B} 
\notag\\
&=\sum_{m,n} \frac{e^{\frac{\beta}{2}(E_{n}+E_{m})}}{m!n!Z(\beta)} \left[a^{\dagger(+)}_{\Omega}\right]^{n}\left[a^{\dagger(-)}_{\Omega}\right]^{n}\left[b^{\dagger(+)}_{\Omega}\right]^{m}\left[b^{\dagger(-)}_{\Omega}\right]^{m}\Ket{0^{(+)},0^{(-)}}^{+}_{B}\Ket{0^{(+)},0^{(-)}}^{-}_{B}.
&\hspace{0.3cm}
\label{eqn307}
\end{align}
\end{widetext}
If the vacuum states are subject to the normalization condition
\begin{equation}
_{B}^{+}\left\langle 0(\beta)|0(\beta)\right\rangle^{+}_{B}=_{B}^{-}\left\langle 0(\beta)|0(\beta)\right\rangle^{-}_{B}=1.
\label{eqn308}
\end{equation}
Such that it is possible to establish that the partition function is
\begin{equation}
Z^{+}(\beta)=Z^{-}(\beta)=\frac{1}{1-e^{-\beta\left|\omega\right|}},
\label{eqn309}
\end{equation}
which makes it possible to obtain the vacuum state as

\begin{equation}
\Ket{0(\beta)}^{+}_{B}=\sqrt{1-e^{-\beta\left|\omega\right|}} \exp\left[e^{-\frac{\beta|\omega|}{2}}a^{\dagger(+)}a^{\dagger(-)}\right]\Ket{0^{(+)},0^{(-)}},
\label{eqn310}
\end{equation}
\begin{equation}
\Ket{0(\beta)}^{-}_{B}=\sqrt{1-e^{-\beta\left|\omega\right|}} \exp\left[e^{-\frac{\beta|\omega|}{2}}b^{\dagger(+)}b^{\dagger(-)}\right]\Ket{0^{(+)},0^{(-)}}.
\label{eqn311}
\end{equation}

And the expected value of the occupation number as
\begin{equation}
\left\langle N\right\rangle^{+}=^{+}_{B}\Bra{0(\beta)}a^{\dagger(+)}a^{(+)}\Ket{0(\beta)}^{+}_{B},
\label{eqn312}
\end{equation}
\begin{equation}
\left\langle \bar{N}\right\rangle^{-}=^{-}_{B}\Bra{0(\beta)}b^{\dagger(+)}b^{(-)}\Ket{0(\beta)}^{+}_{B}.
\label{eqn313}
\end{equation}
To this end, it is necessary to write the temperature-dependent creation and annihilation operators $\beta$, as a Bogoliubov transformation of the form 

\begin{equation}
a^{(+)}=u(\beta)a^{(+)}(\beta)+v(\beta)a^{\dagger(-)}(\beta),
\label{eqn314}
\end{equation}
\begin{equation}
a^{\dagger(+)}=u(\beta)a^{\dagger(+)}(\beta)+v(\beta)a^{(-)}(\beta),
\label{eqn315}
\end{equation}
\begin{equation}
a^{(-)}=u(\beta)a^{(-)}(\beta)+v(\beta)a^{\dagger(+)}(\beta),
\label{eqn316}
\end{equation}
\begin{equation}
a^{\dagger(-)}=u(\beta)a^{\dagger(-)}(\beta)+v(\beta)a^{(+)}(\beta),
\label{eqn317}
\end{equation}
similarly,
\begin{equation}
b^{(+)}=u(\beta)b^{(+)}(\beta)+v(\beta)b^{\dagger(-)}(\beta),
\label{eqn318}
\end{equation}
\begin{equation}
b^{\dagger(+)}=u(\beta)b^{\dagger(+)}(\beta)+v(\beta)b^{(-)}(\beta),
\label{eqn319}
\end{equation}
\begin{equation}
b^{(-)}=u(\beta)b^{(-)}(\beta)+v(\beta)b^{\dagger(+)}(\beta),
\label{eqn320}
\end{equation}
\begin{equation}
b^{\dagger(+)}=u(\beta)b^{\dagger(-)}(\beta)+v(\beta)b^{(+)}(\beta),
\label{eqn321}
\end{equation}
where
\begin{align}
    	u(\beta)&=\frac{1}{\sqrt{1-e^{-\beta\omega}}}=\cosh \theta(\beta),   
\notag\\
	v(\beta)&=\frac{1}{\sqrt{e^{\beta\omega}-1}}=\sinh \theta(\beta).
&\hspace{0.3cm}
\label{eqn322a}
\end{align}
The foregoing makes it possible to obtain the expected value of the occupation number as \cite{Ribeiro_2004,PaniagodeSouza:2002nz,Pulido_2008,Takahashi:1996zn,Biswas:1987cp,Terashima:1999vw,RojasC:2021kws,Arenas:2011be}
\begin{equation}
\left\langle N\right\rangle^{+}=\left\langle \bar{N}\right\rangle^{-}=\frac{1}{e^{\beta|\omega|}-1}.
\label{eqn322}
\end{equation}

\section{MOMENTUM-ENERGY TENSOR FOR A SCALAR FIELD IN BTZ SPACETIME}\label{sec7}

It is known that the tensor $T_{\mu\nu}$ for a scalar field is defined as  \cite{RojasC:2021kws, Shiraishi:1993qnr, Arenas:2011be, PhysRevD.17.1477}
\begin{equation}
T_{\mu\nu}=-\frac{2}{\sqrt{-g}}\frac{\delta ( \mathcal{L}_{M}) }{\delta g^{\mu\nu}},
\label{eqn323}
\end{equation}
where the Lagrangian density $\mathcal{L}_{M}$  of the matter fields is defined by .\eqref{eqn158}. The variation makes it possible to obtain the tensor $T_{\mu\nu}$
\begin{equation}
T_{\mu\nu}=-\frac{2}{\sqrt{-g}}\left[\frac{\delta \sqrt{-g} }{\delta g^{\mu\nu}}\mathcal{L}_{M}  +\sqrt{-g}\frac{\delta \mathcal{L}_{M} }{\delta g^{\mu\nu}}\right],
\label{eqn324}
\end{equation}
where
\begin{equation}
\frac{\delta \sqrt{-g} }{\delta g^{\mu\nu}}=-\frac{1}{2}\sqrt{-g}g_{\mu\nu}.
\label{eqn325}
\end{equation}
Inserting \eqref{eqn325} in \eqref{eqn324}
\begin{align}
T_{\mu\nu}&=-\frac{2}{\sqrt{-g}}\left[-\frac{1}{2}\sqrt{-g}g_{\mu\nu}\mathcal{L}_{M}  +\sqrt{-g}\frac{\delta \mathcal{L}_{M} }{\delta g^{\mu\nu}} \right]
\notag\\
&=g_{\mu\nu}\mathcal{L}_{M}-2\frac{\delta \mathcal{L}_{M}}{\delta g^{\mu\nu}},
&\hspace{0.3cm}
\label{eqn326}
\end{align}

where
\begin{equation}
\frac{\delta \mathcal{L}_{M}}{\delta g^{\mu\nu}}=-\frac{1}{2}\partial_{\mu}\Phi\partial_{\nu}\Phi.
\label{eqn327}
\end{equation}
So, finally,
\begin{align}
T_{\mu\nu}&=-\frac{g^{\mu\nu}}{2}\left[g^{\alpha\beta}\partial_{\alpha}\Phi\partial_{\beta}\Phi  -m^{2}\Phi^{2}\right]+\partial_{\mu}\Phi \partial_{\nu}\Phi 
\notag\\
&=\left[\partial(_{\mu }\partial_{\nu'})-\frac{g^{\mu\nu}}{2}\left(\partial^{\beta}\partial_{\beta'}-m^{2}\right)\right]\Phi^{2}.
&\hspace{0.3cm}
\label{eqn326a}
\end{align}
The Wightman function corresponds to a two-point Green’s function of the form
\begin{equation}
\left\langle T_{\mu\mu}(x,x')\right\rangle=\mathcal{D}_{\mu\nu'}W(x,x'),
\label{eqn327a}
\end{equation}
where
\begin{equation}
\mathcal{D}_{\mu\nu'}=\partial(_{\mu }\partial_{\nu'})-\frac{g^{\mu\nu}}{2}\left(\partial^{\beta}\partial_{\beta'}-m^{2}\right),
\label{eqn328}
\end{equation}
and
\begin{equation}
\Phi^{2}(x,x')=W(x,x').
\label{eqn329}
\end{equation}
Such that $W(x,x')^{+}$, is defined as the Wightman function for positive frequency modes, in other 
\begin{equation}
W(x,x')=\Bra{0}\Phi(\underline{x})\Phi^{*}(\underline{x'})\Ket{0}.
\label{eqn330}
\end{equation}
Under the Killing-Boulware ($KB^{*}$) vacuum state scheme

\begin{equation}
W(x,x')_{KB^{*}}=_{KB^{*}}\Bra{0}\Phi^{(\epsilon)}_{\Omega}(\underline{x})\Phi^{*(\epsilon)}_{\Omega'}(\underline{x}')\Ket{0}_{KB^{*}}.
\label{eqn331}
\end{equation}
And also, under the Hartle-Hawking ($HH^{*}$) vacuum state scheme

\begin{equation}
W(x,x')_{HH^{*}}=_{HH^{*}}\Bra{0}\Psi^{(\epsilon)}_{\Omega}(\underline{x})\Psi^{*(\epsilon)}_{\Omega'}(\underline{x}')\Ket{0}_{HH^{*}}.
\label{eqn332}
\end{equation}
Moreover, considering that the modes of the field have been written as \eqref{eqn237} and \eqref{eqn238}for each of the regions $R,L$, in other words

\begin{equation}
\Phi^{(\epsilon)}_{\Omega}(t,\underline{x})=\sum_{\epsilon,\Omega}{\left[a^{(\epsilon)}_{\Omega}F^{(\epsilon)}_{\Omega}+b^{\dagger (\epsilon)}_{\Omega}F^{*(\epsilon)}_{\Omega}\right]}
\label{eqn333}
\end{equation}

and its conjugate hermitian
\begin{equation}
\Phi^{*(\epsilon)}_{\Omega}(t,\underline{x})=\sum_{\epsilon,\Omega}{\left[a^{\dagger(\epsilon)}_{\Omega}F^{*(\epsilon)}_{\Omega}+b^{(\epsilon)}_{\Omega}F^{(\epsilon)}_{\Omega}\right]},
\label{eqn334}
\end{equation}
where $\epsilon=\pm$. Considering the $KB^{*}$ vacuum state, it is expressed as
\begin{align}
\Ket{0}_{KB^{*}}&=\Ket{0^{(+)},0^{(-)}}^{+}_{KB^{*}}\otimes \Ket{0^{(+)},0^{(-)}}^{-}_{KB^{*}} 
\notag\\
&=\Ket{0^{(\epsilon)}}_{KB^{*}}\otimes\Ket{0^{(-\epsilon)}}_{KB^{*}}.
&\hspace{0.3cm}
\label{eqn335}
\end{align}
And the commutation rules

\begin{equation}
\left[a^{(\epsilon)}_{\Omega},a^{\dagger(\epsilon')}_{\Omega'}\right]=\left[b^{(\epsilon)}_{\Omega},b^{\dagger(\epsilon')}_{\Omega'}\right]=\epsilon\epsilon(\omega)\delta_{\epsilon\epsilon'}\delta_{\Omega\Omega'}.
\label{eqn336}
\end{equation}
From \eqref{eqn331}, \eqref{eqn333} and \eqref{eqn334}

\begin{align}
W(x,x')_{KB^{*}}&=_{KB^{*}}\Bra{0}\Phi^{*(\epsilon)}_{\Omega}(\underline{x})\Phi^{(\epsilon)}_{\Omega'}(\underline{x}')\Ket{0}_{KB^{*}} 
\notag\\
&= \sum_{\epsilon,\Omega,\epsilon',\Omega'}{\epsilon\epsilon(\omega)\delta_{\epsilon\epsilon'}\delta_{\Omega\Omega'}}F^{*(\epsilon)}_{\Omega}F^{(\epsilon)}_{\Omega'}
&\hspace{0.3cm}
\notag\\
&= \sum_{\epsilon,\Omega,\epsilon',\Omega'}F^{*(\epsilon)}_{\Omega}F^{(\epsilon)}_{\Omega'}\Theta(\epsilon\omega),
&\hspace{0.3cm}
\label{eqn337}
\end{align}
where $\epsilon(\omega)=\sgn{\omega}=\Theta{\epsilon \omega}$, $\delta_{\Omega\Omega'}=\delta_{\omega\omega'}\delta_{\mathfrak{m}\mathfrak{m}'}$, $\underline{x}=r,\phi$ and $\Omega=\omega,\mathfrak{m}$. The Wightman function under the  $HH^{*}$  scheme is
\begin{equation}
W(x,x')_{HH^{*}}=_{HH^{*}}\Bra{0}\Psi^{*(\epsilon)}_{\Omega}(\underline{x})\Psi^{(\epsilon)}_{\Omega'}(\underline{x}')\Ket{0}_{HH^{*}}, 
\label{eqn338}
\end{equation}
where the scalar field has been written as
\begin{equation}
\Psi^{(\epsilon)}_{\Omega}(U,V,x)=\sum_{\epsilon, \Omega}\left[d^{(\epsilon)}_{\Omega}G_{\Omega}+f^{\dagger(\epsilon)}_{\Omega}G^{*}_{\Omega}\right]
\label{eqn339}
\end{equation}
and its conjugate hermitian
\begin{equation}
\Psi^{*(\epsilon)}_{\Omega}(U,V,x)=\sum_{\epsilon, \Omega}\left[d^{\dagger(\epsilon)}_{\Omega}G^{*}_{\Omega}+f^{(\epsilon)}_{\Omega}G_{\Omega}\right].
\label{eqn340}
\end{equation}
At this point it should be clarified that there is an equivalence between the $KB^{*}$ modes  \eqref{eqn333},\eqref{eqn334} and the $HH^{*}$  modes \eqref{eqn339},\eqref{eqn340}. Such equivalence is mediated by a Bogoliubov transformation \eqref{eqn233}. Considering the $KB^{*}$ vacuum state, it is expressed as

\begin{align}
\Ket{0}_{HH^{*}}&=\Ket{0^{(+)},0^{(-)}}^{+}_{HH^{*}}\otimes \Ket{0^{(+)},0^{(-)}}^{-}_{HH^{*}} 
\notag\\
&=\Ket{0^{(\epsilon)}}_{HH^{*}}\otimes\Ket{0^{(-\epsilon)}}_{HH^{*}}.
&\hspace{0.3cm}
\label{eqn341}
\end{align}
And the commutation rules

\begin{equation}
\left[d^{(\epsilon)}_{\Omega},d^{\dagger(\epsilon')}_{\Omega'}\right]=\left[f^{(\epsilon)}_{\Omega},f^{\dagger(\epsilon')}_{\Omega'}\right]=\epsilon\epsilon(\omega)\delta_{\epsilon\epsilon'}\delta_{\Omega\Omega'}.
\label{eqn343}
\end{equation}

This makes it possible to estimate
\begin{widetext} 
\begin{align}
   W(x,x')_{HH^{*}}&=_{HH^{*}}\Bra{0}\Psi^{(\epsilon)}_{\Omega}(\underline{x})\Psi^{*(\epsilon)}_{\Omega'}(\underline{x}')\Ket{0}_{HH^{*}} 
\notag\\
	&=\sum_{\epsilon,\Omega,\epsilon',\Omega'}{\epsilon\epsilon(\omega)\delta_{\epsilon\epsilon'}\delta_{\Omega\Omega'}}G^{*(\epsilon)}_{\Omega}G^{(\epsilon)}_{\Omega'}
&\hspace{0.3cm}
\notag\\
	&=\sum_{\epsilon,\Omega,\epsilon',\Omega'}G^{*(\epsilon)}_{\Omega}G^{(\epsilon)}_{\Omega'}\Theta(\epsilon\omega)
&\hspace{0.3cm}
\notag\\
	&=\sum_{\epsilon,\Omega,\epsilon',\Omega'}\left[F^{(\epsilon)}_{\Omega}\cosh \chi+F^{(-\epsilon)}_{\Omega}\sinh \chi\right]\left[F^{*(\epsilon)}_{\Omega'}\cosh \chi+F^{*(-\epsilon)}_{\Omega'}\sinh \chi\right]
&\hspace{0.3cm}
\notag\\
	&=\sum_{\epsilon,\Omega}\left[F^{(\epsilon)}_{\Omega}F^{*(\epsilon)}_{\Omega}\cosh^{2}\chi +F^{(\epsilon)}_{\Omega}F^{*(-\epsilon)}_{\Omega}\cosh \chi\sinh \chi+F^{(-\epsilon)}_{\Omega}F^{*(\epsilon)}_{\Omega}\cosh \chi\sinh \chi+F^{(-\epsilon)}_{\Omega}F^{*(-\epsilon)}_{\Omega}\sinh^{2}\chi\right]\Theta(\epsilon\omega).
&\hspace{0.3cm}
\label{eqn344}
\end{align}
\end{widetext}
Where for \eqref{eqn344}, 
\begin{align}
  F^{(\epsilon)}_{\Omega}&=  F_{\Omega}\Theta_{\epsilon}(x) ,
\notag\\
	F^{*(\epsilon)}_{\Omega}			&=F^{*}_{\Omega}\Theta_{\epsilon}(x),
&\hspace{0.3cm}
\notag\\
F^{(-\epsilon)}_{\Omega}	&=F_{\Omega}\Theta_{-\epsilon}(x),
&\hspace{0.3cm}
\notag\\
	F^{*(-\epsilon)}_{\Omega} &=F^{*}_{\Omega}\Theta_{-\epsilon}(x).
&\hspace{0.3cm}
\label{eqn345}
\end{align}
And products for \eqref{eqn345}

\begin{equation}
F^{(\epsilon)}_{\Omega}	F^{*(-\epsilon)}_{\Omega} 	= 0.
\label{eqn346}
\end{equation}

\begin{equation}
F^{(-\epsilon)}_{\Omega}	F^{*(\epsilon)}_{\Omega} 	= 0.
\label{eqn346a}
\end{equation}

Therefore, it is possible to obtain
\begin{widetext} 
\begin{equation}
   W(x,x')_{HH^{*}}=\sum_{\epsilon,\Omega}\left[F^{(\epsilon)}_{\Omega}F^{*(\epsilon)}_{\Omega}\cosh^{2}\chi +F^{(-\epsilon)}_{\Omega}F^{*(-\epsilon)}_{\Omega}\sinh^{2}\chi\right]\Theta(\epsilon\omega).
\label{eqn347}
\end{equation}
\end{widetext}
The difference between \eqref{eqn347} and \eqref{eqn337} is
\begin{widetext} 
\begin{align}
     W(x,x')_{HH^{*}}- W(x,x')_{KB^{*}}	&=   \left( W_{HH^{*}}- W_{KB^{*}}\right)(x,x')
\notag\\
	&=\sum_{\epsilon,\Omega}\left[F^{(\epsilon)}_{\Omega}F^{*(\epsilon)}_{\Omega}\cosh^{2}\chi +F^{(-\epsilon)}_{\Omega}F^{*(-\epsilon)}_{\Omega}\sinh^{2}\chi- F^{(\epsilon)}_{\Omega}F^{*(\epsilon)}_{\Omega}  \right]           \Theta(\epsilon\omega)
&\hspace{0.3cm}
\notag\\
	&=\sum_{\epsilon,\Omega}\left[F^{(\epsilon)}_{\Omega}F^{*(\epsilon)}_{\Omega}(\cosh^{2}\chi-1) +F^{(-\epsilon)}_{\Omega}F^{*(-\epsilon)}_{\Omega}\sinh^{2}\chi  \right]           \Theta(\epsilon\omega)
&\hspace{0.3cm}
\notag\\
	&=   \sum_{+,\Omega} \Theta(+\omega) \sinh^{2}\chi\left[F^{(+)}_{\Omega}F^{*(+)}_{\Omega}  +F^{(-)}_{\Omega}F^{*(-)}_{\Omega} \right]                 
&\hspace{0.3cm}
\notag\\
	&+\sum_{-,\Omega} \Theta(-\omega) \sinh^{2}\chi \left[F^{(+)}_{\Omega}F^{*(+)}_{\Omega}  +F^{(-)}_{\Omega}F^{*(-)}_{\Omega} \right]. 
&\hspace{0.3cm}
\label{eqn348}
\end{align}
\end{widetext}
It is possible to constrain \eqref{eqn348} to one of the regions, so it is possible to obtain

\begin{widetext} 
\begin{equation}
 \left( W_{HH^{*}}- W_{KB^{*}}\right)(x,x')=\sum_{\Omega} \sinh^{2}\chi\left[F^{*(\epsilon)}_{\Omega}F^{(\epsilon)}_{\Omega}  +F^{*(-\epsilon)}_{\Omega}F^{(-\epsilon)}_{\Omega} \right].
\label{eqn349}
\end{equation}
\end{widetext}
Furthermore, according to \eqref{eqn211}, \eqref{eqn230} and \eqref{eqn322a}. It follows that \eqref{eqn349} is simplified to
\begin{equation}
\left( W_{HH^{*}}- W_{KB^{*}}\right)(x,x')=\sum_{\Omega} \frac{1}{e^{\beta\omega}-1}F^{*(+)}_{\Omega}F^{(+)}_{\Omega},
\label{eqn350}
\end{equation}
where, $F^{*(+)}_{\Omega},F^{(+)}_{\Omega}$are determined by \eqref{eqn235} and \eqref{eqn236}.

From the foregoing, it is feasible to obtain the component $\left\langle T_{00}(x,x')\right\rangle$   as

\begin{widetext} 
\begin{align}
    \partial_{0}\partial_{0'}	\left( W_{HH^{*}}- W_{KB^{*}}\right)(x,x')&=   \partial_{0}\partial_{0'}	\sum_{\Omega} \frac{1}{e^{\beta\omega}-1}\left[F^{*(+)}_{\Omega}F^{(+)}_{\Omega}\right]_{x=x'}
\notag\\
	&= \sum_{\Omega} \frac{1}{e^{\beta\omega}-1}\partial_{0}\partial_{0'}	\left[F^{*(+)}_{\Omega}F^{(+)}_{\Omega}\right]_{x=x'}
&\hspace{0.3cm}
\notag\\
	&=\frac{1}{2}\sum_{\omega} \frac{\omega}{e^{\beta\omega}-1}\sum_{\mathfrak{m}} \varphi^{*}_{\Omega}(r)\varphi(r)_{\Omega},
&\hspace{0.3cm}
\label{eqn351}
\end{align}
\end{widetext}
the sum over $\omega$, in the limit to the continuum can be expressed as an integral
\begin{equation}
\sum_{\omega} \frac{\omega}{e^{\beta\omega}-1}\longrightarrow \int^{\infty}_{-\infty}\frac{1}{e^{\beta\omega}-1}d\omega=2\int^{\infty}_{0}\frac{1}{e^{\beta\omega}-1}d\omega.
\label{eqn352}
\end{equation}

From the foregoing, it follows that
\begin{widetext} 
\begin{equation}
 \partial_{0}\partial_{0'}	\left( W_{HH^{*}}- W_{KB^{*}}\right)(x,x')=\int^{\infty}_{0}\frac{\omega}{e^{\beta\omega}-1}d\omega\sum_{\mathfrak{m}} \varphi^{*}_{\Omega}(r)\varphi(r)_{\Omega}.
\label{eqn353}
\end{equation}
\end{widetext}

Considering the BTZ metric  \eqref{eqn157}, where it follows that for the scalar field it is possible to obtain the Klein-Gordon equation \eqref{eqn160} and whose modes \eqref{eqn161} make it possible to obtain the radial equation determined as \eqref{eqn167} and \eqref{eqn168}. It is possible to consider a new transformation of the radial component of \eqref{eqn161} of the form
\begin{equation}
\varphi(r)_{\Omega}=\frac{1}{\sqrt{rf(r)}}\psi_{\Omega}(r).
\label{eqn354}
\end{equation}

Which makes it possible to obtain
\begin{equation}
\frac{d^{2}\psi(r)_{\Omega}}{dr^{2}}+\mathbf{T^{*}}^{2}\psi(r)_{\Omega}=0
\label{eqn355}
\end{equation}

\begin{equation}
\mathbf{T^{*}}^{2}=\frac{1}{f(r)}\left\{\frac{\omega^{2}}{f(r)}-m^{2}-\frac{\mathfrak{m}^{2}}{r}+B\right\}.
\label{eqn356}
\end{equation}
\begin{equation}
B=\frac{1}{2}\left[-\frac{d}{dr}\left(\frac{df(r)}{dr}-\frac{1}{2}f(r)\right) +\frac{d}{dr}\left(\ln|f(r)^{1/2}|\right)-\frac{f(r)}{2r^{2}}\right]
\label{eqn356a}
\end{equation}
Under the WKB approximation, the component $\varphi(r)_{\Omega}$is

\begin{equation}
\psi_{\Omega}(r)=\frac{1}{\sqrt[4]{4\omega^{2}\mathbf{T^{*}}^{2}}}e^{-i\int \mathbf{T^{*}}dr}.
\label{eqn357}
\end{equation} 
 The combination of \eqref{eqn354} and \eqref{eqn357}, allows for

\begin{equation}
\varphi^{*}_{\Omega}(r)\varphi(r)_{\Omega}=\frac{1}{2\omega r f(r)\mathbf{T^{*}}},
\label{eqn358}
\end{equation}
therefore, it follows that

\begin{widetext} 
\begin{equation}
 \partial_{0}\partial_{0'}	\left( W_{HH^{*}}- W_{KB^{*}}\right)(x,x')|_{x=x'}=\frac{1}{2r}\int^{\infty}_{0}\frac{\omega}{e^{\beta\omega}-1}d\omega \frac{1}{\omega}\sum_{\mathfrak{m}} \frac{1}{ f(r)\mathbf{T^{*}}}.
\label{eqn359}
\end{equation}
\end{widetext}
The sum over $\mathfrak{m}$ in the limit to the continuum is rewritten as

\begin{equation}
\sum_{\mathfrak{m}} \frac{1}{ f(r)\mathbf{T^{*}}}\longrightarrow \int^{\mathfrak{m}_{max}}_{0}\frac{1}{ f(r)\mathbf{T^{*}}}d\mathfrak{m},
\label{eqn360}
\end{equation}
which makes it possible to rewrite \eqref{eqn359}  as

\begin{widetext} 
 \begin{equation}
 \partial_{0}\partial_{0'}	\left( W_{HH^{*}}- W_{KB^{*}}\right)(x,x')|_{x=x'}=\frac{1}{2r}\int^{\infty}_{0}\frac{\omega}{e^{\beta\omega}-1}d\omega \frac{1}{\omega}\int^{\mathfrak{m}_{max}}_{0}\frac{1}{ f(r)\mathbf{T^{*}}}d\mathfrak{m}.
\label{eqn361}
\end{equation}
\end{widetext}

choosing that
\begin{equation}
\mathbf{T^{*}}^{2}=\mathbf{T^{*}}^{2}(r,\omega,\mathfrak{m})|_{\mathfrak{m}=\mathfrak{m}_{max}}=0
\label{eqn362}
\end{equation}
\begin{widetext} 
\begin{equation}
  \mathbf{T^{*}}^{2}_{max}=\frac{1}{f(r)}\left\{\frac{\omega^{2}}{f(r)}-m^{2}-\frac{\mathfrak{m}^{2}_{max}}{r}+B\right\}=0,
\label{eqn363}
\end{equation}
\end{widetext}

in other words,
\begin{widetext} 
\begin{align}
   \frac{\mathfrak{m}^{2}_{max}}{r}	&=  \frac{\omega^{2}}{f(r)}-m^{2}+B
\notag\\
	&=p^{2}.
\label{eqn364}
\end{align}
\end{widetext}
Therefore, it follows that \eqref{eqn362} is simplified to

\begin{equation}
 \mathbf{T^{*}}(r,\omega,\mathfrak{m})|_{\mathfrak{m}=\mathfrak{m}_{max}}=\sqrt{\frac{1}{f(r)}\left[p^{2}-\frac{\mathfrak{m}^{2}}{r}\right]}.
\label{eqn365}
\end{equation}
Inserting \eqref{eqn365} in

\begin{equation}
\int^{\mathfrak{m}_{max}}_{0}\frac{1}{ f(r)\mathbf{T^{*}}}d\mathfrak{m}=\int^{\mathfrak{m}_{max}}_{0}\frac{1}{ f(r)\sqrt{\frac{1}{f(r)}\left[p^{2}-\frac{\mathfrak{m}^{2}}{r}\right]}}d\mathfrak{m},
\label{eqn366}
\end{equation}
And then \eqref{eqn361}

\begin{widetext} 
\begin{align}
    \partial_{0}\partial_{0'}	\left( W_{HH^{*}}- W_{KB^{*}}\right)(x,x')|_{x=x'}  	&=\frac{1}{2r}\int^{\infty}_{0}\frac{\omega}{e^{\beta\omega}-1}d\omega \frac{1}{\omega} \int^{\mathfrak{m}_{max}}_{0}\frac{1}{ f(r)\sqrt{\frac{1}{f(r)}\left[p^{2}-\frac{\mathfrak{m}^{2}}{r}\right]}}d\mathfrak{m}
\notag\\
	&=\frac{1}{2r}\int^{\infty}_{0}\frac{\omega}{e^{\beta\omega}-1}d\omega \frac{1}{\omega} \frac{\sqrt{rf(r)}}{f(r)}\int^{\mathfrak{m}_{max}}_{0}\frac{1}{ \sqrt{\left[p^{2}r-\mathfrak{m}^{2}\right]}}d\mathfrak{m}.
&\hspace{0.3cm}
\label{eqn367}
\end{align}
\end{widetext}
Integrating by $\mathfrak{m}$ is

\begin{widetext} 
\begin{equation}
\int^{\mathfrak{m}_{max}}_{0}\frac{1}{ \sqrt{\left[p^{2}r-\mathfrak{m}^{2}\right]}}d\mathfrak{m}=-\arctan\left[\frac{\mathfrak{m}_{max}\sqrt{p^{2}r+\mathfrak{m}^{2}_{max}}}{p^{2}r-\mathfrak{m}_{max}}\right]=\arctan\left[\frac{\mathfrak{m}_{max}}{\sqrt{p^{2}r-\mathfrak{m}^{2}_{max}}}\right].
\label{eqn368}
\end{equation}
\end{widetext}

It is possible to expand \eqref{eqn368} in a Taylor series

\begin{align}
 \arctan\left[\frac{\mathfrak{m}_{max}}{\sqrt{p^{2}r-\mathfrak{m}^{2}_{max}}}\right]&=\frac{\sqrt{p^{2}r}\mathfrak{m}_{max}}{p^{2}r}+\frac{\sqrt{p^{2}r}\mathfrak{m}^{3}_{max}}{6p^{4}r^{2}} +\ldots 
\notag\\
	&\approx \frac{1}{p\sqrt{r}}\mathfrak{m}_{max}
&\hspace{0.3cm}
\notag\\
	& \approx 1.
&\hspace{0.3cm}
\label{eqn369}
\end{align}

Consequently, \eqref{eqn367}  is simplified to

\begin{widetext} 
\begin{equation}
 \partial_{0}\partial_{0'}	\left( W_{HH^{*}}- W_{KB^{*}}\right)(x,x')|_{x=x'}  =\frac{1}{2r}\int^{\infty}_{0}\frac{\omega}{e^{\beta\omega}-1}d\omega \frac{1}{\omega}\frac{\sqrt{rf(r)}}{f(r)}.
\label{eqn370}
\end{equation}
\end{widetext}

With local energy per mode
\begin{equation}
E=\frac{\omega}{\sqrt{f(r)}}
\label{eqn371}
\end{equation}
and the local temperature
\begin{equation}
T(r)=\frac{T_{H}}{\sqrt{f(r)}},\,\,\,T_{H}=\frac{1}{\beta},
\label{eqn372}
\end{equation}
then
\begin{equation}
\beta\omega=\frac{E}{T(r)}.
\label{eqn373}
\end{equation}
Considering the relativistic energy $E^{2}=m^{2}+p^{2}$, with the condition that $E\gg m$, makes it possible to obtain

\begin{equation}
\frac{\omega d\omega}{f(r)}=pdp.
\label{eqn374}
\end{equation}
Then
\begin{align}
    \partial_{0}\partial_{0'}	\left( W_{HH^{*}}- W_{KB^{*}}\right)(x,x')|_{x=x'} 	&=\int^{\infty}_{0}\frac{E}{e^{E/T(r)}-1} p dE f(r)\Delta 
\notag\\
	&= \int^{\infty}_{0}\frac{E}{e^{E/T(r)}-1} pdp f(r)\Delta 
&\hspace{0.3cm}
\label{eqn375}
\end{align}
where $\Delta=\frac{2\pi}{\omega^{2}\sqrt{r}}$. It is possible to assert that $\Delta|_{r=r_{+}}=\frac{2\pi}{\omega^{2}\sqrt{r_{+}}}\approx const$. This is possible, since the shell is in a meta-stable phase according to \eqref{eqn156} \cite{RojasC:2020qnz,2020Symm...12.2072R}. Moreover, it follows that
\begin{equation}
g^{00}g_{00}+g^{0\beta}g_{0\beta}=1.
\label{eqn376}
\end{equation}
Then $g^{00}g_{00}=1$,  also $g^{00}\partial_{0}=\partial^{0}$, then \eqref{eqn375}
\begin{widetext} 
\begin{align}
   -g^{00} \partial_{0}\partial_{0'}	\left( W_{HH^{*}}- W_{KB^{*}}\right)(x,x')|_{x=x'}  	&=- \partial^{0}\partial_{0'}	\left( W_{HH^{*}}- W_{KB^{*}}\right)(x,x')|_{x=x'} 
\notag\\
	&= \int^{\infty}_{0}\frac{E}{e^{E/T(r)}-1} pdp\, \Delta. 
&\hspace{0.3cm}
\label{eqn377}
\end{align}
\end{widetext}
Considering \eqref{eqn328} with $\mu=\nu=0$

\begin{widetext}
\begin{equation}
\mathcal{D}_{00'}=\left[\partial(_{0 }\partial_{0'})-\frac{g_{00}}{2}\left(\partial^{0}\partial_{0'}-m^{2}\right)\right]\left( W_{HH^{*}}- W_{KB^{*}}\right)(x,x')|_{x=x'}
\label{eqn378}
\end{equation}
\end{widetext}
and
\begin{equation}
\left(\partial^{\beta}\partial_{\beta'}-m^{2}\right)\left( W_{HH^{*}}- W_{KB^{*}}\right)(x,x')|_{x=x'}, 
\label{eqn379}
\end{equation}
where, 
\begin{equation}
\left|\partial_{\mu}\Phi\right|^{2}=\partial_{\mu}\Phi^{*}\partial_{\mu}\Phi.
\label{eqn380}
\end{equation}
In other words, 
\begin{widetext} 
\begin{align}
    	g^{\mu\nu}\partial_{\mu}\Phi^{*}\partial_{\nu}\Phi&=  g^{tt}\partial_{t}\Phi^{*}\partial_{t}\Phi+g^{rr}\partial_{r}\Phi^{*}\partial_{r}\Phi+g^{\phi\phi}\partial_{\phi}\Phi^{*}\partial_{\phi}\Phi 
\notag\\
	&=-\frac{1}{f(r)}\partial_{t}\Phi^{*}\partial_{t}\Phi+f(r)\partial_{r}\Phi^{*}\partial_{r}\Phi+\frac{1}{r^{2}}\partial_{\phi}\Phi^{*}\partial_{\phi}\Phi 
&\hspace{0.3cm}
\notag\\
	&=\left[-\frac{\omega^{2}}{f(r)}\varphi^{2}_{\Omega}(r)+f(r)\left|\frac{\partial\varphi_{\Omega}(r)}{\partial r}\right|^{2}\right]\frac{1}{2\omega}+\frac{1}{r^{2}}\frac{\varphi^{2}_{\Omega}(r)}{2\omega}\mathfrak{m}^{2}.
&\hspace{0.3cm}
\label{eqn381}
\end{align}
\end{widetext}
Where $\Omega=\omega,\mathfrak{m}$  for BTZ, in addition, considering that $\varphi_{\Omega}(r)$ is broken down into the incoming \eqref{eqn180} and outgoing \eqref{eqn180a} modes. Consequently,
 
\begin{widetext} 
\begin{equation}
2\omega \sum_{\mathfrak{m}}g^{\mu\nu}\partial_{\mu}\Phi^{*}\partial_{\nu}\Phi=\sum_{\mathfrak{m}} \left[-\frac{\omega^{2}}{f(r)}\varphi^{2}_{\Omega}(r)+f(r)\left|\frac{\partial\varphi_{\Omega}(r)}{\partial r}\right|^{2}+\frac{\varphi^{2}_{\Omega}(r)}{r^{2}}\mathfrak{m}^{2}\right],
\label{eqn382}
\end{equation}
\end{widetext}
where the term,

\begin{equation}
\left|\frac{\partial\varphi_{\Omega}(r)}{\partial r}\right|^{2}=\frac{d}{dr}\left[\ln\left|\varphi_{\Omega}(r)\right|\right]^{2}\varphi_{\Omega}^{2}(r).
\label{eqn383}
\end{equation}
Therefore,
\begin{widetext} 
\begin{equation}
2\omega \sum_{\mathfrak{m}}g^{\mu\nu}\partial_{\mu}\Phi^{*}\partial_{\nu}\Phi=\sum_{\mathfrak{m}} \varphi_{\Omega}^{2}(r)\left[-\frac{\omega^{2}}{f(r)}+f(r)\frac{d}{dr}\left[\ln\left|\varphi_{\Omega}(r)\right|\right]^{2}     +\frac{\mathfrak{m}^{2}}{r^{2}}\right].
\label{eqn384}
\end{equation}
\end{widetext}
According to \eqref{eqn354} and \eqref{eqn357}
\begin{equation}
\ln\left|\varphi_{\Omega}(r)\right|=-i\int \mathbf{T^{*}} dr-\frac{1}{2}\ln\left|rf(r)\mathbf{T^{*}}\right|+cte
\label{eqn385}
\end{equation}
and
\begin{equation}
\ln\left|\varphi^{*}_{\Omega}(r)\right|=i\int \mathbf{T^{*}} dr-\frac{1}{2}\ln\left|rf(r)\mathbf{T^{*}}\right|+cte.
\label{eqn386}
\end{equation}
Make it possible to write

\begin{equation}
\frac{d}{dr}\left[\ln\left|\varphi_{\Omega}(r)\right|\right]^{2}=\mathbf{T^{*}}^{2}+\frac{1}{4}\frac{d}{dr}\ln\left|rf(r)\mathbf{T^{*}}\right|.
\label{eqn387}
\end{equation}
Which makes it possible to simplify \eqref{eqn384}
\begin{widetext}
\begin{equation}
2\omega \sum_{\mathfrak{m}}g^{\mu\nu}\partial_{\mu}\Phi^{*}\partial_{\nu}\Phi=\sum_{\mathfrak{m}} \varphi_{\Omega}^{2}(r)\left[-m^{2}-\frac{\mathfrak{m}^{2}}{r} -f(r)\frac{\mathfrak{m}^{2}}{r^{2}} +\alpha_{\omega,\mathfrak{m}}(r)\right].
\label{eqn388}
\end{equation}
\end{widetext}
where
\begin{widetext} 
\begin{equation}
\alpha_{\omega,\mathfrak{m}}(r)=-\frac{1}{2}\frac{d}{dr}\left[\frac{df(r)}{dr}+\ln\left|f(r)\right|\right]+f(r)\left[\frac{1-r}{4r}\right]
+\frac{f(r)}{4}\frac{d}{dr}\left[\ln\left|f(r)\mathbf{T^{*}}\right|\right].
\label{eqn389}
\end{equation}
\end{widetext}
Considering the high frequency range and close to the gravitational radius $r=r_{+}$ , then
\begin{equation}
\lim_{r\longrightarrow r_{+}}\alpha_{\omega,\mathfrak{m}}(r)=0,
\label{eqn390}
\end{equation}
In other words \eqref{eqn390},
\begin{widetext} 
\begin{equation}
2\omega \sum_{\mathfrak{m}}g^{\mu\nu}\partial_{\mu}\Phi^{*}\partial_{\nu}\Phi=\sum_{\mathfrak{m}} \varphi_{\Omega}^{2}(r)\left[-m^{2}-\frac{\mathfrak{m}^{2}}{r} -f(r)\frac{\mathfrak{m}^{2}}{r^{2}} \right]+\sum_{\mathfrak{m}} \varphi_{\Omega}^{2}(r)\alpha_{\omega,\mathfrak{m}}(r).
\label{eqn391}
\end{equation}
\end{widetext}
For large values of  $m$
\begin{equation}
\lim_{r\longrightarrow\infty}\left[-m^{2}-\frac{\mathfrak{m}}{r} -f(r)\frac{\mathfrak{m}}{r^{2}} \right]=-m^{2}
\label{eqn392}
\end{equation}
then
\begin{widetext} 
\begin{equation}
2\omega \frac{1}{2}\sum_{\eta=\pm}\sum_{\mathfrak{m}}g^{\mu\nu}\partial_{\mu}\Phi^{*}\partial_{\nu}\Phi= \frac{1}{2}\sum_{\eta=\pm}\sum_{\mathfrak{m}} \varphi_{\Omega}^{2}(r)\left[-m^{2} \right]+ \frac{1}{2}\sum_{\eta=\pm}\sum_{\mathfrak{m}} \varphi_{\Omega}^{2}(r)\alpha_{\omega,\mathfrak{m}}(r),
\label{eqn393}
\end{equation}
\end{widetext}
where the two possible directions of the incoming and outgoing modes have been considered.

\begin{equation}
g^{\mu\nu}\partial_{\mu}\Phi^{*}\partial_{\nu}\Phi+\Phi_{\Omega}^{2}(r)m^{2} =  \Phi_{\Omega}^{2}(r)\frac{\alpha_{\omega,\mathfrak{m}}(r)}{2\omega}.
\label{eqn394}
\end{equation}

Which makes it possible to write
\begin{equation}
g^{\mu\nu}\partial_{\mu}\partial_{\nu}+m^{2}=\frac{\alpha_{\omega,\mathfrak{m}}(r)}{2\omega}
\label{eqn395}
\end{equation}
Then \eqref{eqn350}

\begin{widetext} 
\begin{align}
    \left[g^{\mu\nu}\partial_{\mu}\partial_{\nu}+m^{2}\right]\left( W_{HH^{*}}- W_{KB^{*}}\right)(x,x')	&=  \sum_{\Omega} \frac{1}{e^{\beta\omega}-1}F^{*(+)}_{\Omega}F^{(+)}_{\Omega}\frac{\alpha_{\omega,\mathfrak{m}}(r)}{2\omega}, 
\notag\\
	 \left[\partial^{\mu'}\partial_{\mu}+m^{2}\right]\left( W_{HH^{*}}- W_{KB^{*}}\right)(x,x')	&=\frac{1}{2}\sum_{\omega}\frac{\omega^{-1}}{e^{\beta\omega}-1}\sum_{\mathfrak{m}}F^{*(+)}_{\Omega}F^{(+)}_{\Omega}\alpha_{\omega,\mathfrak{m}}(r).
\label{eqn396}
\end{align}
\end{widetext}
In the limit to the continuum over $\omega$

\begin{widetext} 
\begin{equation}
 \left[\partial^{\mu'}\partial_{\mu}+m^{2}\right]\left( W_{HH^{*}}- W_{KB^{*}}\right)(x,x')=\frac{1}{2} \int^{\infty}_{-\infty} \frac{\omega^{-1}}{e^{\beta\omega}-1}\sum_{\mathfrak{m}}F^{*(+)}_{\Omega}F^{(+)}_{\Omega}\alpha_{\omega,\mathfrak{m}}(r).
\label{eqn397}
\end{equation}
\end{widetext}
Considering \eqref{eqn327}  and \eqref{eqn353} 

\begin{widetext} 
\begin{align}
   \left\langle T_{\mu\nu} (x,x')\right\rangle	&=  \mathcal{D}_{\mu\nu'}W(x,x') 
\notag\\
	&=\left[\partial(_{\mu}\partial_{\nu'})-\frac{g_{\mu\nu}}{2}\left(\partial^{\beta'}\partial_{\beta}-m^{2}\right)\right]\left( W_{HH^{*}}- W_{KB^{*}}\right)(x,x')
&\hspace{0.3cm}
\notag\\
	&=\partial(_{\mu}\partial_{\nu'})\left( W_{HH^{*}}- W_{KB^{*}}\right)(x,x')-\frac{g_{00}}{2}\left(\partial^{\beta'}\partial_{\beta}-m^{2}\right)\left( W_{HH^{*}}- W_{KB^{*}}\right)(x,x').
&\hspace{0.3cm}
\label{eqn398}
\end{align}
\end{widetext}
For the time components of the tensor $T_{\mu\nu}$

\begin{widetext} 
\begin{equation}
  \left\langle T_{00} (x,x')\right\rangle	=\partial(_{0}\partial_{0'})\left( W_{HH^{*}}- W_{KB^{*}}\right)(x,x')-\frac{g_{00}}{2}\left(\partial^{\beta}\partial_{\beta'}-m^{2}\right)\left( W_{HH^{*}}- W_{KB^{*}}\right)(x,x').
\label{eqn399}
\end{equation}
\end{widetext}

\begin{widetext}
\begin{align}
    	g^{00}\left\langle T_{00} (x,x')\right\rangle	&=  \left\langle T^{0}_{0} (x,x')\right\rangle	 
\notag\\
	&=-\partial^{0}\partial_{0'}\left( W_{HH^{*}}- W_{KB^{*}}\right)(x,x')-\left(\partial^{\beta}\partial_{\beta'}-m^{2}\right)\left( W_{HH^{*}}- W_{KB^{*}}\right)(x,x').
\label{eqn400}
\end{align}
\end{widetext}
Where, for \eqref{eqn400}, the first term on the right is determined by \eqref{eqn375}  and is proportional to the frequency $\omega$, while the second term is proportional to $1/\omega$ determined by \eqref{eqn396}. Consequently, in limit of high frequencies, the second term can be disregarded, in other words
\begin{equation}
\left\langle T^{0}_{0} (x,x')\right\rangle	=\sigma(r)=-\int^{\infty}_{0}\frac{E}{e^{E/T(r)}-1}pdp.
\label{eqn401}
\end{equation}
With the condition that $\Delta|_{r_{+}}\approx cte$, allowing us to simplify \eqref{eqn401} even more as

\begin{equation}
\sigma(r)=-\int^{\infty}_{0}\frac{E}{e^{E/T(r)}-1}\frac{2\pi p\,dp}{h^{2}}.
\label{eqn401a}
\end{equation}

Supposing an ideal gas model and $E=pv$

\begin{align}
    P	&=\frac{1}{2}\sigma (r)  
\notag\\
	&=\frac{1}{2}\int^{\infty}_{0}\frac{E}{e^{E/T(r)}-1}\frac{2\pi p\,dp}{h^{2}}
&\hspace{0.3cm}
\notag\\
	&=\frac{1}{2}\int^{\infty}_{0}\frac{pv}{e^{E/T(r)}-1}\frac{2\pi p\,dp}{h^{2}}.
&\hspace{0.3cm}
\label{eqn402}
\end{align}

\section{THERMODYNAMIC ANALYSIS OF A SCALAR FIELD IN BTZ}\label{sec8}
Let partition function of a scalar field in BTZ per mode, as

\begin{equation}
Z_{\Omega}=\sum^{\infty}_{n=0}e^{-\beta n\omega}=\frac{1}{1-e^{\beta \omega}},
\label{eqn403}
\end{equation}
where $\Omega=\omega,\mathfrak{m}$ and also

\begin{equation}
\underline{n}=\left\{n_{\omega}\forall \Omega, \omega>0\right\}.
\label{eqn404}
\end{equation}
Thus, the energy of the field is

\begin{equation}
E_{\underline{n}}=\sum_{\Omega}n_{\Omega}\omega.
\label{eqn405}
\end{equation}
From the foregoing, the partition function for all possible modes near the gravitational radius  $r_{+}$

\begin{align}
    	Z&= \sum_{\underline{n}}e^{-\beta E_{\underline{n}}}  
\notag\\
	&=\prod_{\Omega,\omega>0}Z_{\Omega}
&\hspace{0.3cm}
\notag\\
	&=\prod_{\Omega,\omega>0}\sum^{\infty}_{n=0}e^{-\beta n\omega}
		&\hspace{0.3cm}
\notag\\
	&=\prod_{\Omega,\omega>0}\frac{1}{1-e^{\beta \omega}}.
\label{eqn406}
\end{align}
On the other hand, the entropy is determined as \cite{susskind2005introduction}

\begin{equation}
S=\beta\left\langle H\right\rangle+\ln\left|Z\right|,
\label{eqn407}
\end{equation}
where 
\begin{equation}
\left\langle H\right\rangle=-\frac{\partial}{\partial \beta}\ln\left|Z\right|.
\label{eqn408}
\end{equation}
so
\begin{equation}
\sum_{\Omega}f(\omega) = \int^{\infty}_{0}N(\omega)f(\omega)d\omega .
\label{eqn409}
\end{equation}
Therefore, for the partition function,

\begin{align}
 \ln\left|Z\right| 	&=\sum_{\Omega}\ln\left|Z_{\Omega}\right|
\notag\\
	&=\int^{\infty}_{0}N(\omega)f(\omega)d\omega 
&\hspace{0.3cm}
\notag\\
	&=\sum_{\Omega}f(\omega),
	&\hspace{0.3cm}
\label{eqn410}
\end{align}
where
\begin{equation}
f(\omega)=\ln\left|\frac{1}{1-e^{-\beta\omega}}\right|,
\label{eqn411}
\end{equation}
also, $N(\omega)d\omega$ corresponds to the number of modes $\psi_{\Omega}(\underline{x})$ are in the range $\omega$  and $\omega+d\omega$for $\Omega=\omega,\mathfrak{m}$  and $S$ is associated with the simplified density matrix. Considering \eqref{eqn406}.

\begin{equation}
\ln\left|Z\right|=\sum_{\eta=\pm}\sum_{\mathfrak{m}}\mathfrak{m}\ln\left|\frac{1}{1-e^{-\beta\omega}}\right|.
\label{eqn412}
\end{equation}
The wavenumber is required to be real \eqref{eqn356}

\begin{equation}
\mathbf{T^{*}}^{2}(r,\omega,\mathfrak{m})\geq 0.
\label{eqn413}
\end{equation}

This is under the WKB \eqref{eqn357} approach

\begin{equation}
\psi_{\Omega}(r)=\frac{1}{\sqrt[4]{4\omega^{2}\mathbf{T^{*}}^{2}}}\sin\left[{\int^{r_{\mathfrak{m},\omega}}_{r_{+}+\epsilon}\mathbf{T^{*}}(r,\omega,\mathfrak{m})dr}\right].
\label{eqn414}
\end{equation} 
Which implies that the field modes cancel each other out near the gravitational radius $r_{+}$  when Neumann-Dirichlet boundary conditions are considered

\begin{equation}
\sin\left[{\int^{r_{\mathfrak{m},\omega}}_{r_{+}+\epsilon}\mathbf{T^{*}}(r,\omega,\mathfrak{m})dr}\right]=0
\label{eqn414a}
\end{equation}
\begin{equation}
\int^{r_{\mathfrak{m},\omega}}_{r_{+}+\epsilon}\mathbf{T^{*}}(r,\omega,\mathfrak{m})dr=n\pi,
\label{eqn415}
\end{equation}
such that
\begin{equation}
\mathbf{T^{*2}}(r,\omega,\mathfrak{m})=0,\,\,\,\mathbf{T^{*2}}(r,\omega,\mathfrak{m})\geq 0\longrightarrow r'<r.
\label{eqn416}
\end{equation}
\begin{equation}
r=r_{+}+\epsilon,\,\,r=R\longrightarrow \omega\geq 0,\,\,m\geq 0.
\label{eqn416a}
\end{equation}
where $T^{*2}(r,\omega,\mathfrak{m})$ is defined by \eqref{eqn356}. Taking \eqref{eqn415}

\begin{equation}
n(\omega,\mathfrak{m})=\frac{1}{\pi}\int^{r_{\mathfrak{m},\omega}}_{r_{+}+\epsilon}\mathbf{T^{*}}(r,\omega,\mathfrak{m})dr.
\label{eqn417}
\end{equation}
Considering the variation $\frac{\partial n(\omega,\mathfrak{m})}{\partial \omega}$

\begin{widetext} 
\begin{equation}
\frac{\partial n(\omega,\mathfrak{m})}{\partial \omega}=\frac{1}{\pi}\int^{r_{\mathfrak{m},\omega}}_{r_{+}+\epsilon} \frac{\partial \mathbf{T^{*}}(r(\omega,\mathfrak{m});\omega,\mathfrak{m})}{\partial \omega}dr(\omega,\mathfrak{m})+\frac{1}{\pi}\int^{r_{\mathfrak{m},\omega}}_{r_{+}+\epsilon} \mathbf{T^{*}}(r(\omega,\mathfrak{m});\omega,\mathfrak{m})\frac{\partial}{\partial \omega}\left[dr(\omega,\mathfrak{m})\right],
\label{eqn418}
\end{equation}
\end{widetext}

where
\begin{equation}
dr(\omega,\mathfrak{m})\gg \frac{\partial}{\partial \omega}\left[dr(\omega,\mathfrak{m})\right]
\label{eqn419}
\end{equation}
Therefore, it follows that
\begin{equation}
\frac{\partial n(\omega,\mathfrak{m})}{\partial \omega}\approx \frac{1}{\pi}\int^{r_{\mathfrak{m},\omega}}_{r_{+}+\epsilon} \frac{\partial \mathbf{T^{*}}(r(\omega,\mathfrak{m});\omega,\mathfrak{m})}{\partial \omega}dr(\omega,\mathfrak{m}).
\label{eqn420}
\end{equation}
On the other hand, in the continuum limit for \eqref{eqn412}

\begin{widetext} 
\begin{align}
    \ln\left|Z\right|	&=\int_{n} dn \int_{\mathfrak{m}}d\mathfrak{m}\,\,\mathfrak{m}\,\,\ln\left|\frac{1}{1-e^{-\beta\omega}}\right|   
\notag\\
	&=\int_{n} \frac{dn}{d\omega}d\omega \int_{\mathfrak{m}}d\mathfrak{m}\,\,\mathfrak{m}\,\,\ln\left|\frac{1}{1-e^{-\beta\omega}}\right| 
\notag\\
&\hspace{0.3cm}
\notag\\
	&=\int_{\omega} d\omega \left[ \frac{1}{\pi}\int^{r_{\mathfrak{m},\omega}}_{r_{+}+\epsilon} \frac{\partial \mathbf{T^{*}}(r(\omega,\mathfrak{m});\omega,\mathfrak{m})}{\partial \omega}dr(\omega,\mathfrak{m})\right]\int_{\mathfrak{m}}d\mathfrak{m}\,\,\mathfrak{m}\,\,\ln\left|\frac{1}{1-e^{-\beta\omega}}\right|
&\hspace{0.3cm}
\notag\\
	&=\frac{1}{\pi}\int_{\omega} \int^{r_{\mathfrak{m},\omega}}_{r_{+}+\epsilon} \int_{\mathfrak{m}}\,\,d\omega d\mathfrak{m}dr(\omega,\mathfrak{m})\mathfrak{m}\,\,\ln\left|\frac{1}{1-e^{-\beta\omega}}\right|	
	\frac{\partial \mathbf{T^{*}}(r(\omega,\mathfrak{m});\omega,\mathfrak{m})}{\partial \omega}.
\label{eqn421}
\end{align}
\end{widetext}
Where conditions \eqref{eqn416} and \eqref{eqn416a} are met. Integrating by parts \eqref{eqn421} 

\begin{align}
     \ln\left|Z\right|&=   	\frac{1}{\pi}\int^{r_{\mathfrak{m},\omega}}_{r_{+}+\epsilon} dr \int_{\mathfrak{m}} d\mathfrak{m}\,\,\mathfrak{m}\,\mathbf{T^{*}}\,\ln\left|\frac{1}{1-e^{-\beta\omega}}\right|
\notag\\
	&+	\frac{1}{\pi}\int^{r_{\mathfrak{m},\omega}}_{r_{+}+\epsilon} dr \int_{\mathfrak{m}} d\mathfrak{m}\,\,\mathfrak{m}\int_{\omega}\frac{\beta  \mathbf{T^{*}}}{e^{\beta\omega}-1}d\omega.
\label{eqn422}
\end{align}

Regarding \eqref{eqn422}, the following observations are made
\begin{enumerate}
	\item The partition function contains two contributions: one proportional to the perimeter
	\begin{equation}
\int^{r_{\mathfrak{m},\omega}}_{r_{+}+\epsilon} dr \int_{\mathfrak{m}} d\mathfrak{m}\,\,\mathfrak{m}\,\mathbf{T^{*}}\,\ln\left|\frac{1}{1-e^{-\beta\omega}}\right|.
\label{eqn423}
\end{equation}
	\item And another to the area over the phase space
	\begin{equation}
\int^{r_{\mathfrak{m},\omega}}_{r_{+}+\epsilon} dr \int_{\mathfrak{m}} d\mathfrak{m}\,\,\mathfrak{m}\int_{\omega}\frac{\beta  \mathbf{T^{*}}}{e^{\beta\omega}-1}d\omega.
\label{eqn424}
\end{equation}
	\end{enumerate}
	With the condition that the area contribution is much larger than the perimeter contribution in the partition function \eqref{eqn422}, it follows that
	
	\begin{widetext} 
	\begin{equation}
\int^{r_{\mathfrak{m},\omega}}_{r_{+}+\epsilon} dr \int_{\mathfrak{m}} d\mathfrak{m}\,\,\mathfrak{m}\int_{\omega}\frac{\beta  \mathbf{T^{*}}}{e^{\beta\omega}-1}d\omega \gg \int^{r_{\mathfrak{m},\omega}}_{r_{+}+\epsilon} dr \int_{\mathfrak{m}} d\mathfrak{m}\,\,\mathfrak{m}\,\mathbf{T^{*}}\,\ln\left|\frac{1}{1-e^{-\beta\omega}}\right|.
\label{eqn425}
\end{equation}
\end{widetext}

Which makes it possible to approximate \eqref{eqn422}

\begin{equation}
 \ln\left|Z\right|\approx\frac{1}{\pi}\int^{r_{\mathfrak{m},\omega}}_{r_{+}+\epsilon} dr \int_{\mathfrak{m}} d\mathfrak{m}\,\,\mathfrak{m}\int_{\omega}\frac{\beta  \mathbf{T^{*}}}{e^{\beta\omega}-1}d\omega.
\label{eqn426}
\end{equation}
According to \eqref{eqn356}
\begin{equation}
\frac{1}{f(r)}\left\{\frac{\omega^{2}}{f(r)}-m^{2}-\frac{\mathfrak{m}^{2}}{r}+B\right\}=0
\label{eqn427}
\end{equation}
\begin{equation}
\frac{\mathfrak{m}^{2}}{r}=\frac{L}{r}=\frac{\omega^{2}}{f(r)}-m^{2}+B,
\label{eqn428}
\end{equation}

thus, for \eqref{eqn428}, it is possible to define $L=\mathfrak{m}$ and $L_{max}=\mathfrak{m}_{max}$. In other words,

\begin{equation}
\frac{L_{max}}{r}=\frac{\omega^{2}}{f(r)}-m^{2}+B.
\label{eqn429}
\end{equation}

Therefore,
\begin{equation}
\mathfrak{m}=\pm\sqrt{L},\,\,\,d\mathfrak{m}=\frac{dL}{2\sqrt{L}}.
\label{eqn430}
\end{equation}

then, the integral with respect to m$\mathfrak{m}$in \eqref{eqn426}

\begin{align}
    \int_{\mathfrak{m}} d\mathfrak{m}\,\,\mathfrak{m}\mathbf{T^{*}}	&= \frac{1}{2}  \int_{L} dL\,\,\mathbf{T^{*}}
\notag\\
	&=\frac{1}{2\sqrt{rf(r)}}\int^{L_{max}}_{0}dL\sqrt{L_{max}-L}
&\hspace{0.3cm}
\notag\\
	&=\frac{1}{3\sqrt{rf(r)}}L^{3/2}_{max}.
&\hspace{0.3cm}
\label{eqn431}
\end{align}
Inserting \eqref{eqn431} in \eqref{eqn426}

\begin{align}
    	\ln\left|Z\right|&=  \frac{1}{\pi}\int^{\infty}_{0}d\omega\int^{R}_{r_{+}+\epsilon}dr\left[\frac{\beta}{e^{\beta\omega}-1}\right]\frac{1}{3\sqrt{rf(r)}}L^{3/2}_{max} 
\notag\\
	&=\int^{\infty}_{0}\frac{\beta}{e^{\beta\omega}-1}d\omega\,\,\left[\frac{1}{3\pi}\int^{R}_{r_{+}+\epsilon}dr'\frac{1}{\sqrt{rf(r)}}L^{3/2}_{max}\right].
	\notag\\
	&=\int^{\infty}_{0}\frac{\beta N(\omega)}{e^{\beta\omega}-1}d\omega
	&\hspace{0.3cm}
\label{eqn432}
\end{align}
Taking \eqref{eqn432}  and considering \eqref{eqn429}

\begin{align}
  N(\omega)&=\frac{1}{3\pi}\int^{R}_{r_{+}+\epsilon}dr\frac{1}{\sqrt{rf(r)}}L^{3/2}_{max}  
\notag\\
	&=\frac{1}{3\pi}\int^{R}_{r_{+}+\epsilon}dr\frac{1}{\sqrt{rf(r)}}r^{3/2} \left[\frac{\omega^{2}}{f(r)}-m^{2}+B\right]^{3/2}
&\hspace{0.3cm}
\notag\\
	&=\frac{1}{3\pi}\int^{R}_{r_{+}+\epsilon}dr\frac{1}{\sqrt{rf(r)}}r^{3/2} p^{3}.
&\hspace{0.3cm}
\label{eqn434}
\end{align}
where it possible to define
\begin{align}
    	 N(\omega)&=  \int^{R}_{r_{+}+\epsilon}dr\frac{1}{3\pi}\frac{1}{\sqrt{rf(r)}}r^{3/2} \left[\frac{\omega^{2}}{f(r)}-m^{2}+B\right]^{3/2} 
\notag\\
	&=\int^{R}_{r_{+}+\epsilon}dr N^{*}(\omega)
\label{eqn434w}
\end{align}

\begin{figure}[H]
\centering
		\includegraphics[width=0.45\textwidth]{ 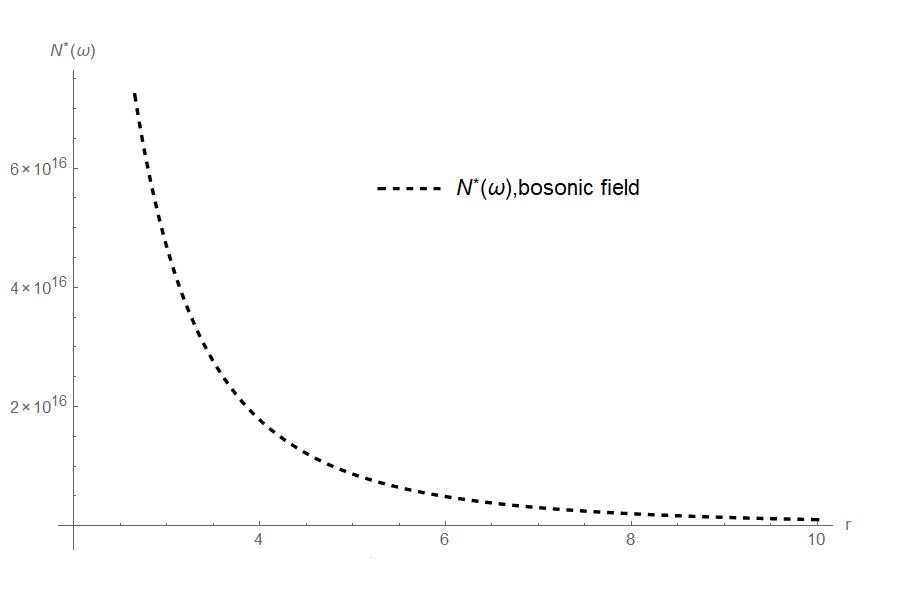}
\caption{Occupation number $N^{*}(\omega)$ for a BTZ spacetime.}
\label{Fig9b}
\end{figure}
On the other hand, the Helmholtz free energy $F$, is related to the partition function $Z$ as
\begin{align}
    	F&=-\frac{1}{\beta}  \ln\left|Z\right| 
\notag\\
	&=-\int^{\infty}_{0}\frac{N(\omega)}{e^{\beta\omega}-1}d\omega.
&\hspace{0.3cm}
\label{eqn433}
\end{align}

In that same direction, the internal energy of the scalar field is
\begin{align}
    U	&=   -\frac{\partial}{\partial \beta}\ln\left|Z\right|
\notag\\
	&= -\frac{\partial}{\partial \beta}\int^{\infty}_{0}\frac{N(\omega)}{e^{\beta\omega}-1}d\omega
&\hspace{0.3cm}
\notag\\
	&= -\int^{\infty}_{0}d\omega N(\omega)\frac{\partial}{\partial \beta}\left[\frac{\beta}{e^{\beta\omega}-1}\right]
&\hspace{0.3cm}
\notag\\
	&=-\int^{\infty}_{0}d\omega N(\omega)\left[-\omega\frac{\partial N(\omega)}{\partial\omega}\right]
	&\hspace{0.3cm}
\notag\\
	&=\int^{\infty}_{0}d\omega \frac{\omega}{e^{\beta\omega}-1}\frac{\partial N(\omega)}{\partial\omega}
		&\hspace{0.3cm}
\notag\\
	&=\int^{\infty}_{0}d\omega \frac{\omega}{e^{\beta\omega}-1}N'(\omega).
\label{eqn435}
\end{align}
Where $N'(\omega)=\frac{\partial N(\omega)}{\partial\omega}$ has been defined. According to \eqref{eqn434}
\begin{align}
    \frac{\partial N(\omega)}{\partial\omega}	&=   \frac{\partial}{\partial\omega} \left[  \frac{1}{3\pi}\int^{R}_{r_{+}+\epsilon}dr\frac{1}{\sqrt{rf(r)}}r^{3/2} \left[\frac{\omega^{2}}{f(r)}-m^{2}+B\right]^{3/2} \right]
\notag\\
	&= \frac{1}{\pi}\int^{R}_{r_{+}+\epsilon}dr\frac{1}{\sqrt{rf(r)}}\frac{r\omega}{f(r)}\sqrt{r\left(\frac{\omega^{2}}{f(r)}-m^{2}+B\right)}.
\label{eqn436}
\end{align}
Inserting \eqref{eqn436} in \eqref{eqn435}
\begin{widetext} 
\begin{align}
    U	&=  \int^{\infty}_{0}d\omega \frac{\omega}{e^{\beta\omega}-1} \frac{1}{\pi}\int^{R}_{r_{+}+\epsilon}dr\frac{1}{\sqrt{rf(r)}}\frac{r\omega}{f(r)}\sqrt{r\left(\frac{\omega^{2}}{f(r)}-m^{2}+B\right)}
\notag\\
	&=\int^{R}_{r_{+}+\epsilon}\frac{2\pi r}{\sqrt{f(r)}}dr\int^{\infty}_{0}\frac{E}{e^{E/T(r)}-1}\frac{2\pi p dp}{h^{2}}
&\hspace{0.3cm}
\notag\\
	&= \int d\mathcal{A}\int^{\infty}_{0}\frac{E}{e^{E/T(r)}-1}\frac{2\pi p dp}{h^{2}}
&\hspace{0.3cm}
\notag\\
	&= \int d\mathcal{A}\,\, \sigma(r)
\label{eqn437}
\end{align}
\end{widetext}
where $\sigma(r)$  is defined by \eqref{eqn401}. The entropy $S$ of the field is obtained as
\begin{widetext} 
\begin{align}
    	S&=\beta[U-F]   
\notag\\
	&=\beta\int^{\infty}_{0}\frac{d\omega}{e^{\beta\omega}-1}[\omega N'(\omega)+N(\omega)]
&\hspace{0.3cm}
\notag\\
	&=\beta\int^{\infty}_{0}\frac{d\omega}{e^{\beta\omega}-1}\frac{\partial}{\partial\omega}[\omega N(\omega)]
	&\hspace{0.3cm}
	\notag\\
	&= \int^{R}_{r_{+}+\epsilon}\frac{2\pi r dr}{\sqrt{f(r)}}\left[\frac{3\beta}{2}\int^{\infty}_{0}\frac{E}{e^{E/T(r)}-1}\frac{2\pi pdp}{h^{2}}\,\, \mathbb{H} \right]
	&\hspace{0.3cm}
	\notag\\
	&=\int^{R}_{r_{+}+\epsilon} d\mathcal{A}\,\, s(r)\,\,\mathbb{H},
\label{eqn438}
\end{align}
\end{widetext}
where 
\begin{equation}
\int^{R}_{r_{+}+\epsilon} d\mathcal{A}=\int^{R}_{r_{+}+\epsilon}\frac{2\pi r dr}{\sqrt{f(r)}}
\label{eqn439}
\end{equation}
is the area integral in BTZ spacetime. In addition,
\begin{equation}
s(r)=\frac{3\beta}{2}\int^{\infty}_{0}\frac{E}{e^{E/T(r)}-1}\frac{2\pi pdp}{h^{2}}. 
\label{eqn440}
\end{equation}
is the field entropy density
\begin{equation}
\mathbb{H}=\frac{2p^{2}}{9\pi\omega(e^{\beta\omega}-1)},
\label{eqn441}
\end{equation}
and is the coupling factor. From the foregoing, it is possible to obtain
\begin{equation}
s(r)=\beta[\sigma(r)+P].
\label{eqn442}
\end{equation}
Let $x=\frac{E}{T}$, then \eqref{eqn440}
\begin{align}
    s(r)	&= \frac{3}{2}T^{3}(r)  \frac{2\pi}{h^{2}}  \int^{\infty}_{0}\frac{x^{2}}{e^{x}-1}dx.
\notag\\
	&=\frac{6\pi\zeta[3]}{h^{2}}T^{2}(r).
\label{eqn443}
\end{align}
where $\zeta[3]$ corresponds to the Riemann Zeta function. Taking \eqref{eqn438} and considering \eqref{eqn210} and \eqref{eqn211} 
\begin{align}
  S&=\int d\mathcal{A}\,\, s(r)
\notag\\
	&=\int\frac{2\pi rdr}{\sqrt{f(r)}}\frac{6\pi\zeta[3]}{h^{2}}T^{2}(r).
	&\hspace{0.3cm}
	\notag\\
	&=\frac{6\pi\zeta[3]}{h^{2}}T^{2}_{\infty}\mathcal{P}\int^{r_{+}+\epsilon}_{r_{+}}\frac{dr}{f(r)^{3/2}},
\label{eqn444}
\end{align}
where $r_{+}=\sqrt{Ml^{2}}$ and $\mathcal{P}=2\pi\sqrt{Ml^{2}}$. Where the metric factor $f(r)$  can be expressed in terms of the surface gravity \eqref{eqn276}
\begin{equation}
f(r)\approx  f'(r_{+})(r-r_{+}),\,\,\,\epsilon=r-r_{+}.
\label{eqn445}
\end{equation}
On the other hand, let the proper distance above the horizon $\alpha$ be
\begin{align}
    \alpha	&=\int^{r_{1}}_{r_{+}}\frac{dr}{\sqrt{f(r)}}   
\notag\\
	&=\int^{r_{1}}_{r_{+}}\frac{dr}{ \sqrt{f'(r_{+})(r-r_{+})}}
&\hspace{0.3cm}
\notag\\
	&=\frac{2}{\sqrt{f'(r_{+})}}\sqrt{r_{1}-r_{+}}
	\notag\\
	&=\frac{2}{\sqrt{f'(r_{+})}}\sqrt{\epsilon},
&\hspace{0.3cm}
\label{eqn446}
\end{align}
which makes it possible to establish \cite{2020Symm...12.2072R, RojasC:2020qnz} 
\begin{equation}
\epsilon=\frac{1}{2}\kappa_{0}\alpha^{2}.
\label{eqn447}
\end{equation}
This makes it possible to evaluate \eqref{eqn444}
\begin{align}
  S&=\frac{6\pi\zeta[3]}{h^{2}}T^{2}_{\infty}\mathcal{P}\int^{r_{+}+\epsilon}_{r_{+}}\frac{dr}{f(r)^{3/2}}
\notag\\
	&=\frac{6\pi\zeta[3]}{h^{2}}\left[\frac{T_{\infty}}{\kappa_{0}/2\pi}\right]^{2}\frac{\mathcal{P}}{4\pi^{2}}\frac{\gamma}{\alpha}
	&\hspace{0.3cm}
	\notag\\
	&=\frac{3\zeta[3]}{2h^{2}\pi}\gamma\left[\frac{T_{\infty}}{\kappa_{0}/2\pi}\right]^{2}\frac{\mathcal{P}}{\alpha},
\label{eqn448}
\end{align}
 where $\gamma=\sqrt{2}(\sqrt{2}-1)$.
 In the case where $S_{\mbox{Ent}}=S_{BH}$, implies that $T_{\infty}=T_{H}$, makes it possible to obtain
\begin{align}
     \alpha 	&=   \frac{3 \zeta[3]\, \gamma \,\mathcal{P} }{4\pi h^{2}} \sqrt{\frac{2G_{3}}{l^{2}M} }
\notag\\
	&= \frac{3 \zeta[3] \,\gamma}{2 \pi h^{2}} \sqrt{2G_{3}}.
\label{eqn449}
\end{align}
Then,
\begin{equation}
S_{BH}=2\pi\sqrt{\frac{Ml^{2}}{2G_{3}}}.
\label{eqn450}
\end{equation}

It is possible, from \eqref{eqn448}  to calculate the specific heat of BTZ as \cite{Rojas:2011ee}
\begin{align}
    C_{V}	&=C_{P}  
\notag\\
	&=T_{\infty} \left(\frac{\partial S}{\partial T_{\infty}}\right)_{P}
&\hspace{0.3cm}
\notag\\
	&= \frac{3\zeta[3]}{h^{2}\pi}\gamma\left[\frac{T_{\infty}}{\kappa_{0}/2\pi}\right]^{2}\frac{\mathcal{P}}{\alpha}
&\hspace{0.3cm}
\label{eqn451}
\end{align}
\section{DISCUSSIONS AND CONCLUSIONS} \label{sec9}

A thermal source of the Bekenstein- Hawking entropy $S_{BH}$  for a BTZ black hole was modeled as an entropy of entanglement of a real massive scalar field. Entropy associated with a thin spherical dust shell, collapsing gravitationally from a specific distance to a radius slightly larger than the gravitational radius, in an $AdS_{3}$ space-time, and described by a FIDO observer.

In sections \ref{sec2} and  \ref{sec3} describe the equation of motion of the contracting shell in the BTZ space-time, measured by an observer comoving to the shell in terms of the proper time $\tau$. Then, we obtained the equation of motion of the shell in the same spacetime, expressed by equation \eqref{eqn156}, in terms of the coordinate time $t$, according to the measurement of a FIDO observer.

Sections \ref{sec5}, \ref{sec6} and \ref{sec7} the quantization of a real massive scalar field over the BTZ spacetime was  developed. Specifically, in a Thermo Field Dynamics context, we calculated the expected value for the component $T_{00}(x,x')$ of the momentum-energy tensor with respect to the Boulware and Hartle-Hawking vacuum states, for the mentioned scalar field. Thus, we show in expression \eqref{eqn401a}, under the WKB approximation, that the thermal entropy is strongly located near the outer surface of the shell, at the collapse limit, according to the FIDO observer. In these terms, we obtained a well-defined energy density, of a hot scalar field.

Section \ref{sec8}  performs a thermodynamic analysis of the thermal environment found in the previous sections. In greater detail, we determined from the partition function corresponding to the hot scalar field, the occupation number $N(\omega)$ in proximity to the outer surface of the shell, which, according to the FIDO observer, is very close to the corresponding event horizon. Then we calculated the internal energy U and the entropy density $s(r)$, expressed by equations \eqref{eqn437}  and \eqref{eqn443} .

At the end of \ref{sec8} , the entropy of entanglement was calculated, resorting to the cutoff introduced by 't Hooft  in his wall model, and corrected by Mukohyama and Israel. This modified model involved a reinterpretation of the vacuum state and of the described object. To an outside observer, the object described is star-like and not a black hole. However, the external observer cannot distinguish between one and the other.

With the previously described collapsing shell model, in agreement with the FIDO observer, a gap is established between the inner surface of the shell and the event horizon, which is identified with $\epsilon$, related to the proper distance above the horizon $\alpha$, in the modified wall model. Thus, the 't Hooft wall coincides with the shell in proximity to the event horizon. The possibility of calculating a microscopic parameter with macroscopic criteria is noteworthy. Thus, we calculated the finite entropy of entanglement given by equation \eqref{eqn448}.

What we show with the gravitational collapse shell model is that the thermal source of the Bekenstein-Hawking entropy, in the Mukohyama-Israel two-source model, based on a complementarity principle \cite{Mukohyama:1998rf}, is external and fully corresponds to the thermal entropy of entanglement.

Finally, we obtain an expression for the Bekenstein-Hawking entropy as a function of the interpretation of the entropy of entanglement, for the case of one species.

The results obtained in this research complement the progress of the study of the Bekenstein-Hawking entropy in the context of asymptotic symmetries, reinforcing the considerations made by Fursaev \cite{Fursaev:2004qz}. An interesting common geometric background between the model developed in this article and the modeling of asymptotic symmetries is the eternal black hole in $AdS_{3}$ \cite{Maldacena:2001kr}. In this background, we present two types of properties related to the thermal character of the Bekenstein-Hawking entropy, intimately associated with two natural boundaries, near the horizon and the asymptotic region when  $r\longrightarrow\infty$. The near-horizon boundary has the same properties of an asymptotically flat static spacetime, so in effective terms, it is possible to consider the Boulware and Hartle-Hawking vacuum states and obtain an entropy related to the Bekenstein-Hawking entropy for the thermal source. The other boundary substantiates the calculations with the $AdS/CFT$ duality, resorting to the metric’s Euclidean properties. The calculation of the Bekenstein-Hawking entropy with this formalism is related to the one presented in this article, resorting to the complementary source of Mukohyama-Israel, based on the Gibbons-Hawking formalism \cite{Mukohyama:1998rf}.


\end{document}